%% file: main.tex
\documentclass[english]{article}
\usepackage[letterpaper,margin=1.25in]{geometry}
\RequirePackage[OT1]{fontenc}
\RequirePackage{graphicx}
\RequirePackage{amsthm}
\RequirePackage{amsmath}
\RequirePackage{yhmath}
\RequirePackage[authoryear]{natbib}
\RequirePackage[colorlinks,citecolor=blue,urlcolor=blue]{hyperref}
\usepackage{algorithm,algorithmic}
\usepackage{amssymb}
\usepackage{footnote}
\usepackage{subfig}
\usepackage{array}
\usepackage{setspace}
\usepackage{comment}
\usepackage{tabularx}
\theoremstyle{plain}

\numberwithin{equation}{section}
\numberwithin{figure}{section}

\usepackage[utf8]{inputenc}
\usepackage[OT1]{fontenc}
\RequirePackage{amsthm}
\RequirePackage{amsmath}
\usepackage{amssymb}
\usepackage{babel}
\usepackage[page,title]{appendix}
\usepackage{comment}
\usepackage{float}
\usepackage{setspace}
\usepackage{lscape} 
\usepackage{xcolor}
\PassOptionsToPackage{normalem}{ulem}
\usepackage{ulem}

\renewcommand{\color}[2][]{\ignorespaces}

\begin{document}
\title{Linear Estimation of Structural and Causal Effects for Nonseparable Panel Data}

    \author{Victor Chernozhukov (MIT),  Ben Deaner (UCL), Ying Gao (UBC)\\ 
    Jerry Hausman (MIT and NBER), Whitney K. Newey (MIT and NBER)\footnote{The present paper formed the basis of the Fisher–Schultz Lecture given by Whitney Newey at the 2023 European Meeting of the Econometric Society in Barcelona. 
This research was supported by NSF Grants 1757140 and 224247. 
Helpful comments were provided by the editor, two referees, I. Fernandez-Val, B. Graham, G. Imbens, and R. Matzkin. 
A replication file including software and documentation is posted on the Econometrica website }
    }

\maketitle

\begin{abstract}
This paper develops linear estimators for structural and causal parameters of   nonseparable models using panel data. These models incorporate unobserved, time-varying, individual heterogeneity, which may be correlated with the regressors.  
Estimation is based on an approximation of a conditional average potential outcome by a linear sieve specification with individual-specific parameters. Effects of interest are estimated by a bias corrected average of individual ridge regressions. 
We demonstrate how this approach can be applied to estimate causal effects, counterfactual consumer welfare, and averages of individual taxable income elasticities. 
We show that the proposed estimator has an empirical Bayes interpretation and possesses a number of other useful properties. We formulate Large-$T$ asymptotics that can accommodate discrete regressors and which bypass partial identification in this case.
We employ the methods to estimate average equivalent variation and deadweight loss for potential price increases using data on grocery purchases.
\end{abstract}

\input{Introduction}

\input{estimation_reorg}

\input{Application_to_demand_estimation}
\input{Asymptoticsetc}

\bibliographystyle{authordate1}
\bibliography{panel_cites}

\appendix

\input{tables_final}

\input{derivative_discussion}
\input{binarychoice}

\input{newproofs}

\end{document}

%% file: introduction.tex
   \section{Introduction}

    Panel data provide a valuable means of identifying and estimating economic effects when there is dependence between variables of interest and unobservable heterogeneity. 
    Specifically, in ``fixed effects''-type structural economic models,  heterogeneity like tastes or technology is time-invariant, whereas  variables of interest change over time. One can exploit this time-variation to identify and estimate the effects of those variables. Similarly, in causal models, panel data can be used to identify and estimate counterfactual effects of interest when treatment varies over time and unobserved confounders do not. 
    
    This paper develops linear methods for identifying and estimating  structural economic parameters or treatment effects for  nonseparable models in panel data.  More precisely, we consider outcome models that are not additively separable in observables and unobservables. 
    These provide very general specifications for economic and causal analysis, with heterogeneity representing tastes and/or technology for economic models or counterfactual outcomes in causal models.

We consider a moderate or large number of time periods or assume that the outcome is a random linear combination of known functions of regressors, a model referred to here as linear random coefficients (LRC). 
	 The basic identifying assumption is a ``fixed effects" condition, referred to henceforth as `time-homogeneity'. This assumption states that the  distribution of heterogeneity in each time period conditional on observed regressors, does not depend on the time period.  This same condition is employed in \cite{Manski1987} for linear index binary choice models, and \cite{Chernozhukov2013} for nonseparable models. The condition  allows for the inclusion of time-effects like secular trends.

    To use linear methods, we approximate a nonseparable expected potential outcome that is a smooth function of variables of interest by the expected potential outcome for a LRC model. 
    The approximation is accomplished by choosing the regressors of the LRC to be a sieve for approximating the average potential outcome as a function of counterfactual values of variables of interest.  
    Least squares over time for each individual is then used to estimate individual specific coefficients for the approximating LRC model. 
   We regularize using ridge regression for each individual, thus allowing the individual coefficients to be weakly or not identified. 
    We bias correct, for ridge regularization, the estimators of effects of interest.

    The effects of interest we consider are averages of individual-specific effects. 
    Examples of these effects given here are average policy or treatment effects, average equivalent variation and deadweight loss for demand, and average tax effects for nonlinear budget sets. 
    We estimate any effect of interest as a bias corrected average of individual-specific linear combinations of the individual ridge regression coefficients.
    For the bias-corrected estimator: A) We derive an empirical Bayes interpretation of the bias corrected ridge estimator in the LRC model and show that B) as the ridge penalty goes to infinity the estimator approaches the fixed effects estimator that imposes that slope coefficients, i.e. coefficients of non-constant regressors, are constant across individuals; C) in the LRC model the estimator is unbiased 
    when true slope coefficients do not vary over individuals; D) in the LRC model,  
    when all individuals have a nonsingular second moment matrix of regressors, i.e., all individual-specific coefficients are identified, the bias of the estimator  goes to zero as the ridge penalty shrinks to zero.

    We a ways of quantifying the extent of regularization across individuals of any effect of interest.
    We compare the chosen linear combination coefficients, corresponding to the effect of interest, with regularized coefficients that come from the debiased ridge estimation.
    We measure the extent of regularization using the distribution across individuals of the Euclidean distance between the chosen linear  combinations and  regularized counterparts.
    We interpret this distance in terms of possible true values of the individual contribution to the object of interest.

   We apply our methods to estimate bounds on average equivalent variation and deadweight loss for consumer demand. 
   Here, panel data approximately controls for price endogeneity resulting from market equilibrium, when time-varying individual heterogeneity is independent of time-varying unobserved supply shocks, as discussed in Section 4. 
   The methods are applied to scanner data with the outcome variable specified as expenditure share and regressors being powers of the log of prices and total expenditure.
   Average equivalent variation and deadweight loss for price increases on soda or milk are estimated.
   
   We develop large sample theory where the number $T$ of time periods grows with the number of individuals and for the LRC model where $T$ is greater than the number of regressors. 
The number of regressors is allowed to grow with $T$ so that the approximation error for a general nonseparable model is small enough for accurate inference. 
   We also allow for nonidentifiability of effects of interest for any 
   $T$ but specify that the identified set shrinks at some power of $T$ as $T$ grows that is sufficient for asymptotic inference using the debiased ridge estimator of a parameter of interest. 
   Conditions for such shrinkage rates for the identified set are known from \cite{Chernozhukov2013}.

  \cite{Chamberlain1982}, \cite{Chamberlain1992}, \cite{Pesaran1995}, \cite{Wooldridge2005}, \cite{Arellano}, and \cite{Graham2012} have previously considered estimation of LRC models. 
  \cite{Manski1987}, \cite{Honore1992}, \cite{Abrevaya2000}, \cite{Chernozhukov2013}, \cite{Hoderlein2012}, \cite{Shi}, 
  \cite{Fernandez2021}, and \cite{Pakes}, have all considered estimation of nonseparable panel models under time-homogeneity. \cite{Altonji2005} considers identification via control functions, as does \cite{Semenova}, while also allowing for sparse, additive individual-specific effects. 
\cite{Bonhomme2015} and \cite{Bonhomme2022} model individual heterogeneity as discrete.
 We innovate in using series approximation of a smooth, nonparametric average  potential outcome, in the use of bias corrected ridge regularization for estimation of average effects, in providing methods for evaluating the extent of regularization across individuals, and in providing asymptotic theory  for our estimators of structural/causal effects for growing $T$ when there is only partial identification for fixed $T$.

 In Section 2, we describe the models and effects we consider. 
 Section 3 explains the sieve approximation, gives the bias corrected, average ridge estimator of structural and causal effects, and describes its properties. 
 Section 4 describes how the general model and methods can be applied to demand analysis.
 Section 5 gives an application to grocery demand data.
 Section 6 gives large sample theory.

%% file: estimation_reorg.tex
\section{Models and Parameters of Interest}

We consider panel data made up of observations across $n$ individuals, indexed by $i$, and  $T$ time periods, indexed by $t$. 
The individuals are drawn independently and identically from some population. 
Each observation consists of a scalar outcome variable $Y_{it}$ and a vector of variables of interest $X_{it}$ that can affect $Y_{it}$.

\color{blue}
We assume a non-separable structural/causal model for $Y_{it}$ of the form 
\begin{equation}
Y_{it}=g(X_{it},\eta_{it}),\text{ }(i=1,...,n;\text{ }t=1,...,T),\label{basemodel}
\end{equation}
where $\eta_{it}$ are individual- and time-specific unobserved variables representing preferences, technology, or  potential outcomes that may be correlated with  $X_{it}$. 
The $\eta_{it}$ may be serially correlated across $t$.
The model is nonseparable in the sense that it allows for general interactions between the observed variables $X_{it}$ and the unobserved heterogeneity $\eta_{it}$.

We specify that the function $g$ has a causal/structural interpretation, that $g(x,\eta_{it})$ is the counterfactual value of the outcome $Y_{it}$ in a counterfactual world in which the vaiables of interest take the value $x$. Thus, $Y_{it}(x):=g(x,\eta_{it})$ is a potential outcome.  

A simple and quite flexible nonseparable model is a linear random coefficients (LRC) model, where there is a known $J\times1$ vector $b(x)=(b_1(x),...,b_J(x))'$ of functions of $x$, with first component $b_1(x)=1$, such that  
\begin{equation}
Y_{it}=g(X_{it},\eta_{it})=b(X_{it})'\eta_{it}.
\end{equation}
Here a potential outcome is $b(x)'\eta_{it}$. Let $X_{i}=(X_{i1}^{\prime},...,X_{iT}^{\prime})^{\prime}$ be the full history of of variables of interest for individual $i$.
Also let $\beta_{it}=E[\eta_{it}|X_i]$ where we suppress dependence of $\beta_{it}$ on $X_i$ for notational convenience. 
Then 
\begin{equation}
E[Y_{it}|X_i]=b(X_{it})'\beta_{it}.
\end{equation}
Here we see that $\beta_{it}$ are regression coefficients so we could try to estimate them in order to estimate counterfactual objects like $b(x)'\beta_{it}$.
The problem is there is only one observation $Y_{it}$ available to do this, which is not enough data to identify $\beta_{it}$.

\color{black}

In this paper we use conditional time-homogeneity of $\eta_{it}$ to identify and estimate objects of interest.
	\theoremstyle{definition} \newtheorem*{A1}{Assumption 1 (Time-Homogeneity)} 
\begin{A1} 
	The distribution of $\eta_{it}$ 
	conditional on $X_{i}$ does not depend on $t$.
\end{A1}

\color {blue}

The identifying power of Assumption 1 for the LRC model is that $\beta_{it}=E[\eta_{it}|X_i]=E[\eta_{i1}|X_i]:=\beta_i$ does not vary over time, so that
\begin{equation*}
E[Y_{it}|X_i]=b(X_{it})'\beta_{i}, (t=1,...,T).
\end{equation*}
Here all $T$ observations observations are available to estimate each $\beta_i$ making identification of $\beta_i$ possible when $T$ is greater than or equal to the number $J$ of regressors in the LRC model.  
\color{black}

Assumption 1 allows for endogeneity where the
conditional distribution of $(\eta_{i1},...,\eta_{iT})'$ given $X_{i}$ may depend on $X_{i}$. 
Such endogeneity is present when $X_{it}$ includes choice or equilibrium values that are determined by the preferences or technology represented by $\eta_{it}.$
It is common to decompose $\eta_{it}$ into time constant  components $\alpha_{i}$ and time-varying components $v_{it}$.
The $\alpha_{i}$ trivially satisfies Assumption 1, so that Assumption 1 only restricts $v_{it}$.
Assumption 1 requires that $v_{it}$  have the same distribution in  each period conditional on the history of regressors $X_{i}$.
 In economic applications, it is always   important to accommodate time-varying preferences and technology as represented by $v_{it}$.  Models that do not incorporate time-varying unobservables may fail to fit the variation in outcomes over time.

Assumption 1 has been imposed in the many previous papers on nonseparable panel data that were referred to in the Introduction.
Assumption 1 is also satisfied in discrete choice, multinomial logit demand models such as those of \cite{Dubois2020}, where $v_{it}$ are logit disturbances in discrete choice demand models.
In demand models $v_{it}$ represents variation in preferences over time for a given individual. 
Such variation could be due to a taste for variety as further discussed in Section 4.
Assumption 1 imposes stability over time on such preferences or technology.
This condition seems important for the nonparametric model, where identifying structural/causal effects while allowing unrestricted correlation between $X_{it}$ and $\eta_{it}$ seems difficult without time homogeneity.
Assumption 1 does allow for systematic variation over time in the outcome $Y_{it}$ via the observable variables of interest $X_{it}$ whose magnitude varies with time. For example, $X_{it}$ may include time trends or seasonal indicators.
It may also be possible to allow for time period specific effects, as was recently done by \cite{Jin2026} in a related model, though that is beyond the scope of this paper.

\color{black}

A main focus and innovation of this paper is estimation of causal and structural effects for the nonparametric, nonseparable model  where the functional form of $g(X_{it},\eta_{it})$ is unknown, the dimension of $\eta_{it}$ is unspecified and may even be infinite, and Assumption 1 is satisfied.
This model does not impose any functional form assumption for the outcome function $g$ and is therefore very general. For example, it allows for discrete outcomes where $g(X_{it},\eta_{it})$ may depend on indicator functions for unknown functions of $X_{it}$ and $\eta_{it}$.

Our object (i.e., parameter) of interest is an average difference in linear combinations of potential outcomes with coefficients specified by the researcher. 
Let $X_{it}^{+}$ and $X_{it}^{-}$ be counterfactual values of variables of interest that can differ from the observed $X_{it}$.
Also let $H_{it}^{+}$ and $H_{it}^{-}$ be coefficients that are specified by the researcher. We estimate objects of the form
\begin{equation}
\theta_0=E\big[\frac{1}{T}\sum_{t=1}^{T}\big(H_{it}^{+}g(X_{it}^{+},\eta_{it})-H_{it}^{-}g(X_{it}^{-},\eta_{it})\big)\big]\label{objinterest}
\end{equation}
In addition, to objects of the form above, we discuss identification and estimation of partial effects in Appendix B.

We assume throughout that the following conditional independence assumption holds. 
\theoremstyle{definition} \newtheorem*{A2}{Assumption 2 (Counterfactuals)} 
\begin{A2} 
$((X_{it}^{+}$,$H_{it}^{+}, X_{it}^{-}, H_{it}^{-}),t=1,...,T)$ is independent of $(\eta_{i1},t=1,...,T)$, conditional on $X_i$.
\end{A2}

\color{blue} Because the variables $X_{it}^{+}$, $X_{it}^{-}$,  $H_{it}^{+}$, and $H_{it}^{-}$ are chosen by the researcher, Assumption 2 can be made to hold by construction. 
For example, it holds if these variables are constructed as  functions of $X_i$ and/or simulated variables that are independent of the data, as is the case in all of the examples we consider. 
However, if these objects are chosen to be functions of $X_i$ and some observables $Z_{it}$ not included in $X_i$, then veracity of  Assumption 2 depends on the statistical relationship between $X_i$, $Z_{it}$, and $\eta_{it}$.

We note that our object of interest $\theta_0$ implicitly depends on the number of time periods $T$, because we do not restrict any of the observable variables in the definition of $\theta_0$ to be stationary over time. For notational convenience we suppress this dependence in what follows, while noting here that it is accounted for in the asymptotic analysis of Section 6.

\color{black}

A number of important policy-relevant objects may be written in the form of $\theta_0$. 
Three examples we consider are average causal effects of alternative treatment regimes, bounds on average equivalent variation and on deadweight loss in demand analysis, and taxable income effects with nonlinear budget sets. In Appendix C we discuss marginal effects in binary choice models.

\subsubsection*{Example 1: Average Effects of Alternative Treatment Regimes}
\color{blue}

To define treatment effects within our model, recall that the potential outcome for a counterfactual level $x$ of the regressors is $Y_{it}(x)=g(x,\eta_{it})$.
Here, the heterogeneity $\eta_{it}$ determines potential outcomes and endogeneity of  $\eta_{it}$ corresponds to correlation between potential outcomes and observed treatments $X_{it}$, similarly to \cite{Imbens}.

\color{black}

Consider two counterfactual treatment regimes. 
In the first, an individual $i$ in period $t$ receives a random treatment ${X}_{it}^{-}$ and in the second they receive treatment ${X}_{it}^{+}$. 
These counterfactual treatments may depend on $X_i$. 
For example, we may wish to compare mean outcomes under the counterfactual assignments ${X}_{it}^{+}$ with the factual treatments, in which case we can set $X_{it}^{-}=X_{it}$. 
The expected average difference between potential outcomes is
\[
\theta_{0}:=E[\frac{1}{T}\sum_{t=1}^{T}\{Y_{it}({X}_{it}^{+})-Y_{it}(
{X}_{it}^{-})\}]=E[\frac{1}{T}\sum_{t=1}^{T}\{g({X}_{it}^{+},\eta_{it})-g(
{X}_{it}^{-},\eta_{it})\}].
\]
A major challenge for estimating such a causal effect is the possibility of unobserved confounding.
That is, there may be latent factors that jointly determine the treatment $X_{it}$ and the outcomes.
The nonparametric nonseparable model we consider allows for the possibility of unobserved confounding. 
Assumption 1 allows us to estimate average treatment effects like $\theta_0$, as explained in Section 3.2.

We can motivate Assumption 1 in this context using a nonparametric structural model for the treatment assignments and the heterogeneity in potential outcomes. Let $\alpha_{i}$ be a vector of time-invariant confounding factors and consider the following model where the time-varying innovations  $\{u_{it}\}_{t=1}^{T}$ and  $\{w_{it}\}_{t=1}^{T}$ are each jointly independent of $\alpha_{i}$:
\begin{align*}
\eta_{it}	=e(\alpha_{i},u_{it}),\,\,
X_{it}	=x(\alpha_{i},w_{it}).
\end{align*}
The first equation decomposes the endogenous heterogeneity in potential outcomes into variation between individuals, captured in $\alpha_i$, and variation over time $u_{it}$. Let us suppose that the innovations $\{u_{it}\}_{t=1}^{T}$ are jointly independent of $\{w_{it}\}_{t=1}^{T}$ and $\alpha_i$. This implies $u_{it}$ is independent of the history of treatments $X_i$. In addition, suppose the marginal distribution of $u_{it}$ (but not necessarily $w_{it}$) is time-invariant.  Under these conditions, Assumption 1 holds.

In this model, the time-invariant factors $\alpha_i$ are akin to fixed effects. They are individual-specific characteristics that explain the confounding between treatments and outcomes and do not vary over time. The condition that the temporal variation in potential outcomes, captured in $u_{it}$, is independent of the history of treatment assignments is akin to strict exogeneity. Unlike in the classic fixed effects model, $\alpha_i$ may enter non-separably into the possibly non-linear model for the outcome $Y_{it}$. Similar panel data treatment effect models were explicitly formulated in \cite{Chernozhukov2013} and \cite{Torgovitsky}.


\subsubsection*{Example 2: Average Equivalent Variation and Deadweight Loss Bounds}
The average equivalent variation and deadweight loss of a price change are important objects of interest in empirical demand analysis.  
Obtaining bounds on these quantities is a crucial step in  assessing the welfare impact of a policy that may alter consumer prices, such as a sales tax. 
Suppose $Y_{it}$ is the expenditure share of some commodity and $X_{it}=(P_{it},Z_{it})$, where  $P_{it}$ is the product price and $Z_{it}$ is a vector of covariates that includes total expenditure $M_{it}$ and the prices of other goods.
 
In order to define the welfare effects of a price change, we must choose an initial price paid by individual $i$ at time $t$, which we denote by ${P}_{it}^{-}$. 
This starting price  may depend on $X_i$.
For example, ${P}_{it}^{-}$ could simply be $P_{it}$, the price paid in period $t$ by individual $i$ for the good. 
Let $\Delta_{it}$ denote the change in the price of the good for individual $i$ in period $t$ that also may depend on $X_i$. 
Let  $\omega_{t}(X_i)$ be some weighting that may depend upon $X_i$.
Using \cite{Hausman}, we obtain bounds on the weighted average equivalent variation from a price change from $P_{it}^{-}$ to $P_{it}^{-}+\Delta_{it}$. 
Let $\pi$ be an upper or lower bound on the income effect for every individual and let
$U_{i}=(U_{i1},...,U_{iT})'$ be a vector of $T$ random variables that are uniformly distributed on $(0,1)$ and independent of the data, i.e. that are simulation draws from the standard uniform distribution.
Taking $g(p,Z_{it},\eta_{it})$ to be the counterfactual expenditure share at price $p$, a bound on the weighted average equivalent variation is 
\begin{align}
	\theta_{EV}=&  E[\frac{1}{T}\sum_{t=1}^{T}H_{it}^{+} g({P}_{it}^{-}+\Delta
	_{it}U_{it},Z_{it},\eta_{it})]\\
	H_{it}^{+} =&  \omega_{t}(X_i)\exp\big(-\pi({P}_{it}^{-}+\Delta_{it}U_{it})\big)\Delta
	_{it}\frac{M_{it}}{{P}_{it}^{-}+\Delta_{it}U_{it}}\label{hdef1}.
\end{align}
If $\pi$ is a lower (upper) bound on the income effect for every individual then, by \cite{Hausman}, $\theta_{EV}$ is an upper (lower) bound on the average over time and individuals of the equivalent variation for a change from ${P}_{it}^{-}$ to ${P}_{it}^{-}+\Delta_{it}$, weighted by $\omega_{t}(X_i)$.
The weights $\omega_{t}(X_i)$ allow us to assess the welfare impact on particular sub-populations, such as those in a low income bracket or with a certain family size. 

A corresponding deadweight loss bound can be obtained by subtracting the weighted average change in final demand as below.
\begin{align}
	\theta_{DWL}=&  \theta_{EV}-E\big[\frac{1}{T}\sum_{t=1}^{T}H_{it}^{-} g({P}_{it}^{-}+\Delta_{it},Z_{it},\eta_{it})\big]\\
	H_{it}^{-} =&\omega_t(X_i)\Delta_{it}\frac{M_{it}}{{P}_{it}^{-}
		+\Delta_{it}}\label{hdef2}
\end{align}
If $\pi$ is a lower (upper) bound on the income effect then $\theta_{0}$ will
be an upper (lower) bound for weighted deadweight loss averaged over all time periods and individuals.

Both $\theta_{EV}$ and $\theta_{DWL}$ are of the form in (\ref{objinterest}). In particular, $\theta_{EV}$ corresponds to the case with $H_{it}^{+}$ defined by (\ref{hdef1}), $H_{it}^{-}=0$, and $X_{it}^{+}=({P}_{it}^{-}+\Delta_{it}U_{it},Z_{it}')'$. 
The deadweight loss shares these choices of $H_{it}^{+}$ and $X_{it}^{+}$, but in this case, $H_{it}^{-}$ is set as in (\ref{hdef2}) and $X_{it}^{-}=({P}_{it}^{-}+\Delta_{it},Z_{it}')'$.
This example is discussed further in Section 5, which provides an application to consumer panel demand data.

\subsubsection*{Example 3: Average Heterogeneous Taxable Income Elasticities}

In structural economic models, an object of interest can be the expectation of a coefficient of an LRC model.
An example is the panel, budget set regression of \cite{Blomquist2024}. 
There an individual specific isoelastic utility function, together with time-varying scale heterogeneity in preferences, leads to an outcome $Y_{it}$, equal to the log of taxable income, that is a nonlinear function of the budget set. 
By integrating out the time-varying scale factor, using linear approximation of certain integrals, and specifying $X_{it}$ to be a piecewise linear budget frontier, an LRC is obtained with $J=3$, $\beta_{i2}$ equal to the taxable income elasticity for individual $i$, and
\begin{equation}
E[Y_{it}|X_i]=\beta_{i1}+b_2(X_{it})\beta_{i2}+b_3(X_{it})\beta_{i3},
\end{equation}
where $\beta_{i2}$ is the taxable income elasticity for individual $i$, $b_2(x)$ is the natural logarithm of the slope of the last budget segment, and $b_3(x)$ is the ln of rato of slopes of the last and first budget segments.
The panel data setting of this paper allows for endogeneity of piecewise linear budget sets, where budget sets may be correlated with preferences. 
The parameter of interest $\theta_0=E[\beta_{i2}]$  can be represented in the form of equation (4) by choosing $H^+_{it}=H^-_{it}=1$ and specifying  $X_{it}=(1,b_2(X_{it}),b_3(X_{it}))',X_{it}^{+}=X_{it}+e_2,$ and $X_{it}^{-}=X_{it}$, where $e_2$ is a unit vector with $1$ in the second position and zeros elsewhere. 

\color{blue}

\section{Linear Approximation and Estimation}

In this Section we discuss approximation of nonparametric nonseparable models by LRC specifications. Thus we obtain a linear approximation to the object of interest $\theta_0$ which motivates a linear estimator. The approximation relies on the fact that $\theta_0$ is linear in the conditional average potential outcome, which is
\begin{equation*}
h(x,X_i):=E[g(x,\eta_{i1})|X_i]=E[g(x,\eta_{it})|X_i],
\end{equation*}
where the second equality follows by Assumption 1. 
The following result gives the formula for $\theta_0$ in terms of $h(x,X_i)$ which motivates our linear approximation and estimation method.

\theoremstyle{plain} \newtheorem*{T1new}{Theorem 1} \begin{T1new}
If Assumptions 1 and 2 are satisfied and if the moments $E[|g(X_{it}^+,\eta_{it})|]$, $E[|H_{it}^+g(X_{it}^+,\eta_{it})|]$, and $E[|H_{it}^-g(X_{it}^-,\eta_{it})|]$ are finite, then   
\begin{equation}
\theta_0=E\big[\frac{1}{T}\sum_{t=1}^{T}\big(H_{it}^{+}h(X_{it}^{+},X_i)-H_{it}^{-}h(X_{it}^{-},                        X_i)\big)\big].\label{objinterest}
\end{equation}
\end{T1new}

\subsection{Approximation}

\color{blue}
Theorem 1 demonstrates that we can approximate $\theta_0$ by approximating $h(x,X_i)$.
We can approximate $h(x,X_i)$ by the average potential outcome for the LRC model, which is
\begin{equation*}
b(x)'\beta_i=b(x)'E[\eta_{it}|X_i]=E[b(x)'\eta_{it}|X_i].
\end{equation*}
The idea is that if $b(x)$ is a rich enough vector of approximating functions and $h(x,X_i)$ is a smooth enough function of $x$, uniformly in $X_i$, then $b(x)'\beta_{i}$ will approximate $h(x,X_i)$ as a function of $x$ for some $\beta_i$.

\color{blue}

In this approximation the function $h(x,X_i)$ is being approximated by $b(x)'\beta_i$ separately for each value of $X_i$.
As $X_i$ varies the coefficients $\beta_i$ are allowed to vary so that $b(x)'\beta_i$ remains a good approximation of $h(x,X_i)$.
Such approximations are known to exist when $x$ is bounded and $h(x,X_i)$ is continuously differentiable up to order $d$ with derivatives bounded, uniformly in $x$ and $X_i$. 
Such uniform approximation results are often referred to as Jackson theorems, following \cite{Jackson1911} where approximation rates were obtained for power series and scalar $x$. 
Jackson theorems for multivariate $x$, power series, splines, and other approximating functions are given, for example, in  \cite{DeVore1993}.
We provide a more formal analysis of the approximation error in Section 6. 

One could also consider the LRC model $b(x)'\eta_{it}$ as an approximation to the nonparametric model $g(x,\eta_{it})$, as was done in a previous version of this paper.
Approximating $g(x,\eta_{it})$ is more difficult because in important examples of interest, such as those with discrete $Y_{it}$, $g(x,\eta_{it})$ will not be smooth in $x$.
In contrast, the average potential outcome $h(x,X_i)$ can be very smooth in $x$ even when $Y_{it}$ is discrete, as long as some elements of $\eta_{it}$ are continuously distributed with smooth pdf, because $h(x,X_i)=E[g(x,\eta_{it})|X_i]$ integrates over $\eta_{it}$.  
For this reason, and because our objects of interest depend on just $h(x,X_i)$, we focus on approximating $h(x,X_i)$.


\color{black}

The approximation of $h(x,X_i)$ leads to a corresponding approximation for $\theta_0$.
The approximation $h(x,X_i)\approx b(x)'\beta_i$ implies  $h(X_{it}^+,X_i)\approx b(X_{it}^+)'\beta_i$ and $h(X_{it}^-,X_i)\approx b(X_{it}^-)'\beta_i$.
Plugging these approximations into Theorem 1 gives
 \begin{equation}	\theta_{0}\approx E[a_{i}^{\prime}\beta_i],\,\text{ }a_{i}=\frac{1}{T}\sum_{t=1}^{T}\{H_{it}^{+}b({X}_{it}^{+})-H_{it}^{-}b({X}_{it}^{-})\}.\label{tapprox}%
\end{equation}
Thus, $\theta_0$ is approximately an expectation of the inner product of a known vector $a_i$ with $\beta_i$.
Also, this approximation is exact in the LRC model where $h(x,X_i)=b(x)'\beta_i$.
Consequently, an estimator of $\theta_0$ can be constructed by forming the inner product of the known $a_i$ with an estimator of $\beta_i$ and averaging over individuals, as described in the next subsection.

The approximating LRC model is a random parameters specification like that of \cite{newey2025}. We innovate in approximating a nonparametric, nonseparable model by one that is linear in known functions of $x$.

\color{black}

\subsection{Estimation}
\color{blue}
To estimate $\theta_0$ using equation (13) we need an estimator $\hat\beta_i$ of $\beta_i$ for each individual $i$. 
To construct these estimators we use the approximate linearity of  $E[Y_{it}|X_i]=h(X_{it},X_i)\approx b(X_{it})'\beta_i$ in $b(X_{it})$ by regressing $Y_{it}$ on $b(X_{it})$ over all time periods $(t=1,...,T)$ to form $\hat\beta_i$.
From the literature on series estimation of conditional means   (e.g. Gallant, 1981), it is known that such a linear regression can be used to estimate functions of a conditional mean under specific identification and regularity conditions.
We apply this insight to the panel data setting by using linear regression for each individual over all $t$ and then averaging across individuals.
In Section 6 we make this intuition precise by giving regularity conditions for consistency and asymptotic normality of the resulting $\hat\theta$ for large $T$.
\color{black} 

In practice, there could be high multicollinearity in this regression, particularly if $T$ is not much larger than the number $J$ of basis functions.
Here we address this problem using individual specific ridge regression.
To describe these ridge regressions let $Y_i := (Y_{i1},...,Y_{iT})'$, $B_i :=[b(X_{i1}),...,b(X_{iT})]'$, $Q_{i}=B_{i}^{\prime}B_{i}/T,$ $D_{i}$ be a diagonal matrix with $0$ as its upper left entry and all other diagonal entries strictly positive, and $\lambda$ a positive constant. 
A ridge regression estimator of ${\beta}_i$ is defined as
\begin{equation}
\hat{\beta}_{i}=(Q_{i}+\lambda D_{i})^{-1}B_{i}^{\prime}Y_{i}/T. \label{ridgedef}
\end{equation}
The zero in the top left entry of $D_i$ ensures that we do not penalize the intercept in the ridge regression. 
By allowing $D_i$ to be individual-specific we can accommodate individual-level re-scaling of the regressors. \color{blue}In our empirical application we simply set $D_i$ to be the identity matrix with its upper left entry set to zero.\color{black}

We note here that perfect multicollinearity in the regression for individual $i$, where $Q_i$ is singular, is an identification problem and not just a computational issue.
If $Q_i$ is singular then the population least squares coefficients $\beta_i$ from regressing $E[Y_{it}|X_i]$ on $b(X_{it})$ are not unique, and hence not identified.
We return to this important identification issue in Section 3.5 that follows.

These individual ridge estimators are biased, as usual for ridge regression.
It is possible to mitigate this ridge bias in the estimation of  $\theta_0=E[a_i'{\beta}_i]$. \color{blue}
Let $A_{i}$ denote a square $J$-dimensional matrix with $a_{i}^{\prime}$ as its first row and the remaining rows consisting of $J-1$ distinct rows of the identity matrix chosen so that $\frac{1}{n}\sum_{i=1}^{n}A_{i}$ is non-singular. 
In our application we simply use the last $J-1$ rows of the identity. \footnote{\color{blue}Other choices of $A_i$ can also be used. What is necessary is that $A_i$ is square, a function of $a_i$, constructed so that there is a fixed vector $v$ so that $a_{i}'=v'A_{i}$, and $\frac{1}{n}\sum_{i=1}^{n}A_{i}$ is non-singular. Our theoretical results apply for any such $A_i$ and the asymptotic variance we derive is not affected by the choice of $A_i$.\color{black}}
\color{black}
Also, let
\begin{equation}
W_{i}=(Q_{i}+\lambda D_{i})^{-1}Q_{i}. \label{eq:Wdef}%
\end{equation}
The debiased average ridge estimator of $\theta_{0}$ is then
\begin{equation}
\hat{\theta}=\bar{a}^{\prime}(\overline{AW})^{-1}\overline{A\beta},\text{
}\bar{a}=\frac{1}{n}\sum_{i=1}^{n}a_{i},\text{ }\overline{AW}=\frac{1}{n}%
\sum_{i=1}^{n}A_{i}W_{i},\text{ }\overline{A\beta}=\frac{1}{n}\sum_{i=1}%
^{n}A_{i}\hat{\beta}_{i} \label{theta}%
\end{equation}
An estimator $\hat{V}$ for the asymptotic variance of $\sqrt{n}(\hat{\theta}-\theta_0)$ can be obtained via the delta method as
\begin{equation}
\hat{V}=\frac{1}{n}\sum_{i=1}%
^{n}\hat{\psi}_{i}^2,\text{ }
\hat{\psi}_{i}=(a_{i}-\bar{a})^{\prime}(\overline{AW})^{-1}\overline{A\beta
}+\bar{a}^{\prime}(\overline{AW})^{-1}A_{i}[\hat{\beta}_{i}-W_{i}%
(\overline{AW})^{-1}\overline{A\beta}]\label{varest}
\end{equation}

Example 3 illustrates that parameters of interest may include elements of the vector $E[\beta_i]$. 
Each component of this vector has the form $E[a_i'\beta_i]$ where $a_i$ is a unit vector.  
When $a_i$ is constant, the formula for the debiased estimator simplifies so that $A_i$ cancels out. 
Thus we obtain an estimator $\hat{\theta}$ of $E[\beta_i]$ and a corresponding estimator $\hat{V}$ of the asymptotic variance, for $\overline{W}:=
\sum_{i=1}^{n}W_{i}/n$,

\begin{equation}
\hat{\theta}=\overline{W}^{-1}\frac{1}{n}\sum_{i=1}%
^{n}\hat{\beta}_{i},\text{ } \hat{V}=\frac{1}{n}\sum_{i=1}%
^{n}\hat{\psi}_{i}\hat{\psi}_{i}',\text{ }  
\hat{\psi}_{i}=\overline{W}^{-1}
(\hat{\beta}_{i}-W_{i}\hat{\theta}) \label{exampleest}
\end{equation}
This estimator has a straight-forward interpretation. 
The term $\sum_{i=1}^{n}\hat{\beta}_{i}/n$ is the sample average of individual ridge estimates that suffers from ridge bias. 
Multiplying by $\overline{W}^{-1}$ effectively undoes the ridge bias on average.
Specifically, $\hat\theta$ has Property C mentioned in the Introduction, being unbiased in the LRC model when  $\beta_{ij}$ does not vary with $i$ for $j>1$. In the next subsection we discuss this and other properties of $\hat\theta$.

\color{blue}
The debiased average ridge estimator belongs to a general class of regularized panel estimators that includes \cite{Graham2012}. See the discussion of Property C in Section 6.1 for details.

\subsection{Summary of Properties}

The debiased panel ridge estimator has several interesting characteristics that help explain its form and how it may be used and interpreted. 
Here we provide a brief  summary of these properties, with a more formal discussion deferred to Section 6 along with our asymptotic analysis.

\subsubsection*{Property A: Empirical Bayes Interpretation}
The average coefficient estimator $\hat\theta$ of equation (\ref{theta}) can be interpreted as an empirical Bayes estimator.
 Suppose we assume an LRC specification holds and the corresponding residuals are iid normally distributed, and that each $\beta_i$ has a normal prior with common nonzero mean $\bar{\beta}$.
For each $i$, let $\beta_i^{\text{Post}}$ be the posterior mode for each individual parameter $\beta_i$ with $D_i$ being directly proportional to the prior variance-covariance matrix for $\beta_i$. A Bayesian estimator of $E[\beta_i]$ can then be constructed as $\sum_{i=1}^n  \beta_i^{\text{Post}}/n$. This estimator depends on the prior mean $\bar{\beta}$ and the empirical Bayes approach is to use the data to determine this hyper-parameter. Suppose one chooses the unique $\bar{\beta}$ that ensures $\sum_{i=1}^n  \beta_i^{\text{Post}}/n=\bar{\beta}$, which can be understood as a `self-consistency' restriction. The resulting estimator coincides with our estimator $\hat{\theta}$.

\subsubsection*{Property B: Convergence to Fixed Effects with Large Penalty $\lambda$}
As the penalty parameter $\lambda$ diverges to infinity, the estimator $\hat{\theta}$ converges towards a  fixed-effects estimator $\frac{1}{n}\sum_{i=1}^n a_i' \hat{\beta}_{FE,i}$ where $\hat{\beta}_{FE,i}$ is a fixed effects estimator with individually-varying intercept and constant slopes,when $D_i$ does not vary with $i$.

\subsubsection*{Property C: Unbiased When Slope Coefficients Do Not Vary with Individuals}

For the LRC model the debiased ridge estimator of equation (18) is unbiased when $\beta_{ij}$ does not vary with $i$ for $j>1$. 
Note that this holds under `exogenous slopes', that is, if $\eta_{itj}$ is independent of $X_i$ for $j>1$.   
This property holds regardless of the choice of $\lambda$ and applies in cases in which $Q_i$ is singular for some individuals so that $\beta_i$ is not identified.

\subsubsection*{Property D: Convergence Under Small Penalty}
As $\lambda$ shrinks to zero, $\hat\theta$ converges to the average of the product of $a_i$ and the individual OLS estimate of $\beta_i$ if $Q_i$ is non-singular for all individuals.\\
\\

Properties B and D together demonstrate that the choice of penalty parameter $\lambda$ allows for a smooth transition between the two extremes of fixed effects and average of individual OLS estimates.

\color{blue}
\subsection{Guidance for Applications}

In order to construct the estimate $\hat{\theta}$, in the case of nonparametric $g(x,\eta)$ one must choose the vector of basis functions $b(\cdot)$. 
In our empirical application, we use polynomial basis functions (powers and interactions of the regressors) and report results for polynomial bases of various degrees. The number of regressors in our application is moderately large,  so we use relatively low order polynomials. Our most flexible specification is log-linear in a subset of regressors, and for other regressors we include powers and interactions up to order three. In cases with fewer regressors, we suggest the use of cubic spline bases (\cite{DeVore1993}).

The estimator $\hat{\theta}$ given in the formula  (\ref{theta}) depends on two regularization parameters. The first of these is the matrix $D_i$. We suggest researchers take $D_i$ to be the identity matrix with its upper-left entry set to zero, as we do in our empirical application. 

For the penalty parameter $\lambda$, we suggest researchers report results for a range of values. One effect of the debiasing strategy is that, by counteracting the shrinkage of ridge towards zero, it tends to reduce the sensitivity of the final estimates to the choice of the penalty parameter. The problem of developing a data-driven method for selecting $\lambda$ is a problem for future research. Nonetheless, a version of Lepski's method (see e.g., \cite{Birge2001}) may provide a useful heuristic. To apply the method, one selects $\lambda$ to be the largest value such that for any $\lambda'\leq\lambda$, the absolute difference in the corresponding estimates is no greater than four times the standard error under $\lambda'$.

\subsection{Evaluating the Extent of Identification} 

Ridge regression and debiasing affect the contribution of each observation to $\hat\theta$. 
To see this in the LRC, note that by  Assumption 2 and i.i.d. individual data,
\[
E[\hat\theta|X_1,...,X_n]=\frac{1}{n}\sum_{i=1}^{n}\hat{a}_i'\beta_{i},\hat{a}_i'=\bar{a}'(\overline{AW})^{-1}A_iW_i. 
\]
Here we see that $\hat\theta$ is (conditionally) unbiased for the regularized weighted average $\frac{1}{n}\sum_{i=1}^{n}\hat{a}_i'\beta_{i}$ rather than the correct $\frac{1}{n}\sum_{i=1}^{n}a_i'\beta_{i}$.

For example consider the average slope estimator $\hat\theta_2$ in the LRC for $J=2$, where $b(x)=(1,b_2(x))'$ and $a_i=(0,1)'$.
Let $\tilde{Q}_i$ denote the sample variance over $t$ of $b_2(X_{it})$, $w_i=\tilde{Q_i}/
[\lambda+\tilde{Q_i}]$, and $\omega_i=w_i/{\sum_{k=1}^{n}w_k}$ be nonnegative weights that sum to $1$. Then it is straightforward to show that $\hat{a}_i=(0,\omega_i)'$, so that
$E[\hat\theta_2|X_1,...,X_n]={\sum_{i=1}^{n}\omega_i\beta_{i2}}$.
Here we see that the expectation of the debiased ridge estimator is a weighted average of individual specific coefficients $\beta_{i2}$, with larger weight $\omega_i$ given to observations where $\beta_{i2}$ is more strongly identified, in the sense that $\tilde{Q}_i$ is larger, and zero weight given to observations where $\beta_{i2}$ is not identified, i.e. where $\tilde{Q_i}=0$ (and so $Q_i$ is singular).

In general,  $\hat{a}_i$ differs from the corresponding $a_i$ depending on the size of $\lambda$ and $Q_i$. 
If every $Q_i$ is nonsingular then as $\lambda$ shrinks each $\hat{a}_i$ will converge to  $a_i$ for every $i$, as in Property D and shown in Section 6. 
If some $Q_i$ are singular, then each $\hat{a}_i$ need not converge to $a_i$ and there can be especially sharp differences when $Q_i$ is singular. 
In the example given in the previous paragraph, each $\hat{a}_i$ with $Q_i$ singular converges to $(0,0)'$ and each $\hat{a}_i$ with nonsingular $Q_i$ converges to $(0,1)'$. 
More generally, when $Q_i$ is singular and $a_i$ is not in the identified set of linear combinations of $\beta_i$, the limit of $\hat{a}_i$ will not be $a_i$ as $\lambda$ shrinks. 
Thus, the extent of nonidentification of $a_i'\beta_i$ across $i$ should be indicated by differences between $\hat{a}_i$ and $a_i$ across observations for small $\lambda$.

We can use the Euclidean norm $\|\hat{a}_i-a_i\|$ to evaluate the extent of identificatioin in applications. 
Note that
\[
|\hat{a}_i'\beta_i-a_i'\beta_i|\leq\|\hat{a}_i-a_i\|\|\beta_i\|.
\]
Therefore, a quantity that measures the effect of possible singularity of $Q_i$ on the contribution of the $ith$ observation to the estimated effect $\hat\theta$ of interest is
\[
\zeta_i=\|\hat{a}_i-a_i\|
\|\hat\beta\|/n\hat\sigma,
\]
where $\hat\beta$ is the esimator of the average coefficient vector and $\hat\sigma$ is the standard error of $\hat\theta$.
A quantile plot of this object may aid in assessing both the degree of nonsingularity of $Q_i$ over the all $i$ as well as the strength of identification for particular individuals. 

One can also upper-bound the conditional bias due to regularization by\footnote{To derive this bound, first note that $\frac{1}{n}\sum_{i=1}^{n}\hat{a}_i'E[\beta_{i}]=\frac{1}{n}\sum_{i=1}^{n}a_i'E[\beta_{i}]$ and thus $\frac{1}{n}\sum_{i=1}^n \hat{a}_i'\beta_i-\frac{1}{n}\sum_{i=1}^n a_i'\beta_i=\frac{1}{n}\sum_{i=1}^n (\hat{a}_i-a_i')(\beta_i-E[\beta_i])$, then apply Cauchy-Schwarz.}
\[\bigg|E[\hat\theta|X_1,...,X_n]-\frac{1}{n}\sum_{i=1}^n a_i'\beta_i\bigg|\leq \sqrt{\frac{1}{n}\sum_{i=1}^n \|\hat{a}_i-a_i\|^2}\sqrt{ \frac{1}{n}\sum_{i=1}^n \|\beta_i-E[\beta_i]\|^2}.\]
The first square root on the RHS can be calculated directly, and the second may be bounded according to  one's a priori beliefs about the variation in $\beta_i$. Indeed, in recent work, \cite{kwon2025} use a priori bounds on the variance of heterogeneous coefficients to obtain bias-aware confidence intervals in a related setting.

%% file: Application_to_demand_estimation.tex
\section{Nonparametric, Nonseparable Demand Models for Panel Data}

 The nonseparable, nonparametric model $Y_{it}=g(X_{it},\eta_{it})$ (of equation \eqref{basemodel}) provides a very general specification of individual demand. 
 In the application, we take the outcome variable $Y_{it}$ to be the expenditure share on a class of goods, which has long been a useful specification, as in \cite{Deaton1980}, \cite{Chaudhuri2006}, and \cite{hsiao2021}, but other choices of outcome variable will also do. 
 Discrete choice is included as a special case where $Y_{it}$ is the number of units of a particular good purchased by an individual in time period $t$ and the outcome model is specified analogous to that in Section 3.1.
 
 The model allows unobserved heterogeneity $\eta_{it}$ to affect demand in very general ways.
 The $\eta_{it}$ is allowed to be infinite-dimensional corresponding to stochastic revealed preference as in \cite{Richter1990}, \cite{McFadden2005}, and \cite{Kitamura}, with demand restricted to be single valued. 
 Such choice specifications have been considered by \cite{Lewbel2001}, \cite{Blomquista}, \cite{Blundell2014}, \cite{Hoderlein2014}, \cite{Bhattacharya}, \cite{Dette2016}, and \cite{Hausman}. 
 In addition, $\eta_{it}$ may include product specific unobserved characteristics as in \cite{Berry1994} and \cite{Berry1995}.
 The presence of such could create correlation across individuals in $\eta_{it}$.
 Alternatively, if $\eta_{it}$ is tastes by an individual for unobserved product characteristics and preferences are independent across individuals, then correlation across individuals need not be present.

 To help this model relate to existing demand models, it is helpful to decompose the heterogeneity $\eta_{it}$ into a component $\alpha_i$ that does not vary with $t$ and a time-varying component $v_{it}$.
 This decomposition is common in panel demand models, including discrete choice, as in \cite{Chamberlain1984}.
 Here, $\alpha_i$ represents preference features that are stable over time for a given individual, while $v_{it}$ allows some time variation in demand.
 For example, $v_{it}$ could represent a taste for variety that is not observable to the econometrician. 
 Tastes for variety could also be incorporated by including functions of $t$ in $b(X_{it})$.
 Generally, it is quite common to incorporate time-varying heterogeneity as represented by $v_{it}$ in nonlinear panel data models.

An important feature of panel demand data is that prices are common across consumers and are determined in market equilibrium. 
As a result, prices will generally be endogenous in being related to individual preferences.
Restrictions on $v_{it}$ mitigate potential price endogeneity. 
If $v_{it}$ is i.i.d. over time and independent of unobserved supply shocks, and there are many consumers in the market, then bias from price endogeneity will be small, as shown by \cite{moon2026}.
Intuitively, price effects will be (nearly) identified from the movement of prices over time because supply shocks are independent of time variation in preferences. 
This independence seems plausible when $v_{it}$ is a stochastic taste for variety of an individual and variation in supply is due to cost shocks.
Also, \cite{Hausmana} found that the use of prices from other markets as instruments did not change demand estimates for scanner data, providing evidence that relying on time variation in prices for identification of price effects is consistent with scanner data. 

The interpretation of $v_{it}$ as preference heterogeneity means that preferences are allowed to change over time, even being correlated over time, in order to represent a taste for variety.
Demand specifications with time-varying, unobserved preference effects $v_{it}$ are common in panel data, discrete choice demand being a prime example. 
The presence of $v_{it}$ helps demand and other models fit the data better. 
It allows for departures from the weak axiom of revealed preference in the choice of an individual over time, as has been found in empirical work, for example, \cite{crawford2019}.

Time-varying preferences have little effect on the interpretation of welfare calculations. 
The average equivalent variation and deadweight loss calculations just average over time. 
As such, they estimate the expected value of welfare integrated over as $v_{it}$, similar to welfare estimates for discrete choice panel data.
Such time-average welfare measures are  consistent with utility maximization over time if there are no dynamic linkages in goods. 
Of course, such preferences are not consistent with stockpiling models like that of \cite{Hendel2006}. 
In the application, we take one month as the time unit and focus on goods with little potential for stockpiling to avoid this concern. 

Allowing for zero demand is important in demand modeling. 
For example, consumer data that considers alcohol or tobacco consumption will have many individuals with zero consumption. 
Including zero demand observations in the data correctly accounts for zero consumption in calculations of average equivalent variation and deadweight loss, as shown in  \cite{Hausman}, Theorem 3. 
Intuitively, there is no effect of a price change on the welfare of a consumer who never purchases a product, and the average is also correct when the product is only purchased sometimes.
The nonseparable, nonparametric specification gives the demand equations flexibility to allow for zeros while being consistent with utility maximization.

An observation arising from economic theory is that often, but not always, the policy question of interest depends on only one, or a very few, price effects. 
For example, estimation of individual welfare effects typically depends only on the own price effect when all other prices are held constant   (\cite{Hausman1981}).
Also, small cross-price effects will mitigate market equilibrium effects of changing one price. 
Price changes for one good will shift demand for other goods by small amounts, so that the equilibrium welfare effect from changing only one price can be well approximated by the effect of just that price on average demand. 

Computational simplicity is an important virtue of demand analysis in panel data based on linear approximation we give. 
Average equivalent variation and deadweight loss are estimated by a debiased average of individual specific linear combinations of ridge regressions.
Simulation is used to approximate the integrals in the welfare estimates.
Simple inference is based on the independence of estimates across individuals.
All of these features make this approach to demand estimation simple to implement, even in very large data sets.

\section{Application to Scanner Data}

We apply our methods to estimate price elasticities for groceries and to analyze the impact of counterfactual tax changes on consumer welfare. In this context, the outcome variable $Y_{it}$ is the share of expenditure on a particular class of goods, and the regressors $X_{it}$ include the natural log of prices and total expenditure. 
Our specification generalizes the popular Almost Ideal Demand System (AIDs) of \cite{Deaton1980a} to approximate a nonparametric, fully nonseparable demand model as we describe in Section 4.

Given this specification, our debiasing method has important implications for our elasticity  estimates, and consequently, our estimates of counterfactual welfare. In the absence of debiasing, ridge regression tends to shrink parameters to zero. Therefore, in an AIDS-type specification, ridge would shrink the own price elasticity towards $-1$, the cross-price elasticities towards $0$, and the expenditure elasticity towards $1$. The debiasing mitigates the effects of shrinkage and results in estimates that are less sensitive to the choice of penalty parameter.
However, we note that shrinkage of the cross-price elasticities may be appropriate in consumer demand panel datasets, where small cross-price effects re often found in the literature \cite{Burda2008} and \cite{Burda2012}. Indeed, cross-sectional OLS regressions in our empirical setting recover small cross-price elasticities, as we report in Table IV in Appendix A.

Rather than estimate demand for particular products, we instead focus on the demand for classes of goods. In effect, we model demand at an intermediate level of multi-stage budgeting to estimate welfare effects of price changes for good types. 
Here, the consumer decides how much to spend on a class of goods based on individual- and type-specific second-order flexible price indices, and on the total expenditure on all included classes of goods.

 Modeling demand for good types can be justified by certain conditions on the separability of preferences, as in, for example, \cite{Gorman1959}, \cite{Gorman1981}, \cite{Deaton1980a}, and \cite{Blundell2000}. 
 An alternative motivation relies on  statistical aggregation for the many prices into a price index which is independent of consumer preferences, as in \cite{Hoderlein2012a}.
 For the intermediate level of commodities we consider (e.g., soda), it may be  important to allow for more general substitution patterns across the dissimilar kinds of goods. 
 The flexibility in allowing for general cross-price effects provided by the AIDS demand system of \cite{Deaton1980a}, may be useful here as it is in \cite{Chaudhuri2006} and \cite{hsiao2021}.
    
We use NielsenIQ retail scanner data to construct price indices, and the NielsenIQ Homescan Panel to track purchases and household characteristics.\footnote{The empirical work is
researchers' own analyses calculated (or derived) based in part on data from
Nielsen Consumer LLC and marketing databases provided through the NielsenIQ
Datasets at the Kilts Center for Marketing Data Center at The University of
Chicago Booth School of Business. The conclusions drawn from the NielsenIQ
data are those of the researcher(s) and do not reflect the views of NielsenIQ.
NielsenIQ is not responsible for, had no role in, and was not involved in
analyzing and preparing the results reported herein.}

The data include $2585$ households with Houston-area ZIP codes in the years 2010-2014. The number of monthly observations for each household ranges from $1$ to $60$, and we restrict our analysis to the households included for at least $12$ months.\footnote{We checked for differences in results between using all households and the 2197 that were present for at least a year and found no statistically significant differences. The insensitivity to panel length suggests that attrition bias does not play a large role in this data.}

We construct the price indices for each consumer from data on the monthly total expenditures per good category, and on the quantity purchased per month. The
original data contained time-stamps for purchases. The price indices span 15 aggregated groups of goods: soda, milk, soup, water, butter, cookies,
eggs, orange juice, ice cream, bread, chips, salad, yogurt, coffee, and
cereal. As in \cite{Burda2008} and \cite{Burda2012}, we chose these groups
because they made up a relatively large proportion of total grocery
expenditure. The data also includes demographics such as race, marital status,
household size and composition, and employment status. 

The price index for each group of goods is computed as a weighted geometric average of the
actual purchase prices (expenditure divided by quantity) over all purchases made by the household in the month, with weights
equal to the proportion of expenditure on a specific item associated with a
unique item code. The price index $P_{g,it}$ for household $i$ at time $t$,  for the $g^{th}$ group of goods is
specified by
\[
\ln(P_{g,it})=\sum_{j=1}^{J_{g}}w_{gj,it}\ln(P_{gj,it}),
\]
where $j$ denotes a particular item code, $J_{g}$ is the number of codes for
the $g^{th}$ commodity, $w_{gj,it}$ is the proportion of expenditure on commodity
$g$ that is spent on code $j$, and $P_{gj,it}$ is expenditure by household $i$ on code $j$ divided
by quantity of code $j$ in month $t$. This is a T\"ornqvist price index, which was
shown by \cite{Diewert1976} to be exact for a quadratic utility specification, and
a second-order approximation to the exact price index for any utility.
\cite{Deaton1980} (pp. 132-133) showed that
with weak separability, this price index appears in share equations for a
Rotterdam demand specification (i.e., log quantity as a linear function of log
prices and log expenditure) and suggest that it could lead to a good
approximation when prices within a group tend to move together.

The price indices may be endogenous because the amount spent on a particular
item in a group of goods is a choice of the consumer. Price endogeneity could
be particularly important when a group of goods contains commodities of
varying quality, such as organic and non-organic milk, or fresh and frozen
orange juice. As we discuss in the previous section, our approach can accommodate such endogeneity, provided that the unobserved heterogeneity satisfies the time-invariance condition formalized in Assumption 1.

As stated above, we construct price indices using prices actually paid by each household. Including zero expenditures makes it necessary to impute price indices for time periods where an individual purchased none of a particular good. If a household had purchased the good before, then price indices are imputed as the most recent price faced by the household in a past purchase. Rarely, a good is never purchased prior to a given month, in which case its imputed price is the average price of the same good within a subset of stores similar to those at which the household shops.\footnote{Specifically, we group retailers in the Houston area into $4$ categories, and assign households to their most-visited retailer category each year. Then we construct monthly price indices for each retailer category and each good, which are used to fill in missing prices.} The frequency of household-month observations with zero total expenditures varies by good: for some goods, most households record purchases each month, while other goods, such as orange juice and ice cream, are purchased more infrequently. Our analysis focuses on estimating demand for the goods for which we have the most reliable data, namely, soda and milk.




The inclusion of prices for all $15$ categories of goods allows estimation of cross-price demand effects. This gives us $16$ price and expenditure regressors. This is too large a number of regressors for standard nonparametric estimation, such as kernel regression, where it is thought to be impractical to use more than five or six regressors. For panel estimation, $16$ regressors may also be excessively large. The large number of regressors with small coefficients for the many cross-price effects motivates our use of ridge regularization.

In total, our analysis uses $86,122$ observations across the households. As a baseline, we consider the log-linear AIDS-type specification below. 
\begin{equation}
Y_{it} = \alpha_i + \gamma_i \log \text{Exp}_{it} + \sum_{g} \beta_{g,i} \log P_{g,it} +u_{it}\label{specification}
\end{equation}
where $Y_{it}$ is the share of expenditure by household $i$ in month $t$ on a particular class of goods. $Exp_{it}$ is that household's total monthly expenditure over the $15$ categories of goods, and $P_{g,it}$ is the household's price index for good $g$ in that month. $\alpha_i$, $\gamma_i$, and $\beta_{g,i}$ are individual-specific coefficients, and $u_{it}$ a time-varying residual. We estimate separate models for soda and milk, with no restrictions that the coefficients in each case are the same.

In order to more precisely approximate a possibly non-linear and non-separable underlying demand model, in some of our analyses we enrich the specification (\ref{specification}) by including some powers and interactions of log prices and total expenditure.



Table I contains elasticity estimates for both soda and milk. We employ the model (\ref{specification}) and compare three methods for estimation. These are cross-sectional OLS, fixed-effects estimates,  individual-specific ridge without debiasing, and estimates that employ our debiased individual-specific ridge method.

In order to perform individual-ridge, we must select the matrix $D_i$ in the formula (\ref{ridgedef}). We let $D_i$ be the identity matrix with its first diagonal entry set to zero. We carry out ridge using two alternative choices for the penalty parameter $\lambda$. As a robustness check, we carry out the analysis with and without the inclusion of seasonal dummy variables. Seasonal variation in both price and tastes could be problematic for our analysis, as it suggests that heterogeneity in preferences is time-varying given prices, which would contradict Assumption 1.

\begin{table}[ht]
    \caption{Estimates of own-price elasticity, with (top) and without (bottom) season dummies.}
    \label{tab:my_label}
    \centering
    \begin{tabular}{c|cccc|cc}
    & OLS & FE & Ridge $0.05$ & Ridge $0.0005$ & DBR $0.05$ & DBR $0.0005$\\
    \hline
soda & -0.795 & -0.815 & -0.829 & -0.790 & -0.775 & -0.777 \\  
 & (0.003) & (0.004) & (0.007) & (0.016) & (0.009) & (0.016) \\
  milk & -1.206 & -0.607 & -0.843 & -0.480 & -0.445 & -0.349 \\ 
  & (0.012) & (0.016) & (0.011) & (0.038) & (0.046) &  (0.037) \\
  \hline
  soda & -0.795 & -0.815 & -0.823 & -0.768 & -0.770 & -0.756 \\
  & (0.003) & (0.003) & (0.007) & (0.017) &  (0.009) & (0.017)\\
  milk & -1.206 & -0.608 & -0.838 & -0.454 & -0.454 & -0.347 \\
  & (0.012) & (0.016) & (0.008) & (0.041) & (0.041) & (0.041) \\
    \end{tabular}
    
    	\begin{minipage}[t]{1\columnwidth}%
		{\scriptsize{}\color{blue}The columns respectively contain estimates from cross-sectional OLS, Fixed Effects, individual-ridge without debiasing and penalty parameters $0.05$ and $0.0005$, and our debiased ridge estimates (abbreviated to `DBR') with those same penalties. These methods are used to estimate the average of the coefficient on log own-price in specification (\ref{specification}). We obtain elasticities by dividing these average coefficient estimates by the average (over all individuals and time periods) of the expenditure share of the relevant good and subtracting unity. Bootstrap standard errors are given in parentheses below each estimate.\color{black}}
	\end{minipage}{\scriptsize\par}
\end{table}
\color{blue}
The columns in Table I respectively contain own-price elasticity estimates obtained using cross-sectional OLS, Fixed Effects, individual-ridge without debiasing and penalty parameters, and our debiased ridge estimates. For the latter two methods, we present estimates for two different values of the penalty parameter, namely $0.05$ and $0.0005$. To account for dependence between the coefficient estimates and the mean expenditure share, standard errors are calculated by bootstrap.\color{black}

In all cases, the estimates are  insensitive to the inclusion of seasonal dummies. The ridge estimates without debiasing are sensitive to the choice of penalty parameter, particularly in the case of milk.  As we discuss above, in our specification, the shrinkage associated with ridge will tend to bias the elasticity estimates towards $-1$. Indeed, when we do not debias, the elasticity estimates from ridge are closer to $-1$ when we employ a higher penalty than with a smaller penalty. This is particularly striking for milk. By contrast, when we debias, which mitigates the shrinkage associated with ridge, we obtain elasticity estimates that are much less sensitive to the choice of penalty.

The elasticity estimates from our debiased ridge method roughly align with those found in the previous literature (see, for example, the meta-analysis of \cite{Andreyeva2010}).

\color{blue}
Compared with cross-sectional OLS and Fixed Effects, the individual ridge estimates relax the assumption of homogeneous coefficients (and intercepts, in the case of OLS). For sufficiently small values of $\lambda$,  this relaxation will typically result in estimates with greater standard errors. On the other hand, if $\lambda$  is sufficiently large, the slope estimates will be shrunk strongly towards zero, generally leading to smaller standard errors than those of the Fixed Effects estimates, and possibly the OLS estimates as well. Indeed, in Table 1 the individual ridge elasticity estimates with a small value of $\lambda$  have greater standard errors than OLS and Fixed Effects, but for the larger value, the milk elasticity standard errors are smaller for individual ridge. Shrinkage towards zero can be a  major source of bias, and our debiasing method acts to counter the overall shrinkage towards zero. As such, for sufficiently large values of $\lambda$, the debiasing will typically increase standard errors. Indeed, in Table 1, for the larger value of $\lambda$, the debiasing is associated with substantially increased standard errors for the milk elasticities. However, for the smaller value of $\lambda$, this effect is less strong. In fact, in Table 1, the debiased ridge estimates with a small $\lambda$ have standard errors that are no greater, and in one case smaller, than without debiasing.

\color{blue}
We apply our methods to estimate an upper bound  on the average equivalent variation consumer surplus and deadweight loss from a $10\%$ increase in price for both soda and milk, while excluding time periods for each individual where the larger price was outside the range of prices in the observed data. This increase is relative to the actual price faced by each household in a particular period. The bound follows the formula in \cite{Hausman},  as detailed in Example 2 in Section 2. The formula requires that we impose a lower bound on the income effect. We take our lower bound to be $0$ which corresponds to the assumption that soda and milk are normal goods. This lower bound on the income effects corresponds to an upper bound on the welfare loss.
\color{black}

In order to provide some distributional analysis, we estimate the welfare bounds separately for households in three different income groups. In particular, for those whose household income (averaged over all periods for which there is data on that household) is in the bottom quartile, top quartile, and for all households.

Tables II and III contain our estimation results for the welfare upper bounds. We apply our analysis for both the log-linear specification (\ref{specification}) and a cubic specification, which supplements the regressors in the linear model with all powers and interactions of the log own-price and total expenditure up to order three. We provide results for various choices of the penalty parameter $\lambda$. The welfare estimates have been annualized; that is, the numbers represent the welfare change over the course of a year.

\begin{table}[ht]
\caption{Soda Welfare Upper Bounds}
\centering
\begin{tabular}{cccc|ccc}
\hline 
 & \multicolumn{3}{c|}{\uline{Deadweight Loss (Linear)}} & \multicolumn{3}{c}{\uline{Deadweight Loss (Cubic)}}\tabularnewline
 & \multicolumn{3}{c|}{Income Quartiles} & \multicolumn{3}{c}{Income Quartiles}\tabularnewline
\cline{2-7} \cline{3-7} \cline{4-7} \cline{5-7} \cline{6-7} \cline{7-7} 
$\lambda$ & Upper & Lower & All & Upper & Lower & All\tabularnewline
\hline
$0.05$ & 0.367 & 0.407 & 0.399 & 0.365 & 0.408 & 0.398\tabularnewline
 & (0.028) & (0.033) & (0.013) & (0.030) & (0.035) & (0.014)\tabularnewline
$0.0005$ & 0.359 & 0.407 & 0.394 & 0.404 & 0.409 & 0.400\tabularnewline
 & (0.029) & (0.041) & (0.015) & (0.042) & (0.041) & (0.020)\tabularnewline
\end{tabular}

\centering
\begin{tabular}{cccc|ccc}
\hline 
 & \multicolumn{3}{c|}{\uline{Consumer Surplus (Linear)}} & \multicolumn{3}{c}{\uline{Consumer Surplus (Cubic)}}\tabularnewline
 & \multicolumn{3}{c|}{Income Quartiles} & \multicolumn{3}{c}{Income Quartiles}\tabularnewline
\cline{2-7} \cline{3-7} \cline{4-7} \cline{5-7} \cline{6-7} \cline{7-7} 
$\lambda$ & Upper & Lower & All & Upper & Lower & All\tabularnewline
\hline 
$0.05$ & 10.13 & 10.54 & 10.64 & 10.12 & 10.58 & 10.66\tabularnewline
 & (0.682) & (0.707) & (0.281) & (0.680) & (0.707) & (0.280)\tabularnewline
$0.0005$ & 10.15 & 10.57 & 10.66 & 10.09 & 10.59 & 10.66\tabularnewline
 & (0.683) & (0.707) & (0.281) & (0.679) & (0.709) & (0.282)\tabularnewline
\end{tabular}

    	\begin{minipage}[t]{1\columnwidth}%
		{\scriptsize{}\color{blue} Upper bounds on soda deadweight loss and consumer surplus calculated using the formulas in Example 2. We use a lower bound of $0$ on the income effect. Figures are calculated separately for households in three  income groups: `Upper', whose average household income  is in the top quartile, `Lower' for those in the bottom quartile, and `All', which includes all households. Results are provided for the `Linear' (in logs)  specification (\ref{specification}) and a `Cubic' specification, which supplements the regressors in the log-linear model with all powers and interactions of the log own-price and total expenditure up to order three. Standard errors calculated using the formula (\ref{varest}) are given in parentheses.\color{black}}
	\end{minipage}{\scriptsize\par}
\end{table}

\begin{table}[ht]
\caption{Milk Welfare Upper Bounds}
\centering
\begin{tabular}{cccc|ccc}
\hline 
 & \multicolumn{3}{c|}{\uline{Deadweight Loss (Linear)}} & \multicolumn{3}{c}{\uline{Deadweight Loss (Cubic)}}\tabularnewline
 & \multicolumn{3}{c|}{Income Quartiles} & \multicolumn{3}{c}{Income Quartiles}\tabularnewline
\cline{2-7} \cline{3-7} \cline{4-7} \cline{5-7} \cline{6-7} \cline{7-7} 
$\lambda$ & Upper & Lower & All & Upper & Lower & All\tabularnewline
\hline 
$0.05$ & 0.178 & 0.148 & 0.158 & 0.142 & 0.140 & 0.121\tabularnewline
 & (0.017) & (0.014) & (0.009) & (0.026) & (0.023) & (0.017)\tabularnewline
$0.0005$ & 0.120 & 0.108 & 0.120 & 0.190 & 0.172 & 0.136\tabularnewline
 & (0.024) & (0.021) & (0.013) & (0.043) & (0.046) & (0.031)\tabularnewline
\end{tabular}

\centering
\begin{tabular}{cccc|ccc}
\hline 
 & \multicolumn{3}{c|}{\uline{Consumer Surplus (Linear)}} & \multicolumn{3}{c}{\uline{Consumer Surplus (Cubic)}}\tabularnewline
 & \multicolumn{3}{c|}{Income Quartiles} & \multicolumn{3}{c}{Income Quartiles}\tabularnewline
\cline{2-7} \cline{3-7} \cline{4-7} \cline{5-7} \cline{6-7} \cline{7-7} 
$\lambda$ & Upper & Lower & All & Upper & Lower & All\tabularnewline
\hline 
$0.05$ & 8.09 & 6.73 & 7.41 & 8.15 & 6.71 & 7.43\tabularnewline
 & (0.490) & (0.390) & (0.170) & (0.495) & (0.394) & (0.171)\tabularnewline
$0.0005$ & 8.15 & 6.76 & 7.44 & 8.12 & 6.70 & 7.43\tabularnewline
 & (0.495) & (0.394) & (0.171) & (0.494) & (0.396) & (0.173)\tabularnewline
\end{tabular}

    	\begin{minipage}[t]{1\columnwidth}%
		{\scriptsize{}\color{blue} Upper bounds on milk deadweight loss and consumer surplus calculated using the formulas in Example 2. We use a lower bound of $0$ on the income effect. Figures are calculated separately for households in three  income groups: `Upper', whose average household income  is in the top quartile, `Lower' for those in the bottom quartile, and `All', which includes all households. Results are provided for the `Linear' (in logs)  specification (\ref{specification}) and a `Cubic' specification, which supplements the regressors in the log-linear model with all powers and interactions of the log own-price and total expenditure up to order three. Standard errors calculated using the formula (\ref{varest}) are given in parentheses.\color{black}}
	\end{minipage}{\scriptsize\par}
\end{table}

The estimates of average  deadweight loss and consumer surplus for the full set of households are remarkably stable, both between the linear and cubic specifications, and for different values of the penalty parameter. In part, this may reflect the tendency of  debiasing to mitigate the shrinkage induced by regularization, and thus to reduce sensitivity to the choice of penalty parameter $\lambda$.

We estimate that the deadweight loss from a price increase for soda is markedly higher than for milk. This is not surprising given that milk, unlike soda, is a staple food, and so demand for this product may be relatively inelastic. Indeed, this aligns with our elasticity estimates in Table I. 

 \cite{Harding2017} analyze the role of prices in determining food purchases and nutrition and estimate the impact of taxes on nutrition and individual welfare. \cite{Allcott2019} and \cite{Dubois2020} have also considered the welfare effects of taxing soda. Like \cite{Dubois2020}, our panel approach estimates individual-specific demands. Our approach is simpler in that it is based on continuous demand modeling and individual ridge regression with total expenditure included in the demand function. Also, our application averages over on-the-go and larger store purchases as well as over individuals that purchase soda and those that do not. We obtain substantially larger estimates of average equivalent variation than their compensating variation which is to be expected because we model household demand and they model individual demand.

Figure 1 plots the quantiles of the nonidentification measure $\zeta_i$ from Section 3.5. \color{black}.

\begin{figure}[ht]
\centering
	\includegraphics[scale=0.23]{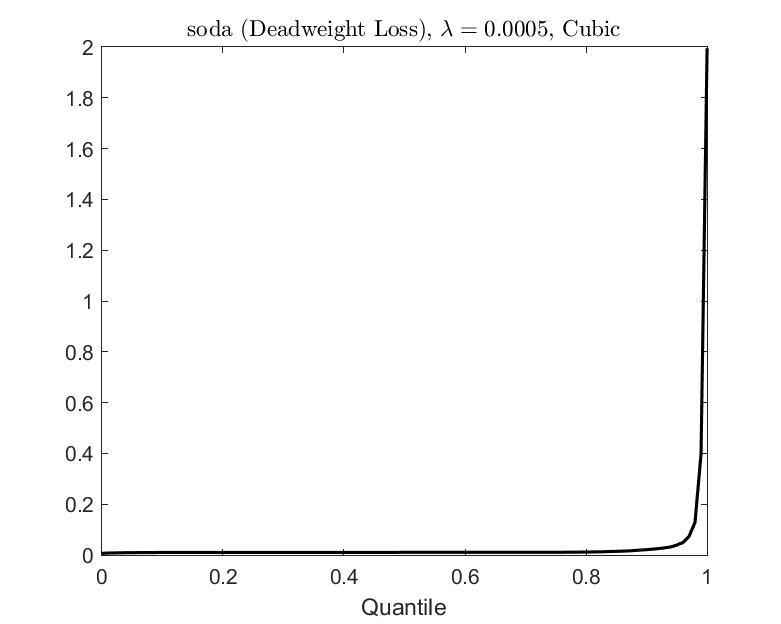}\includegraphics[scale=0.23]{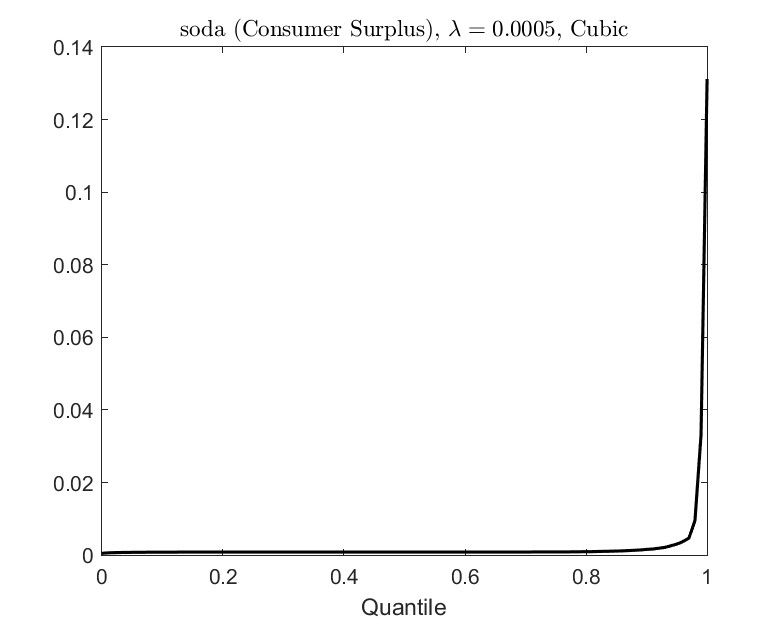}
 \includegraphics[scale=0.23]{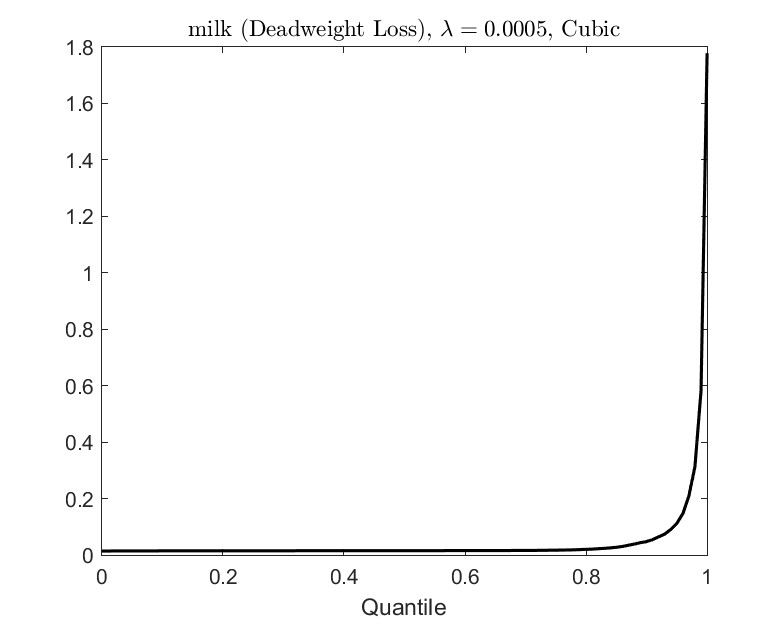}\includegraphics[scale=0.23]{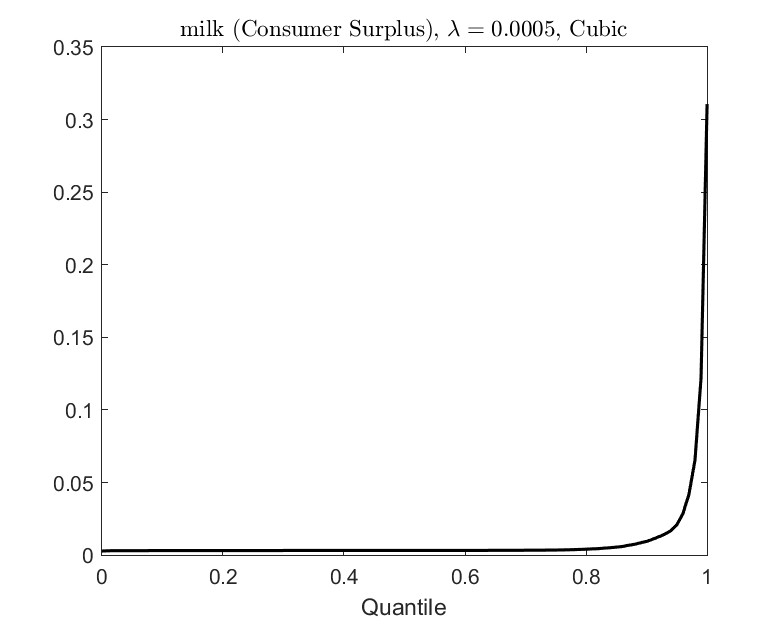}
	\caption{Scaled Distance Between Implied and True $a_i$}	
\end{figure}

We see from the figures that for a large proportion of individuals, estimates the $\zeta_i$ is small. This is particularly clear in the case of our consumer surplus estimates, for which the quantiles below 90\% are almost zero. 

%% file: Asymptoticsetc.tex
\section{Theoretical Results}

We now turn to a formal analysis of the properties of our estimation
procedure. For this purpose, we will be explicit about allowing the number of periods for which we have observations to vary between individuals. In particular, we let $T_i$ denote the number of periods for which we observe data on individual $i$. In addition, we  explicitly define the approximation
error that results from the use of an LRC approximation. Recall that
we employ an approximation $h(x,X_i)\approx b(x)'\beta_i$. To explicitly 
define $\beta_i$, we suppose that $\frac{1}{T_{i}}\sum_{t=1}^{T_{i}}E[b(X_{it})b(X_{it})']$
is non-singular. For each fixed value $\mathbb{X}$ in the support of $X_i$,
define the function $\beta(\mathbb{X})$ as follows.
\[
\beta(\mathbb{X})=E[\frac{1}{T_{i}}\sum_{t=1}^{T_{i}}b(X_{it})b(X_{it})']^{-1}E[\frac{1}{T_{i}}\sum_{t=1}^{T_{i}}b(X_{it})h(X_{it},\mathbb{X})].
\]
That is, $\beta(\mathbb{X})$ is the vector of coefficients from a best
approximation of $h(\cdot,\mathbb{X})$ by a linear combination of the basis
functions $b(\cdot)$. Then we let $\beta_i:=\beta(X_i)$. If an LRC model holds (that is, if $g(x,\eta_{it})=b(x)'\eta_{it}$), then under Assumption 1, the definition above ensures that $\beta_i=E[\eta_{it}|X_i]$, which coincides with the discussion of the LRC model in Section 2. 

We define the approximation error $r(x,X_i)$
as,
\[
r(x,X_i)=h(x,X_i)-b(x)'\beta_i.
\]
We then let $r_{it}=r(X_{it},X_i)$. Note that if outcomes follow an LRC model, then $r(\cdot,X_i)$ is identically zero. We also  define a residual $u_{it}$ by,
\[u_{it}=Y_{it}-h(X_{it},X_i).\]
Let $r_i$ and $u_i$ be the length-$T_i$  column vectors whose $t$-th entries are $r_{it}$ and $u_{it}$ respectively, then we obtain the model
\begin{equation}Y_{i}	=B_i\beta_{i}+u_{i}+r_{i},\,E[u_{i}|X_{i}]=0,\label{model2}
\end{equation}
where $r_i=0$ if the LRC model holds.

It is helpful to introduce some notation. \color{blue}Throughout, for a vector $v$, $\|v\|$ is its Euclidean norm, and for a matrix $A$, $\|A\|$ is the operator norm defined by $\|A\|=\sup_{v\neq0}\|Av\|/\|v\|$. \color{black}  Let $D_{1,i}$ be equal to $D_{i}$ but with the first row and column
removed (recall that by definition, the first row and column of $D_i$ contain only zeros). $J$ is the length of the vector $b(X_{it})$, and
let $B_{1,it}$ denote $b(X_{it})$ with its first entry (the constant)
removed. Finally, define $\bar{Y}_{i}=\frac{1}{T_{i}}\sum_{t=1}^{T}Y_{it}$,
$\bar{B}_{i}=\frac{1}{T_{i}}\sum_{t=1}^{T}B_{1,it}$, and $\tilde{Q}_{i}=\frac{1}{T_{i}}\sum_{t=1}^{T}B_{1,it}B_{1,it}'-\bar{B}_{i}\bar{B}_{i}'$.  

Recall that Section 3 defines $D_i$ to be diagonal with the first diagonal entry zero and the others non-zero. In this section, we allow for more general choices of $D_i$: we retain the condition that the first row and column of $D_i$ contain only zeroes, but unless we state otherwise, $D_{1,i}$ can be any strictly positive-definite matrix. We implicitly assume throughout that the estimator (\ref{theta}) is well defined, that is, $\overline{AW}$ is non-singular.

\subsection{Properties of the Estimator}

Before we turn to the asymptotic behavior of our estimation procedures, we elaborate on a number of notable properties of the estimator outlined in Section 3. In particular, we consider its motivation as an empirical Bayes estimator,  its limiting behavior under large and small values of the penalty parameter, and the sense in which the estimator eliminates regularization bias.

\subsubsection*{Property A: Empirical Bayes Interpretation}

The estimator $\hat{\theta}$ can be expressed as an empirical Bayes
procedure. The empirical Bayes approach imposes a prior on the individual
coefficient vectors $\{\beta_i\}_{i=1}^n$ in (\ref{model2}) and the mean of this prior is estimated jointly with
the individual-level coefficients. 

To be more precise, the debiased ridge estimator is a Bayesian maximum a posteriori (MAP) estimate. The likelihood corresponds to the model (\ref{model2}) in which $r_i=0$ (as in an LRC model) and the residuals $u_{it}$ are iid Gaussian. The prior for the individual-specific coefficients $\{\beta_i\}_{i=1}^n$ is Gaussian and independent across individuals.

It is well-known that standard ridge regression estimates can be expressed as MAP estimates in which the prior is Gaussian with mean zero. What distinguishes our approach is the manner in which the prior mean for $\beta_i$ is determined by the data. In particular, the prior mean $\bar{\beta}$, is pinned down by the  restriction
\begin{equation}\frac{1}{n}\sum_{i=1}^{n}A_{i}{\beta}^{\text{Post}}_{i} =\frac{1}{n}\sum_{i=1}^{n}A_{i}\bar{\beta},\label{constraint}
\end{equation}
where $\{{\beta}^{\text{Post}}_{i}\}_{i=1}^n$ is the posterior mode for $\{{\beta}_{i}\}_{i=1}^n$.  Thus the prior mean for the individual coefficients is fixed by imposing that the prior mode and posterior modes of $\frac{1}{n}\sum_{i=1}^{n}A_{i}{\beta}_i$ are identical. Loosely speaking, it ensures that observing the data does not lead us to update (i.e., improve) upon our prior for $\frac{1}{n}\sum_{i=1}^{n}A_{i}{\beta}_{i}$.

To be yet more precise, consider a Gaussian conditional
likelihood for the outcomes $Y_{it}|X_{i}\overset{iid}{\sim}N(b(X_{it})'\beta_i,\sigma^2)$
and prior for the individual slope parameters $\beta_i\overset{iid}{\sim}N(\bar{\beta},\Sigma)$, 
where $\bar{\beta}$ is the prior mean and $\Sigma$ is the prior variance-covariance
matrix. Given this likelihood and prior, the posterior  density $f^{\text{Post}}$ satisfies the expression below:
\begin{equation}
ln\big(f^{\text{Post}}(\beta_{1},\beta_{2},...,\beta_{n})\big)\propto -\sum_{i=1}^{n}[\frac{1}{T_{i}}\|Y_{i}-B_{i}\beta_{i}\|^{2}+\sigma(\beta_{i}-\bar{\beta})^{\prime}\Sigma^{-1}(\beta_{i}-\bar{\beta})]\label{eq:empb1}
\end{equation}

The parameters $\{\beta_{i}\}_{i=1}^{n}$ that maximize the above (given a fixed $\bar{\beta}$)
are the MAP estimates. In order to obtain an empirical Bayes estimate, we estimate the prior
mean $\bar{\beta}$ from the data by imposing equation (\ref{constraint}).

\theoremstyle{plain} \newtheorem*{P6}{Proposition 1} 
\begin{P6} 
 The estimator $\hat{\theta}$ in (\ref{theta}) can be written as $\hat{\theta}=\frac{1}{n}\sum_{i=1}^{n}a_{i}'{\beta}_{i}^{\text{Post}}$
where $\{{\beta}_{i}^{\text{Post}}\}_{i=1}^n$ and $\bar{\beta}$ 
jointly solve the equation (\ref{constraint}), and equation (\ref{eq:ridgen}) below:
\begin{align}
\{{\beta}_{i}^{\text{Post}}\}_{i=1}^{n}= & \arg\max_{\{\beta_{i}\}_{i=1}^{n}}-\sum_{i=1}^{n}[\frac{1}{T_{i}}\|Y_{i}-B_{i}\beta_{i}\|^{2}+\lambda(\beta_{i}-\bar{\beta})^{\prime}D_{i}(\beta_{i}-\bar{\beta})].\label{eq:ridgen}
\end{align}
\end{P6}
The objective (\ref{eq:ridgen}) is a monotone transformation of a
Bayesian posterior (\ref{eq:empb1}) when $D_{i}=\frac{\sigma}{\lambda}\Sigma^{-1}$. Thus $\{{\beta}_{i}^{\text{Post}}\}_{i=1}^n$ are MAP estimates of the individual slopes. The prior is fixed by the second equation.\footnote{Note that we take $D_i$ to have first row and columns composed of zeroes, and so $D_i$ is singular. As such, strictly speaking we use flat `improper' prior for the individual intercepts.}
 In the special case in which $A_i$ is the identity, solving the two equations above is equivalent to maximizing the objective in (\ref{eq:ridgen}) jointly over both $\beta_i$ and $\bar{\beta}$. That is, $\bar{\beta}$ can be obtained by
\begin{align*}
\{\bar{\beta},{\beta}_{i}^{\text{Post}}\}_{i=1}^{n}= & \arg\max_{\bar{\beta},\{\beta_{i}\}_{i=1}^{n}}-\sum_{i=1}^{n}[\frac{1}{T_{i}}\|Y_{i}-B_{i}\beta_{i}\|^{2}+\lambda(\beta_{i}-\bar{\beta})^{\prime}D_{i}(\beta_{i}-\bar{\beta})].
\end{align*}

\subsubsection*{Property B: Convergence to Fixed Effects with Large Penalty}

As the penalty parameter grows to infinity, our estimator converges to a plug-in fixed effects or generalized fixed effects estimator. To state this formally, let us first note that the standard fixed effects estimate $\hat{\beta}_{FE,i}$ may be expressed as follows. The first component of this vector is an  individual
intercept given by $\bar{Y}_{i}-\bar{B}_{i}'\hat{\beta}_{FE,1}$, where $\hat{\beta}_{FE,1}$ is 
a vector of shared slope parameters and constitutes the remaining components of $\hat{\beta}_{FE,i}$. The slopes are given by
\[
\hat{\beta}_{FE,1}=(\frac{1}{n}\sum_{i=1}^{n}\tilde{Q}_{i})^{-1}\frac{1}{n}\sum_{i=1}^{n}\frac{1}{T_{i}}\sum_{t=1}^{T_{i}}(B_{1,i,t}-\bar{B}_{i})Y_{it}.
\]

$\hat{\beta}_{FE,i}$ is a special case of a generalized fixed-effects
estimator $\hat{\beta}_{GFE,i}$. Again, the first component of $\hat{\beta}_{GFE,i}$
is an individual intercept, in this case  $\bar{Y}_{i}-\bar{B}_{i}'\hat{\beta}_{GFE,1}$,
where $\hat{\beta}_{GFE,1}$ is a vector of shared slopes.
Let $G_{i}$ be a non-singular weighting matrix, then the corresponding vector of 
slope parameters $\hat{\beta}_{GFE,1}$ is defined as follows: 
\[
\hat{\beta}_{GFE,1}=(\frac{1}{n}\sum_{i=1}^{n}G_{i}\tilde{Q}_{i})^{-1}\frac{1}{n}\sum_{i=1}^{n}\frac{1}{T_{i}}\sum_{t=1}^{T_{i}}G_{i}(B_{1,i,t}-\bar{B}_{i})Y_{it}.
\]

A plug-in fixed-effects estimate of $\theta_{0}$ is given by $\frac{1}{n}\sum_{i=1}^{n}a_{i}'\hat{\beta}_{FE,i}$, 
and plug-in generalized fixed-effects estimator by $\frac{1}{n}\sum_{i=1}^{n}a_{i}'\hat{\beta}_{GFE,i}$.

\theoremstyle{plain} \newtheorem*{P3}{Proposition 2} 
\begin{P3} 

$\lim_{\lambda\to\infty}\hat{\theta}=\frac{1}{n}\sum_{i=1}^{n}a_{i}'\hat{\beta}_{GFE,i}$
where $G_{i}=D_{1,i}^{-1}$. If $D_{i}$ does not vary with $i$, then
$\lim_{\lambda\to\infty}\hat{\theta}=\frac{1}{n}\sum_{i=1}^{n}a_{i}'\hat{\beta}_{FE,i}$.
\end{P3}
The proposition states that as the penalty parameter grows towards
infinity, the debiased panel ridge estimator converges to the plug-in
generalized fixed-effects estimator with $G_{i}$ equal to the inverse of $D_{1,i}$. In the special case in which $D_{i}$ does not depend on $i$, this is identical to the standard plug-in fixed effects estimator.

\subsubsection*{Property C: No Regularization Bias Under Exogenous Effects}

Our estimator is based on individual-level ridge regressions. Ridge estimates typically suffer from `regularization' bias. The form of our estimator (\ref{theta}) is designed to mitigate, and in some cases entirely eliminate, regularization bias. In particular, in the case in which $\beta_i$ is constant, apart from the intercept, and there is no approximation error.  

For some insight into the bias properties of the estimator, it is helpful
to compare our method with a plug-in estimator based on individual-specific
OLS. An individual OLS estimate of $\beta_{i}$ is
given below, where $Q_{i}^{\dagger}$ is the pseudo-inverse of $Q_{i}$
and is well-defined even if $Q_{i}$ is singular:
\[
\tilde{\beta}_{i}=Q_{i}^{\dagger}B_{i}'Y_{i}/T_i
\]

A plug-in OLS estimator of $\theta_{0}$ is then $\frac{1}{n}\sum_{i=1}^{n}a_{i}'\tilde{\beta}_{i}$. Suppose Assumptions 1 and 2 hold. 
If $Q_{i}$ is non-singular for all $i$, and the mean of $a_{i}'\tilde{\beta}_{i}$
is finite, then the plug-in OLS estimator is unbiased (up to approximation
error) for $\theta_{0}$. This is because, in the absence of approximation
error, $\tilde{\beta}_{i}$ is a conditionally (on $X_i$)
unbiased estimate of $\beta_{i}$. Moreover, by Assumption 2, $\tilde{\beta}_{i}$ is independent of $a_i$ conditional on $X_i$.  Therefore, if $E[|a_{i}'\tilde{\beta}_{i}|]<\infty$
then we can apply the law of iterated expectations and obtain:
\[
E[a_{i}'\tilde{\beta}_{i}]=E\big[a_{i}'E[\tilde{\beta}_{i}|X_{i},a_i]\big]=E\big[a_{i}'E[\tilde{\beta}_{i}|X_{i}]\big]=E[a_{i}'\beta_{i}].
\]

If individuals are drawn independently and identically from the population,
then consistency of the plug-in OLS estimator follows by the law of
large numbers. However, if $Q_{i}$ is singular with positive probability,
this argument fails because an $\tilde{\beta}_{i}$ is (in general)
biased when $Q_{i}$ is singular. Moreover, the assumption that $E[|a_{i}'\tilde{\beta}_{i}|]$
is finite is crucial. If this moment is infinite, then one cannot apply the law of iterated expectations, nor the law of large numbers. The first moment may be infinite even if $Q_{i}$ is non-singular
almost surely. \cite{Graham2012}  acknowledge that the finite
mean condition may fail, particularly if the number of regressors
is close to the number of time periods. In the case of $a_{i}$ constant,
this situation coincides with the case in which the information bound
derived in \cite{Chamberlain1992} is infinite, and thus regular estimation
is impossible with the number of time periods fixed.

In contrast to OLS, individual-specific ridge estimates have finite
expectation under weak conditions. Proposition 3 provides conditions under which  the expectation of $a_{i}'\hat{\beta}_{i}$ is finite, where $\hat{\beta}_i$ is the individual ridge regression estimate defined in Section 3. The proposition applies even if $Q_{i}$ is singular with positive probability. 

\theoremstyle{plain} \newtheorem*{P1}{Proposition 3} \begin{P1}
Suppose $\lambda>0$, $D_{1,i}$ has eigenvalues bounded below by $c>0$, $\|B_{i}\|$ and $\|a_{i}\|$ are uniformly bounded, and $E[|Y_{it}|]$ is finite. Then $E[|a_{i}'\hat{\beta}_{i}|]<\infty$.
\end{P1}

As we discuss in Section 3, individual ridge estimates are conditionally
biased, even in the absence of approximation error. This in turn suggests
that the sample average $\frac{1}{n}\sum_{i=1}^{n}a_{i}'\hat{\beta}_{i}$
is biased for $E[a_{i}'\beta_{i}]$, even if $r_{it}$ in (\ref{model2}) is identically zero (i.e., if an LRC holds). This motivates our debiasing
strategy. Proposition 4 shows that if effects are exogenous, then the debiased
estimator is exactly unbiased up to approximation error. Note that
the theorem holds for any $\lambda>0$ and also applies when $Q_{i}$
is singular with positive probability. This contrasts with the average of plug-in OLS, which is in general conditionally biased when $Q_i$ is singular with positive probability. By `effects are exogenous', we mean that $\beta_{i,2}$ is mean independent of $X_i$, where $\beta_{i,2}$ is the subvector of $\beta_i$ formed by removing its first component.

\theoremstyle{plain}
\newtheorem*{P2}{Proposition 4} 
\begin{P2} 
Suppose Assumptions 1 and 2 hold, $r_{it}=0$ almost surely, and $\beta_{i,2}$ is mean independent of $X_i$. In addition, 
suppose $D_i$ is a function of  $X_i$. If $\frac{1}{n}\sum_{i=1}^{n}A_{i}W_{i}$
is non-singular then
\[
E[\hat{\theta}|X_{1},X_{2},...,X_{n}]=\frac{1}{n}\sum_{i=1}^{n}E[a_{i}'\beta_{i}|X_{1},X_{2},...,X_{n}].
\]
If the above holds and $E[|\hat{\theta}|]<\infty$ then $E[\hat{\theta}]=E[a_{i}'\beta_{i}]$.
\end{P2}

Proposition 4 requires that $\beta_{i,2}$ is mean independent of $X_i$. The first component of $\beta_i$ is unrestricted. We do not need to restrict this component because we do not penalize the intercept in our individual-ridge regressions (this is why the first row and column of $D_i$ are composed of zeros). A sufficient (but not necessary) condition for the whole vector $\beta_i$ to be mean independent of $X_i$ is that $\eta_{it}$ is independent of $X_i$.

If we strengthen the condition in Proposition 4 so that the entire vector $\beta_i$ is mean independent of $X_i$, then the proposition applies to a general class of estimators. Consider
that we can rewrite the estimator (\ref{theta}) as follows:
\begin{equation}
\hat{\theta}=\bar{a}^{\prime}(\overline{AW})^{-1}\overline{AW\beta},\text{ }\bar{a}=\frac{1}{n}\sum_{i=1}^{n}a_{i},\text{ }\overline{AW}=\frac{1}{n}\sum_{i=1}^{n}A_{i}W_{i},\text{ }\overline{AW\beta}=\frac{1}{n}\sum_{i=1}^{n}A_{i}W_{i}\tilde{\beta}_{i}.\label{general_form}
\end{equation}

If we replace $W_{i}:=(Q_{i}+\lambda D_{i})^{-1}Q_{i}$ with some
other conformable matrix that depends only on the regressors, then we obtain an alternative estimate of
$\theta_{0}$. Consider the special case in which $a_{i}$ is constant, and let $W_{i}$ be an indicator that the determinant of $Q_{i}$
exceeds a cut-off $h$, then the formula yields the estimator of Graham and Powell
(absent adjustment for time-effects). If $\beta_i$ is mean independent of $X_i$, Proposition 2 applies for any estimator of the form above so long as  $W_{i}Q_{i}^{\dagger}Q_{i}=W_{i}$, which 
holds both for our choice of $W_{i}$ as well as that of Graham and
Powell.

\subsubsection*{Property D: Convergence Under Small Penalty}

Suppose that for each $i$, the matrix $Q_{i}$ is non-singular. Then, as $\lambda$
goes to zero, the matrix $(Q_{i}+\lambda D_{i})^{-1}$ converges to
$Q_{i}^{-1}$, for every $i$. As such, our estimate converges to
the plug-in average of individual OLS estimates $\frac{1}{n}\sum_{i=1}^{n}a_{i}'\tilde{\beta}_{i}$. Note the contrast with Property A. As $\lambda\to\infty$ our estimator
converges to a plug-in average of (generalized) fixed
effects estimates, with individual-specific intercepts and shared
slope parameters. If $Q_{i}$
is non-singular for all $i$, then as $\lambda\to0$, the estimator converges instead to the
plug-in average of OLS estimates, which have both individual-specific intercepts and slopes. The choice of
$\lambda$ thus allows us to smoothly transition between these two estimators.

When $Q_{i}$ is singular for some individuals, our estimator generally
does not converge to plug-in individual OLS, but nonetheless, it has an interpretable limit. Proposition 5 considers the special case in which one of the regressors, denoted by $Z_{it}$, is discretely distributed. Because $Z_i$ is discrete, it  may be constant over time for some individual $i$, in which case $Q_i$ is singular.

\theoremstyle{plain} \newtheorem*{P4}{Proposition 5} 
\begin{P4} 
 Suppose $b(X_{it})=(1,Z_{it},B_{2,it}')'$, where $Z_{it}$ is a discrete scalar and the vector $B_{2,it}$ is continuously distributed. Let $C_{i}$ be equal to $1$ if $Z_{i1}=Z_{i2}=...=Z_{iT_{i}}$ and
zero otherwise, and let $\hat{p}=\frac{1}{n}\sum_{i=1}^{n}(1-C_{i})$.

Suppose i. $D_i$ is diagonal, ii. if $C_i= 0$ then  $Q_{i}$ is non-singular, and iii. the submatrix of $\tilde{Q}_{i}$ formed
by removing its first row and column is non-singular. Define  estimates 
\begin{align*}
\tilde{\beta}_{i,1}^{*} & =\bar{Y}_{i}-\bar{Z}_{i}\tilde{\beta}_{i,2}^{*}-\bar{B}_{2,i}'\tilde{\beta}_{i,3}\\
\tilde{\beta}_{i,2}^{*} & =(1-C_{i})\tilde{\beta}_{i,2}+C_{i}\frac{1}{\hat{p}n}\sum_{j=1}^{n}(1-C_{j})\tilde{\beta}_{j,2},
\end{align*}
where $\tilde{\beta}_{i,2}$ and $\tilde{\beta}_{i,3}$ are respectively the individual OLS 
coefficients on $Z_{it}$ and $B_{2,it}$.
Let $\tilde{\beta}_{i}^{*}=(\tilde{\beta}_{i,1}^{*},\tilde{\beta}_{i,2}^{*},\tilde{\beta}_{i,3}')'$. 
 Then $\lim_{\lambda\to0}\hat{\theta}=\frac{1}{n}\sum_{i=1}^{n}a_{i}'\tilde{\beta}_{i}^{*}$.
\end{P4}

The limit in Proposition 5 differs from plug-in individual OLS in that
the OLS coefficient on $Z_{it}$ is replaced with the alternative
estimate $\tilde{\beta}_{i,2}^{*}$, and the intercept is adjusted
accordingly. If $Z_{it}$ varies for individual
$i$, then $\tilde{\beta}_{i,2}^{*}$ is equal to the individual
OLS estimate $\tilde{\beta}_{i,2}$. However, if $Z_{it}$ does not vary, then $\tilde{\beta}_{i,2}^{*}$ is equal to $\frac{1}{\hat{p}n}\sum_{i=1}^{n}(1-C_{i})\tilde{\beta}_{i,2}$, 
which is the average of the OLS coefficients among the individuals
for whom $Z_{it}$ does vary. In other words, for individuals without
variation in $Z_{it}$, we impute the value of this coefficient as
the average among individuals for whom $Z_{it}$ varies. 

\theoremstyle{plain} \newtheorem*{P5}{Proposition 6} 
\begin{P5} 
Define $\tilde{B}_{it}=D_{1,i}^{-1/2}(B_{1,it}-\bar{B}_{i})$ and
let $\tilde{B}_{i}=(\tilde{B}_{i1},\tilde{B}_{i2},...,\tilde{B}_{iT_{i}})'$.
Define the projection matrix $P_{i}=D_{1,i}^{-1/2}\tilde{B}_{i}^{\dagger}\tilde{B}_{i}D_{1,i}^{1/2}$
and the following vectors of coefficients
\begin{align*}
\tilde{\beta}_{i,2}^{\circ} & =D_{1,i}^{-1/2}\tilde{B}_{i}^{\dagger}Y_{i}/T_{i}\\
\tilde{\beta}_{i,2}^{*} & =\tilde{\beta}_{i,2}^{\circ}+(I-P_{i})(\frac{1}{n}\sum_{j=1}^{n}P_{j})^{-1}\frac{1}{n}\sum_{j=1}^{n}\tilde{\beta}_{j,2}^{\circ}.
\end{align*}
In addition, let $\tilde{\beta}_{i,1}^{*}=(\bar{Y}_{i}-\bar{B}_{i}'\tilde{\beta}_{i,2}^{*})$
and define $\tilde{\beta}_{i}^{*}=(\tilde{\beta}_{i,1}^{*},{\tilde{\beta}_{i,2}^{*}}{}')'$/
Then $\lim_{\lambda\to0}\hat{\theta}=\frac{1}{n}\sum_{i=1}^{n}a_{i}'\tilde{\beta}_{i}^{*}$.
\end{P5}

Proposition 6 considers the small $\lambda$ limit of our estimator in the general case. If $Q_{i}$ has full rank, then $\tilde{\beta}_{i}^{*}=\tilde{\beta}_{i}$, the individual OLS estimate.
To interpret $\tilde{\beta}_{i}^{*}$ when $Q_{i}$
is singular, first note that $P_{i}$ is the orthogonal (with respect to the inner-product $\langle a,b\rangle:=a'D_{i}b$) projection
onto the range of $\tilde{B}_{i}'$. 
Therefore, $I-P_{i}$ is the orthogonal projection onto the null space of $\tilde{B}_{i}$.
If $Q_{i}$ is singular, this null space is non-trivial. This is problematic because, by construction, $(I-P_{i})\tilde{\beta}_{i,2}^{\circ}=0$. That
is, the projection of the coefficients $\tilde{\beta}_{i,2}^{\circ}$
onto this subspace is zero, and so, loosely speaking, this part of $\tilde{\beta}_{i,2}^{\circ}$ is missing. The second term in the definition of $\tilde{\beta}_{i,2}^{*}$
adjusts for this by, in effect, replacing the missing part of $\tilde{\beta}_{i,2}^{\circ}$
with the projection of $(\frac{1}{n}\sum_{i=1}^{n}P_{i})^{-1}\frac{1}{n}\sum_{i=1}^{n}\tilde{\beta}_{i,2}^{\circ}$
onto this subspace. In the special case in Proposition 5, when $Q_{i}$
is singular, the null space consists simply of the vectors proportional to
$(1,0,0,...,0)'$, and projecting onto this space yields the coefficients on $Z_{it}$.

\subsection{Consistency and Asymptotic Normality}

In order to establish the statistical properties of our estimator, we impose
some additional conditions. In the assumptions below, inequalities
involving random variables are understood to hold almost surely. Throughout, we define $\delta:=E[\|(Q_{i}+\lambda D_{i})^{-1}\|]$. \color{blue}Note that $\delta$ may change with the sample size. This is because it depends directly on $\lambda$, which we let shrink as the sample grows, and $Q_i$ which changes with the dimension of the vector of basis functions $b(\cdot)$ and the number of time series observations $T_i$. Similarly, $a_i$, $A_i$, $B_i$, $r(\cdot,\cdot)$, $u_i$, and $\beta_i$ all depend on the dimension of the series approximation, and therefore, they too may change with the sample size.\color{black}

\theoremstyle{definition} \newtheorem*{A3}{Assumption 3 (Consistency)} 
\begin{A3} 
For some scalar $0<c<\infty$, i. $\|A_{i}\|,\|a_{i}\|,\|B_i\|\leq c$,
ii. $\|(\frac{1}{n}\sum_{i=1}^{n}A_{i})^{-1}\|\leq c$, iii. $\|\beta_{i}\|\leq c$,
iv. $\|E[u_{i}u_{i}'|X_{i}]\|\leq c$, v. $H_{t}^{-}(\cdot)$ and
$H_{t}^{+}(\cdot)$ are uniformly bounded, vi. $\sup_{x,\mathbb{X}}|r(x,\mathbb{X})|\leq\ell$
with $\ell\to0$,  vii. $T_{i}\geq T$, viii. The first row and
column of $D_{i}$ contains only zeros, and the eigenvalues of $D_{1,i}$
are bounded above by $c$ and below by $1/c$. 
\end{A3}

\theoremstyle{definition} \newtheorem*{A4}{Assumption 4 (Asymptotic Normality)} 
\begin{A4} 
For some finite constants
$c,\xi,v,q>0$ such that $(v-2)(q-2)>4$, i. $\frac{1}{T}\sum_{t=1}^{T}E[u_{it}^{v}]\leq c$,
ii. $1/c\leq Var(a_{i}'\beta_{i})$, and iii. we have
\[
n^{\frac{1}{v}+\frac{1}{q}-\frac{1}{2}}E[\|(Q_{i}+\lambda D_{i})^{-1}\|^{q/2}]^{1/q}(\delta/T)^{\xi}=o(1).
\]
\end{A4}

\theoremstyle{definition} \newtheorem*{A5}{Assumption 5 (Remainders)} 
\begin{A5}
$\lambda\delta,\ell\sqrt{\delta},\ell=o(\sqrt{\frac{1}{n}})$,
$\frac{\delta}{T}=O(1)$, and $\frac{J\lambda^{2}\delta^{3}}{T}=o(1)$.
\end{A5}

Assumption 3 i. and ii. impose conditions on $a_{i}$, $A_{i}$, and $B_i$ which are chosen directly by the
researcher. 3.iii imposes that the individual-specific mean parameter
$\beta_{i}$ is bounded in norm. iv. concerns the conditional second
moments of $u_{i}$. The condition allows for serial correlation in $u_{it}$, but it restricts the correlation so that the operator norm
of $E[u_{i}u_{i}'|X_{i}]$ remains bounded as $T$ grows. In the special case in which $u_{i}$ is serially uncorrelated and homoskedastic, $\|E[u_{i}u_{i}'|X_{i}]\|$ equals the constant variance of $u_{it}$, and so, the condition holds. 3.v restricts
$H_{t}^{-}(\cdot)$ and $H_{t}^{+}(\cdot)$. 3.vi is a condition on the sieve approximation
error. Conditions of this form hold for many choices of sieve space
used in practice under smoothness conditions on $h(\cdot,\mathbb{X})$,
see e.g., \cite{DeVore1993} for examples. 3.vii imposes that the number of time periods
$T_{i}$, which can vary between individuals, is bounded below by
some $T$. 3.viii is a weak condition on $D_{i}$ which is chosen by the researcher.

Assumption 4 ensures a normal limiting
distribution via a Lyapunov
condition. 
The assumption restricts the $v$-th moment of one random variable and the $q$-th
moment of another. $q$ and $v$ must be strictly positive and satisfy $(v-2)(q-2)>4$, which implies that $v,q>2$. The  conditions trade-off in the sense that if $v$ is large, then $q$ need not be much larger than $2$ and 
vice versa. 4.i bounds the average $v$-th moment of $u_{it}$.
4.ii states that the variance of the individual-specific approximation $a_{i}'\beta_{i}$, 
 is bounded below \color{blue} which requires that either  $a_i$ or $\beta_i$ varies between individuals. If the condition fails, then the estimator may achieve faster than $\sqrt{n}$-convergence, in which case the asymptotic variance that we derive is degenerate. This issue is studied in a related setting by \cite{fernandezval2025}, who show that it results in conservative inference for plug-in methods. A similar phenomenon is also considered in \cite{Pesaran}\color{black}. 4.iii requires
that a particular sequence is $o(1)$. Each entry in the sequence is a product of
three terms. The first is $n^{\frac{1}{v}+\frac{1}{q}-\frac{1}{2}}$.
The conditions on $v$ and $q$ imply that this term goes to zero with $n$. The second term
 is the $q/2$-th moment of $\|(Q_{i}+\lambda D_{i})^{-1}\|$
raised to the power $1/q$, the third is $\delta/T$ raised
to the power $\xi$. Note that $\xi$ can be any strictly positive
constant. As such, in the case of $\delta/T\to\infty$, 4.iii
holds for a sufficiently large choice of $\xi$ so long as the $q$-th
moment of $\|(Q_{i}+\lambda D_{i})^{-1}\|$ grows (at most) at a polynomial rate with $T$.

Assumption 5 restricts the rates at which various
sequences converge to zero. It ensures some terms in the asymptotic
expansion of the estimation error are second order.\\

Theorem 2 provides general asymptotic theory for our estimator
(\ref{theta}). It applies both when regressors are continuous or discrete or a combination of the two. The result
follows from the more general result in Lemma 1 in the appendix, which applies to any estimator
of the form (\ref{general_form}) such that 
$W_{i}Q_{i}^{\dagger}Q_{i}=W_{i}$.
\theoremstyle{plain} \newtheorem*{T1}{Theorem 2 ( Asymptotics)} 
\begin{T1}
Suppose Assumptions 1, 2, and 3 hold and $\lambda\delta\to0$.

\textbf{a. (Consistency)} 

\begin{align*}
\hat{\theta}-\theta_{0} & =O_{p}\bigg(\sqrt{\frac{1}{n}}+\sqrt{\frac{\delta}{nT}}+\lambda\delta+\lambda\delta\sqrt{\frac{J\delta}{nT}}+\ell(1+\sqrt{\delta})\bigg).
\end{align*}

\textbf{b. (Asymptotic Normality)}

In addition, if Assumptions 4 and 5 hold, then $\hat{\theta}-\theta_{0}=O_{p}(n^{-1/2})$
and $\sqrt{n}\sigma^{-1}(\hat{\theta}-\theta_{0})\sim^{a}N(0,1)$,
where $\sigma^{2}$ is given by:
\begin{equation}
\sigma^{2}=E[|\frac{1}{T_{i}}a'_{i}W_{i}Q_{i}^{\dagger}B_{i}'u_{i}|^{2}]+Var(a_{i}'\beta_{i}).\label{avar}
\end{equation}
\end{T1}

Theorem 2 shows the crucial role of $\delta$ in the asymptotic
behavior of the estimator. To understand what $\delta$ represents, suppose for simplicity that $D_{i}$ is the identity matrix.
Then $\|(Q_{i}+\lambda D_{i})^{-1}\|$ is equal
to $(\mu_{\text{min}}(Q_{i})+\lambda)^{-1}$, where $\mu_{\text{min}}(Q_{i})$
is the smallest eigenvalue of $Q_{i}$. As such, if $Q_{i}$ is close
to singular, then $\|(Q_{i}+\lambda D_{i})^{-1}\|$ is close to $1/\lambda$.
In the extreme case, if $Q_{i}$ is singular with probability $p$,
then $p/\lambda\leq\delta$. Therefore, the condition $\lambda\delta\to0$
is only possible if $p$ shrinks to zero with the sample size, which
generally requires that $T$ grows with $n$. On the other hand, if $E[1/\mu_{\text{min}}(Q_{i})]$ is bounded by a finite constant, 
then $\delta\leq E[1/\mu_{\text{min}}(Q_{i})]$, uniformly over
$\lambda$. It follows that $\lambda\delta=o(n^{-1/2})$, so long as $\lambda$
shrinks sufficiently quickly to zero.

To examine this in more detail, we consider two extreme cases below.
In the first case, captured in Corollary 1, we suppose that $E[\mu_{\min}(\tilde{Q}_{i})^{-1}]$
is bounded above, where $\mu_{\min}(\tilde{Q}_{i})$ is the smallest eigenvalue of $\tilde{Q}_{i}$. This is only possible if all regressors are continuously
distributed and $T>J$. In the absence of approximation
error (so that $\ell=0$ with $J<T$ fixed), root-$n$ consistency
and asymptotic normality do not require that $T$ grows with the sample
size. Indeed, if $E[\mu_{\min}(\tilde{Q}_{i})^{-1}]$
is finite, then the efficiency bound in
\cite{Chamberlain1992} is finite and root-$n$ regular estimation is possible.

\theoremstyle{plain} \newtheorem*{C1}{Corollary 1 (Continuous Case)} 
\begin{C1}
Let $\|Q_{i}\|\leq c$. Suppose Assumptions 1, 2, and 3 hold. If $E[\mu_{\min}(\tilde{Q}_{i})^{-1}]$ is bounded
above and $\lambda\to0$, then:
\begin{align*}
\hat{\theta}-\theta_{0} & =O_{p}\bigg(\sqrt{\frac{1}{n}}+\lambda+\lambda\sqrt{\frac{J}{nT}}+\ell\bigg).
\end{align*}

If in addition Assumption 4 holds, $\ell,\lambda=o(\sqrt{\frac{1}{n}})$
and $\frac{J\lambda^{2}}{T}=o(1)$, then $\hat{\theta}-\theta_{0}=O_{p}(n^{-1/2})$
and $\sqrt{n}\sigma^{-1}(\hat{\theta}-\theta_{0})\sim^{a}N(0,1)$,
where $\sigma^{2}$ is given by (\ref{avar}).
\end{C1}

At the other extreme, Corollary 2 applies in the case of a single
binary regressor. We assume that for a given individual $i$, the
probability $X_{it}=1$ is given by $\pi_{i}\in(0,1)$ and that, conditional
on $\pi_{i}$, the regressor is independent over time. In this case, 
$\tilde{Q}_{i}$ must be singular with positive probability because
for an individual $i$, the regressor is constant with positive probability. However, as $T_{i}$ grows, this probability shrinks to zero at a rate that depends on the distribution of $\pi_i$.

\theoremstyle{plain} \newtheorem*{C2}{Corollary 2 (Binary Case)} 
\begin{C2}
Suppose Assumptions 1, 2, and 3 hold, and 
$b(X_{it})=(1,X_{it})'$ where $X_{it}$ is binary. Suppose $P(X_{it}=1|\pi_{i})=\pi_{i}$
and the entries of the sequence $\{X_{it}\}_{t=1}^{T_i}$ are jointly
independent conditional on $\pi_{i}$. Let $\pi_{i}$ admit a probability
density $f_{\pi}$ so that $f_{\pi}(\pi)\leq C(1-\pi)^{\omega}\pi{}^{\omega}$
where $\omega>0$. Then if $\lambda\to 0$ and $T\to\infty$ we have:
\begin{align*}
\hat{\theta}-\theta_{0} & =O_{p}\bigg(\lambda+T^{-(1+\omega)}+\sqrt{\frac{1}{n}}+\sqrt{\frac{1}{nT}}+\sqrt{\frac{T^{-(2+\omega)}}{\lambda n}}\bigg).
\end{align*}

In addition, if $\frac{T^{-(1+\omega)}}{\lambda}=O(1)$, $T^{-(1+\omega)},\lambda=o(\sqrt{1/n})$,
and Assumptions 4.i and 4.ii hold with $q/2<\omega$, then $\hat{\theta}-\theta_{0}=O_{p}(n^{-1/2})$
and $\sqrt{n}\sigma^{-1}(\hat{\theta}-\theta_{0})\sim^{a}N(0,1)$,
where $\sigma^{2}$ is given by (\ref{avar}).
\end{C2}

Corollary 2 establishes root-$n$ consistency only under the condition that $T^{-(1+\omega)}=o(\sqrt{1/n})$. Thus the rate at which $T$ must grow with $n$ depends on the rate at which $f_{\pi}(\pi)$ goes to zero as $\pi$ goes to zero or one. If $f_{\pi}(\pi)$ goes quickly to zero, then the probability $X_{it}$ is constant over time goes to zero quickly as $T$ grows, and so $T$ need not increase rapidly with $n$. This phenomenon is considered in \cite{Chernozhukov2013} and tied to the rate at which the identified set shrinks with $T$.

Results for other cases, for example, with both discrete and continuous regressors, may also be obtained from Theorem 2. As in the proofs of Corollaries 1 and 2, it would suffice to derive a convergence rate for $\delta$ under suitable assumptions.

%% file: tables_final.tex
\section{Additional Empirical Results}
Table IV contains cross-sectional OLS expenditure, own-price, and cross-price elasticity estimates for both soda and milk. The figures are estimated using the baseline model in Section 5 without seasonal dummies. Standard errors are to the right of the coefficient and elasticity estimates. For soda the cross-price elasticity with greatest magnitude is for butter, at $-0.2380$ and most of the estimates have magnitude below $0.1$. For milk, all cross-price elasticities have magnitude less than $0.2$ and many have magnitude below $0.01$.
\begin{table}[h]
\caption{OLS cross-price elasticities for soda and milk, 2010-2014}
\begin{tabular}{r|rrrr|rrrr}
\tabularnewline
 &  \multicolumn{4}{c|}{Soda} & \multicolumn{4}{c}{Milk}\tabularnewline
  \hline
 & Coeff & s.e & Elast & s.e.  & Coeff & s.e. & Elast & s.e. \\ 
  \hline
  exp & 0.0208 & 0.0010 & 1.1452 & 0.0067 & -0.0044 & 0.0008 & 0.9608 & 0.0073 \\ 
  soda & 0.0295 & 0.0005 & -0.7945 & 0.0033  & -0.0024 & 0.0004 & -0.0216 & 0.0034 \\ 
  soup & -0.0030 & 0.0012 & -0.0212 & 0.0084 & -0.0008 & 0.0009 & -0.0071 & 0.0078 \\ 
  water & 0.0007 & 0.0005 & 0.0051 & 0.0034 & 0.0002 & 0.0003 & 0.0022 & 0.0030 \\ 
  butter & -0.0340 & 0.0011 & -0.2370 & 0.0079 & 0.0003 & 0.0008 & 0.0024 & 0.0069  \\ 
  cookies & -0.0031 & 0.0007 & -0.0215 & 0.0051 & -0.0042 & 0.0005 & -0.0373 & 0.0046 \\ 
  eggs & -0.0123 & 0.0019 & -0.0858 & 0.0130 & 0.0003 & 0.0013 & 0.0030 & 0.0120 \\ 
  oj & 0.0113 & 0.0012 & 0.0789 & 0.0083 & -0.0006 & 0.0008 & -0.0054 & 0.0071 \\ 
  ice cream & 0.0113 & 0.0015 & 0.0788 & 0.0104 & -0.0184 & 0.0011 & -0.1630 & 0.0093 \\ 
  bread & -0.0274 & 0.0013 & -0.1911 & 0.0093  & -0.0104 & 0.0009 & -0.0927 & 0.0081 \\ 
  chips & 0.0050 & 0.0014 & 0.0350 & 0.0100 & -0.0047 & 0.0011 & -0.0417 & 0.0095 \\ 
  milk & 0.0136 & 0.0017 & 0.0947 & 0.0119   & -0.0232 & 0.0013 & -1.2058 & 0.0117\\ 
  salad & -0.0151 & 0.0010 & -0.1049 & 0.0069 & 0.0062 & 0.0007 & 0.0553 & 0.0063 \\ 
  yogurt &  -0.0051 & 0.0007 & -0.0354 & 0.0046 & 0.0018 & 0.0005 & 0.0160 & 0.0043 \\ 
  coffee & -0.0004 & 0.0009 & -0.0031 & 0.0060 & 0.0013 & 0.0006 & 0.0112 & 0.0053\\ 
  cereal & 0.0067 & 0.0015 & 0.0466 & 0.0108 & -0.0107 & 0.0011 & -0.0953 & 0.0097 \\ 
   \hline
\end{tabular}
\end{table}

%% file: derivative_discussion.tex
\section{Estimating Average Partial Effects}
Here we discuss an extension of our analysis to the Average Partial
Effect (APE). We leave the formal extension of our asymptotic results
to APE estimates as a problem for future work. 

For simplicity, we take $X_{it}$ to be a scalar, but the discussion
may be adapted straight-forwardly to vector-valued $X_{it}$. Recall
that our formal results apply to objects of the form,

\begin{equation}
E\big[\frac{1}{T}\sum_{t=1}^{T}\big(H_{it}^{+}g(X_{it}^{+},\eta_{it})-H_{it}^{-}g(X_{it}^{-},\eta_{it})\big)\big].\label{eq:objest}
\end{equation}
The APE is defined as
\begin{align*}
APE & :=E\big[\frac{1}{T}\sum_{t=1}^{T}\frac{\partial}{\partial x}g(x,\eta_{it})|_{x=X_{it}}\big],
\end{align*}
which can be written as a limit of objects of the form in (\ref{eq:objest}).
To see this, let $\epsilon>0$, then setting $X_{it}^{+}=X_{it}+\epsilon$
and $H_{it}^{+}=H_{it}^{-}=\frac{1}{\epsilon}$, then (\ref{eq:objest})
becomes,
\[
E\big[\frac{1}{T}\sum_{t=1}^{T}\frac{1}{\epsilon}\big(g(X_{it}+\epsilon,\eta_{it})-g(X_{it},\eta_{it})\big)\big].
\]
Assuming we can swap integration and differentiation, letting $\epsilon\to0$
recovers the APE:
\[
APE=\lim_{\epsilon\to0}E\big[\frac{1}{T}\sum_{t=1}^{T}\frac{1}{\epsilon}\big(g(X_{it}+\epsilon,\eta_{it})-g(X_{it},\eta_{it})\big)\big].
\]

Under Assumption 1, and assuming one can swap integration and differentiation,
we can rewrite the APE as,

\begin{align*}
APE & =E\big[\frac{1}{T}\sum_{t=1}^{T}\frac{\partial}{\partial x}h(x,X_{i})|_{x=X_{it}}\big],
\end{align*}
where $h(x,X_{i}):=E[g(x,\eta_{i1})|X_{i}]$, as in Section 3.1. Applying
the approximation to the structural function $h(x,X_{i})\approx b(x)'\bar{\beta_{i}}$
from Section 3.1, we obtain
\begin{align*}
APE & \approx E\big[a_{i}'\bar{\beta}_{i}\big],\,a_{i}:=\frac{1}{T}\sum_{t=1}^{T}\frac{\partial}{\partial x}b(x)|_{x=X_{it}}.
\end{align*}
Estimation of the APE may then be carried out by applying the formulas
in Section 3.2, using the choice of $a_{i}$ above. Note that for
this $a_{i}$ to be well-defined, the basis functions $b(\cdot)$
must all be differentiable. For the case in which $X_{it}$ is a vector,
one can simply replace $\frac{\partial}{\partial x}b(x)$ in the above
with the vector of derivatives of $b(x)$ with respect to the relevant
component of $x$.

%% file: binarychoice.tex
\section{Example 5: Binary Choice}
Consider a choice model where $Y_{it}$ is binary.
For example, $Y_{it}$ may indicate the purchase of a particular product. 
Suppose that the structural function $g$ in model (\ref{basemodel}) has the  form
\[
g(X_{it},\eta_{it})=1\{\Delta U(X_{it},\eta_{2it})-{\eta}_{1it}>0\},
\]
where $\Delta U(X_{it},\eta_{2it})$ is a utility difference and $\eta_{1it}$ is an additive, continuously distributed utility shock.
A parameter of interest in this example is an average effect of changing the regressors on the binary outcome, such as 
\[
\theta_{0}:=E[\frac{1}{T}\sum_{t=1}^{T}\{g({X}_{it}^{+},\eta_{it})-g(
{X}_{it}^{-},\eta_{it})\}].
\]
In binary choice models such objects are often referred to as marginal effects. 

It is useful for estimation of marginal effects that this model has a conditional choice probability that is smooth in possible $X_{it}$ .
Suppose that for $t=1$ the random variable $\eta_{1i1}$ has conditional CDF $G(v|X_i,\eta_{2i1})$ given $X_i$ and $\eta_{2i1}$. 
Then integrating over the conditional distribution of $\eta_{1i1}$ gives 
\[
E[g(x,\eta_{i1})|X_{i}]=E[G(\Delta U(x,\eta_{2i1}))|X_i,\eta_{2i1})|X_i].
\] 
Here $G(\Delta U(x,\eta_{2it})|X_i,\eta_{2i1})$ is the conditional choice probability that will be smooth in $x$ as long as $G(\nu||X_i,\eta_{2i1})$ is smooth in $\nu$ and $\Delta U(x,\eta)$ is smooth in $x$. 
Then, under appropriate regularity conditions $h(x,X_i)$, a conditional expectation of a smooth function of $x$, will also be smooth in $x$.

This example shows how the conditional average structural function may be smooth in $x$ even when the potential outcome is not.
This example is straightforward to generalize to other cases where the distribution of the outcome variable is partly discrete.
Smoothness in $x$ of the average potential outcome is important for the approach to estimation we give, to which we now turn. 

\color{black}

%% file: newproofs.tex
\section{Proofs and Supporting Lemmas}

\begin{proof}[Proof of Theorem 1]
By Assumptions 1 and 2, 
\begin{equation}
E[g(X_{it}^+,\eta_{it})|X_{i},H_{it}^+,X_{it}^+]=h(X_{it}^+,X_i).
\end{equation}
Then by the law of iterated expectations
\begin{align}
E[H_{it}^+g(X_{it}^+,\eta_{it})]=E\big[H_{it}^+E[g(X_{it}^+,\eta_{it})|X_{i},H_{it}^+,X_{it}^+]\big]=
E[H_{it}^+h(X_{it}^+,X_i)].
\end{align}
It follows similarly that $E[H_{it}^-g(X_{it}^-,\eta_{it})] =
E[H_{it}^-h(X_{it}^-,X_i)]$.
The conclusion then follows by plugging these equations in the expression for $\theta_0$. \end{proof}

\begin{proof}[Proof of Proposition 1]
From the FOCs for the optimization problem in the proposition we get
\[
{\beta}_{i}^{\text{Post}}=(Q_{i}+\lambda D_{i})^{-1}\frac{1}{T_{i}}B_{i}'Y_{i}+(Q_{i}+\lambda D_{i})^{-1}\lambda D_{i}\bar{\beta}.
\]
Substituting for ${\beta}_{i}^{\text{Post}}$ into $\frac{1}{n}\sum_{i=1}^{n}A_{i}{\beta}^{\text{Post}}_{i} =\frac{1}{n}\sum_{i=1}^{n}A_{i}\bar{\beta}$,  we see
\[
\frac{1}{n}\sum_{i=1}^{n}A_{i}(Q_{i}+\lambda D_{i})^{-1}\frac{1}{T_{i}}B_{i}'Y_{i}+\frac{1}{n}\sum_{i=1}^{n}A_{i}(Q_{i}+\lambda D_{i})^{-1}\lambda D_{i}\bar{\beta}={\frac{1}{n}\sum_{i=1}^{n}A_{i}\bar{\beta}}.
\]
Solving for $\bar{\beta}$ and simplifying yields
\[
\big(\frac{1}{n}\sum_{i=1}^{n}A_{i}(Q_{i}+\lambda D_{i})^{-1}Q_{i}\big)^{-1}\frac{1}{n}\sum_{i=1}^{n}A_{i}(Q_{i}+\lambda D_{i})^{-1}\frac{1}{T_{i}}B_{i}'Y_{i}=\ensuremath{\bar{\beta}}.
\]
Substituting back into $\frac{1}{n}\sum_{i=1}^{n}A_{i}{\beta}^{\text{Post}}_{i} =\frac{1}{n}\sum_{i=1}^{n}A_{i}\bar{\beta}$ we have
\begin{align*}
\frac{1}{n}\sum_{i=1}^{n}A_{i}{\beta}_{i}^{\text{Post}} & =\big(\frac{1}{n}\sum_{i=1}^{n}A_{i}\big)\big(\frac{1}{n}\sum_{i=1}^{n}A_{i}(Q_{i}+\lambda D_{i})^{-1}Q_{i}\big)^{-1}\frac{1}{n}\sum_{i=1}^{n}A_{i}(Q_{i}+\lambda D_{i})^{-1}\frac{1}{T_{i}}B_{i}'Y_{i}\\
 & =\frac{1}{n}\sum_{i=1}^{n}A_{i}(\overline{AW})^{-1}\overline{A\beta}.
\end{align*}
Multiplying both sides by a row vector whose first entry is one and
with remaining entries zero gives the result.
\end{proof}

\begin{proof}[Proof of Proposition 2]
Let $a_{0,i}$ denote the first component of $a_{i}$ and $a_{1,i}$
the vector that contains the remaining components.

Define the block
matrices $\tilde{W}_{i}:=\begin{pmatrix}1 & \bar{B}_{i}'(\tilde{Q}_{i}/\lambda+D_{1,i})^{-1}D_{1,i}\\
0 & (\tilde{Q}_{i}/\lambda+D_{1,i})^{-1}\tilde{Q}_{i}
\end{pmatrix}$, $\tilde{A}_{i}:=\begin{pmatrix}a_{0,i} & a_{1,i}'/\lambda\\
0 & I
\end{pmatrix}$, and $\tilde{L}_{i}=\begin{pmatrix}1 & -\bar{B}_{i}'(\tilde{Q}_{i}/\lambda+D_{1,i})^{-1}/\lambda\\
0 & (\tilde{Q}_{i}/\lambda+D_{1,i})^{-1}
\end{pmatrix}$. Then with some work, one can show that:
\begin{align}
\hat{\theta}= & \frac{1}{n}\sum_{i=1}^{n}a_{i}'\big(\frac{1}{n}\sum_{i=1}^{n}\tilde{A}_{i}\tilde{W}_{i}\big)^{-1}\frac{1}{n}\sum_{i=1}^{n}\tilde{A}_{i}\tilde{L}_{i}\begin{pmatrix}1 & 0\\
-\bar{B}_{i} & I
\end{pmatrix}\frac{1}{T_{i}}B_{i}'Y_{i}\label{eq:firstep}
\end{align}

Now, using that $D_{1,i}$ is non-singular, it is easy to see that
$\lim_{\lambda\to\infty}\tilde{W}_{i}=\begin{pmatrix}1 & \bar{B}_{i}'\\
0 & D_{1,i}^{-1}\tilde{Q}_{i}
\end{pmatrix}$, $\lim_{\lambda\to\infty}\tilde{A}_{i}=\begin{pmatrix}a_{0,i} & 0\\
0 & I
\end{pmatrix}$, and $\lim_{\lambda\to\infty}\tilde{L}_{i}=\begin{pmatrix}1 & 0\\
0 & D_{1,i}^{-1}
\end{pmatrix}$. Passing the limits we get that $\lim_{\lambda\to\infty}\hat{\theta}$ is equal to the expression below:
\begin{align}
 & \lim_{\lambda\to\infty}\frac{1}{n}\sum_{i=1}^{n}a_{i}'\big(\frac{1}{n}\sum_{i=1}^{n}\tilde{A}_{i}\tilde{W}_{i}\big)^{-1}\frac{1}{n}\sum_{i=1}^{n}\tilde{A}_{i}\tilde{L}_{i}\begin{pmatrix}1 & 0\\
-\bar{B}_{i} & I
\end{pmatrix}\frac{1}{T_{i}}B_{i}'Y_{i}\nonumber \\
= & \frac{1}{n}\sum_{i=1}^{n}a_{i}'\begin{pmatrix}\frac{1}{n}\sum_{i=1}^{n}a_{0,i} & \frac{1}{n}\sum_{i=1}^{n}a_{0,i}\bar{B}_{i}'\\
0 & \frac{1}{n}\sum_{i=1}^{n}D_{1,i}^{-1}\tilde{Q}_{i}
\end{pmatrix}^{-1}\frac{1}{n}\sum_{i=1}^{n}\begin{pmatrix}a_{0,i} & 0\\
0 & D_{1,i}^{-1}
\end{pmatrix}\begin{pmatrix}1 & 0\\
-\bar{B}_{i} & I
\end{pmatrix}\frac{1}{T_{i}}B_{i}'Y_{i}\label{eq:eqbig}
\end{align}
Multiplying out, we have:
\begin{align*}
 & \frac{1}{n}\sum_{i=1}^{n}\begin{pmatrix}a_{0,i} & 0\\
0 & D_{1,i}^{-1}
\end{pmatrix}\begin{pmatrix}1 & 0\\
-\bar{B}_{i} & I
\end{pmatrix}\frac{1}{T_{i}}B_{i}'Y_{i}
=  \begin{pmatrix}\frac{1}{n}\sum_{i=1}^{n}\frac{1}{T_{i}}\sum_{t=1}^{T_{i}}a_{0,i}Y_{it}\\
\frac{1}{n}\sum_{i=1}^{n}\frac{1}{T_{i}}\sum_{t=1}^{T_{i}}D_{1,i}^{-1}(B_{1,i,t}-\bar{B}_{i})Y_{it}
\end{pmatrix}.
\end{align*}

Applying the formula for the inverse of a block matrix and simplifying:
\begin{align*}
 & \frac{1}{n}\sum_{i=1}^{n}a_{i}'\begin{pmatrix}\frac{1}{n}\sum_{i=1}^{n}a_{0,i} & \frac{1}{n}\sum_{i=1}^{n}a_{0,i}\bar{B}_{i}'\\
0 & \frac{1}{n}\sum_{i=1}^{n}D_{1,i}^{-1}\tilde{Q}_{i}
\end{pmatrix}^{-1}\\
= & \begin{pmatrix}1 & \frac{1}{n}\sum_{i=1}^{n}a_{1,i}'\end{pmatrix}\begin{pmatrix}1 & -(\frac{1}{n}\sum_{i=1}^{n}a_{0,i}\bar{B}_{i}')(\frac{1}{n}\sum_{i=1}^{n}D_{1,i}^{-1}\tilde{Q}_{i})^{-1}\\
0 & (\frac{1}{n}\sum_{i=1}^{n}D_{1,i}^{-1}\tilde{Q}_{i})^{-1}
\end{pmatrix}
\end{align*}
Substituting into (\ref{eq:eqbig}) and multiplying, we get:
\begin{align*}
\lim_{\lambda\to\infty}\hat{\theta} =&\begin{pmatrix}1 & \frac{1}{n}\sum_{i=1}^{n}a_{1,i}'\end{pmatrix}\begin{pmatrix}1 & -(\frac{1}{n}\sum_{i=1}^{n}a_{0,i}\bar{B}_{i}')(\frac{1}{n}\sum_{i=1}^{n}D_{1,i}^{-1}\tilde{Q}_{i})^{-1}\\
0 & (\frac{1}{n}\sum_{i=1}^{n}D_{1,i}^{-1}\tilde{Q}_{i})^{-1}
\end{pmatrix}\\
\times&\begin{pmatrix}\frac{1}{n}\sum_{i=1}^{n}\frac{1}{T_{i}}\sum_{t=1}^{T_{i}}a_{0,i}Y_{it}\\
\frac{1}{n}\sum_{i=1}^{n}\frac{1}{T_{i}}\sum_{t=1}^{T_{i}}D_{1,i}^{-1}(B_{1,i,t}-\bar{B}_{i})Y_{it}
\end{pmatrix}\\
  =&\frac{1}{n}\sum_{i=1}^{n}a_{i}'\begin{pmatrix}\bar{Y}_{i}-\bar{B}_{i}'\hat{\beta}_{GFE,1}\\
\hat{\beta}_{GFE,1}
\end{pmatrix}.
\end{align*}

In the special case of $D_{1,i}=D_{1,1}$, for all $i$ we have 
\begin{align*}
\hat{\beta}_{GFE,1} & =(\frac{1}{n}\sum_{i=1}^{n}D_{1,1}^{-1}\tilde{Q}_{i})^{-1}\frac{1}{n}\sum_{i=1}^{n}\frac{1}{T_{i}}\sum_{t=1}^{T_{i}}D_{1,1}^{-1}(B_{1,i,t}-\bar{B}_{i})Y_{it}\\
 & =(\frac{1}{n}\sum_{i=1}^{n}\tilde{Q}_{i})^{-1}\frac{1}{n}\sum_{i=1}^{n}\frac{1}{T_{i}}\sum_{t=1}^{T_{i}}(B_{1,i,t}-\bar{B}_{i})Y_{it} =\hat{\beta}_{FE,1}.
\end{align*}
\end{proof}

\begin{proof}[Proof of Proposition 3]
Recall the formula for $\hat{\beta}_{i}$ is $
\hat{\beta}_{i}  =(Q_{i}+\lambda D_{i})^{-1}B_{i}'Y_{i}/T_{i}
$. 
By the properties of the matrix norm and Holder's inquality:
\[
|a_{i}'\hat{\beta}_{i}|\leq\|a_{i}\|\|(Q_{i}+\lambda D_{i})^{-1}\|\|B_{i}\|\|Y_{i}\|/T_{i}.
\]
We can decompose $(Q_{i}+\lambda D_{i})^{-1}$ into the product of
block matrices as follows:
\begin{align*}
(Q_{i}+\lambda D_{i})^{-1}= & \begin{pmatrix}1 & -\bar{B}_{i}'\\
0 & I
\end{pmatrix}\begin{pmatrix}1 & 0\\
0 & (\tilde{Q}_{i}+\lambda D_{1,i})^{-1}
\end{pmatrix}\begin{pmatrix}1 & 0\\
-\bar{B}_{i} & I
\end{pmatrix}.
\end{align*}
By the properties of the Euclidean matrix norm we then have:
\begin{align*}
\|(Q_{i}+\lambda D_{i})^{-1}\| & \leq(1+\|\bar{B}_{i}\|)^{2}(1+\|(\tilde{Q}_{i}+\lambda D_{1,i})^{-1}\|)\\
 & \leq(1+\|B_{i}\|)^{2}(1+\|(\tilde{Q}_{i}+\lambda D_{1,i})^{-1}\|).
\end{align*}
Now, note that by the properties of the matrix norm
\begin{align*}
\|(\tilde{Q}_{i}+\lambda D_{1,i})^{-1}\| & =\|D_{1,i}^{-1/2}(D_{1,i}^{-1/2}\tilde{Q}_{i}D_{1,i}^{-1/2}+\lambda I)^{-1}D_{1,i}^{-1/2}\|\\
 & \leq\|D_{1,i}^{-1/2}\|^{2}\|(D_{1,i}^{-1/2}\tilde{Q}_{i}D_{1,i}^{-1/2}+\lambda I)^{-1}\|\\
 & \leq\frac{1}{\lambda}\|D_{1,i}^{-1/2}\|^{2}.
\end{align*}

Combining everything so far, we get:
\[
|a_{i}'\hat{\beta}_{i}|\leq\|a_{i}\|(1+\|B_{i}\|)^{3}(1+\frac{1}{\lambda}\|D_{1,i}^{-1/2}\|^{2})\|Y_{i}\|/T_{i},
\]
and so, since $\|a_{i}\|$, $\|B_{i}\|$, and $\|D_{1,i}^{-1/2}\|$
are bounded almost surely and $\lambda>0$, there is a constant $C$
so that $
|a_{i}'\hat{\beta}_{i}|  \leq C\|Y_{i}\|/T_{i}
  \leq C\frac{1}{T_{i}}\sum_{t=1}^{T_{i}}|Y_{it}|
$. It follows that $E[|a_{i}'\hat{\beta}_{i}|]\leq C\frac{1}{T_{i}}\sum_{t=1}^{T_{i}}E[|Y_{it}|]$,
which is finite because $E[|Y_{it}|]$ is finite by supposition.
\end{proof}

\begin{proof}[Proof of Proposition 4]
Recall that under Assumption 1, we have the model
\begin{equation}
Y_{i}=B_{i}\beta_{i}+u_{i}+r_{i},\,E[u_{i}|X_{i}]=0.\label{eq:decomp1}
\end{equation}
Write $\beta_{i}=(\beta_{1,i},\beta_{2,i}')'$, where $\beta_{1,i}$
is the scalar first component of the vector $\beta_{i}$. If $\beta_{2,i}$ (which contains all but the first component of $\beta_i:=\beta(X_i)$), 
is mean independent of $X_{i}$, then $\beta_{2,i}$ does not vary
with $i$, and so we can write $\beta_{2,i}=\beta_{2}$.

Now,
using the general formula for $\hat{\theta}$, and (\ref{eq:decomp1})
with $r_{i}=0$, we have:
\begin{align*}
E[\hat{\theta}|X_{1},X_{2},...,X_{n}] & =E\bigg[\frac{1}{n}\sum_{i=1}^{n}a_{i}'\big(\frac{1}{n}\sum_{i=1}^{n}A_{i}W_{i}\big)^{-1}\frac{1}{n}\sum_{i=1}^{n}A_{i}W_{i}Q_{i}^{\dagger}B_{i}'Y_{i}/T_{i}\bigg|X_{1},...,X_{n}\bigg]\\
 & =E\bigg[\frac{1}{n}\sum_{i=1}^{n}a_{i}'\big(\frac{1}{n}\sum_{i=1}^{n}A_{i}W_{i}\big)^{-1}\frac{1}{n}\sum_{i=1}^{n}A_{i}W_{i}Q_{i}^{\dagger}Q_{i}\beta_{i}\bigg|X_{1},...,X_{n}\bigg]\\
 & +E\bigg[\frac{1}{n}\sum_{i=1}^{n}a_{i}'\big(\frac{1}{n}\sum_{i=1}^{n}A_{i}W_{i}\big)^{-1}\frac{1}{n}\sum_{i=1}^{n}A_{i}W_{i}Q_{i}^{\dagger}B_{i}'u_{i}/T_i\bigg|X_{1},...,X_{n}\bigg].
\end{align*}
For the first term on the RHS of the final equality, note that $W_{i}=W_{i}Q_{i}^{\dagger}Q_{i}$
and so this term simplifies:
\begin{align}
 & E\bigg[\frac{1}{n}\sum_{i=1}^{n}a_{i}'\big(\frac{1}{n}\sum_{i=1}^{n}A_{i}W_{i}\big)^{-1}\frac{1}{n}\sum_{i=1}^{n}A_{i}W_{i}Q_{i}^{\dagger}Q_{i}\beta_{i}\bigg|X_{1},...,X_{n}\bigg]\nonumber \\
= & E\bigg[\frac{1}{n}\sum_{i=1}^{n}a_{i}'\big(\frac{1}{n}\sum_{i=1}^{n}A_{i}W_{i}\big)^{-1}\frac{1}{n}\sum_{i=1}^{n}A_{i}W_{i}\beta_{i}\bigg|X_{1},...,X_{n}\bigg].\label{eq:s1part}
\end{align}

Now, we will show that for our particular choice of $W_i$, if $\beta_{2,i}$ is constant, then 
\[
\frac{1}{n}\sum_{i=1}^{n}a_{i}'\big(\frac{1}{n}\sum_{i=1}^{n}A_{i}W_{i}\big)^{-1}\frac{1}{n}\sum_{i=1}^{n}A_{i}W_{i}\beta_{i}=\frac{1}{n}\sum_{i=1}^{n}a_{i}'\beta_{i}.
\]
For more general choices of $W_i$ the above clearly continues to hold if the entire $\beta_i$ is constant.

To see this, let $a_{0,i}$ denote the first component of $a_{i}$
and $a_{1,i}$ the vector that contains the remaining components.
Note that $W_{i}=\begin{pmatrix}1 & \bar{B}_{i}'\big(I-(\tilde{Q}_{i}+\lambda D_{1,i})^{-1}\tilde{Q}_{i}\big)\\
0 & (\tilde{Q}_{i}+\lambda D_{1,i})^{-1}\tilde{Q}_{i}
\end{pmatrix}$, and so, writing things in terms of block matices and multiplying out
the product we see that
\begin{align*}
  A_{i}W_{i}\beta_{i}
= & \begin{pmatrix}a_{0,i} & a_{1,i}'\\
0 & I
\end{pmatrix}\begin{pmatrix}1 & \bar{B}_{i}'\big(I-(\tilde{Q}_{i}+\lambda D_{1,i})^{-1}\tilde{Q}_{i}\big)\\
0 & (\tilde{Q}_{i}+\lambda D_{1,i})^{-1}\tilde{Q}_{i}
\end{pmatrix}\begin{pmatrix}\beta_{i,1}\\
\beta_{2,i}
\end{pmatrix}\\
= & \begin{pmatrix}a_{0,i}\beta_{i,1}+\big(a_{0,i}\bar{B}_{i}'+(a_{1,i}'-a_{0,i}\bar{B}_{i}')(\tilde{Q}_{i}+\lambda D_{1,i})^{-1}\tilde{Q}_{i}\big)\beta_{2,i}\\
(\tilde{Q}_{i}+\lambda D_{1,i})^{-1}\tilde{Q}_{i}\beta_{2,i}
\end{pmatrix}.
\end{align*}

Define $\beta_{1}^{*}:=\frac{\frac{1}{n}\sum_{i=1}^{n}a_{0,i}\beta_{i,1}}{\frac{1}{n}\sum_{i=1}^{n}a_{0,i}}$.
From the above we get:
\begin{align*}
 & \frac{1}{n}\sum_{i=1}^{n}A_{i}W_{i}\beta_{i}\\
= & \begin{pmatrix}\frac{1}{n}\sum_{i=1}^{n}a_{0,i}\beta_{1}^{*}+\frac{1}{n}\sum_{i=1}^{n}\big(a_{0,i}\bar{B}_{i}'+(a_{1,i}'-a_{0,i}\bar{B}_{i}')(\tilde{Q}_{i}+\lambda D_{1,i})^{-1}\tilde{Q}_{i}\big)\beta_{2,i}\\
\frac{1}{n}\sum_{i=1}^{n}(\tilde{Q}_{i}+\lambda D_{1,i})^{-1}\tilde{Q}_{i}\beta_{2,i}
\end{pmatrix}\\
= & \frac{1}{n}\sum_{i=1}^{n}A_{i}W_{i}\begin{pmatrix}\beta_{1}^{*}\\
\beta_{2,i}
\end{pmatrix}.
\end{align*}
Using $\beta_{2,i}=\beta_{2}$ we then have
\begin{align*}
\frac{1}{n}\sum_{i=1}^{n}a_{i}'\big(\frac{1}{n}\sum_{i=1}^{n}A_{i}W_{i}\big)^{-1}\frac{1}{n}\sum_{i=1}^{n}A_{i}W_{i}\beta_{i}  =\frac{1}{n}\sum_{i=1}^{n}a_{i}'\begin{pmatrix}\beta_{1}^{*}\\
\beta_{2}
\end{pmatrix}
  =\frac{1}{n}\sum_{i=1}^{n}a_{i}'\beta_{i}.
\end{align*}
Now, substituting into (\ref{eq:s1part}), we see that
\begin{align*}
 & E\bigg[\frac{1}{n}\sum_{i=1}^{n}a_{i}'\big(\frac{1}{n}\sum_{i=1}^{n}A_{i}W_{i}\big)^{-1}\frac{1}{n}\sum_{i=1}^{n}A_{i}W_{i}Q_{i}^{\dagger}Q_{i}\beta_{i}\bigg|X_{1},...,X_{n}\bigg]
=  E[\frac{1}{n}\sum_{i=1}^{n}a_{i}'\beta_{i}|X_{1},...,X_{n}].
\end{align*}
For the second term, note that by Assumption 2 and independence of
the observations, $u_{i}$ is jointly independent of $a_{i}$ (and
thus $A_{i}$) conditional on $X_{1},X_{2},...,X_{n}$, and so:
\begin{align*}
 & E\bigg[\frac{1}{n}\sum_{i=1}^{n}a_{i}'\big(\frac{1}{n}\sum_{i=1}^{n}A_{i}W_{i}\big)^{-1}\frac{1}{n}\sum_{i=1}^{n}A_{i}W_{i}Q_{i}^{\dagger}B_{i}'u_{i}/T_i\bigg|X_{1},...,X_{n}\bigg]\\
= & \frac{1}{n}\sum_{i=1}^{n}E\bigg[\big(\frac{1}{n}\sum_{i=1}^{n}a_{i}\big)'\big(\frac{1}{n}\sum_{i=1}^{n}A_{i}W_{i}\big)^{-1}A_{i}W_{i}Q_{i}^{\dagger}B_{i}'/T_i\bigg|X_{1},...,X_{n}\bigg]E[u_{i}|X_{1},...,X_{n}]\\
= & \frac{1}{n}\sum_{i=1}^{n}E\bigg[\big(\frac{1}{n}\sum_{i=1}^{n}a_{i}\big)'\big(\frac{1}{n}\sum_{i=1}^{n}A_{i}W_{i}\big)^{-1}A_{i}W_{i}Q_{i}^{\dagger}B_{i}'/T_i\bigg|X_{1},...,X_{n}\bigg]E[u_{i}|X_{i}]\\
= & 0,
\end{align*}
where we have also used that $W_{i}$ is a function of $X_{i}$. So in all
\begin{align*}
E[\hat{\theta}|X_{1},...,X_{n}] & =\frac{1}{n}\sum_{i=1}^{n}E[a_{i}'\beta_{i}|X_{1},...,X_{n}].
\end{align*}
If $E[|\hat{\theta}|]<\infty$ then the law of iterated expectations
holds $E[\hat{\theta}]=E\big[E[\hat{\theta}|X_{1},...,X_{n}]\big]$.
Taking expectations of both sides of the above we get $
E[\hat{\theta}]=E[a_{i}'\beta_{i}]
$.
\end{proof}

\begin{proof}[Proof of Proposition 5]

If $C_{i}=1$ then $Z_{i}$ does not vary, and so $\tilde{Q}_{i}=\begin{pmatrix}0 & 0\\
0 & \hat{Q}_{1,i}
\end{pmatrix}$ where $\hat{Q}_{1,i}$ is $\tilde{Q}_i$ with its first row and column removed. By supposition $\hat{Q}_{1,i}$ is
non-singular and $D_{i}$ is diagonal. Let $d_{1,i}$ be the first
element of the leading diagonal of $D_{1,i}$. From the above we see
that in this case 
\[
(\tilde{Q}_{i}+\lambda D_{1,i})^{-1}=\begin{pmatrix}\frac{1}{\lambda d_{1,i}} & 0\\
0 & (\hat{Q}_{1,i}+\lambda\tilde{D}_{1,i})^{-1}
\end{pmatrix}.
\]

We can write $W_{i}$ as $W_{i}=\begin{pmatrix}1 & \lambda\bar{B}_{i}'(\tilde{Q}_{i}+\lambda D_{1,i})^{-1}D_{1,i}\\
0 & (\tilde{Q}_{i}+\lambda D_{1,i})^{-1}\tilde{Q}_{i}
\end{pmatrix}$. Substituting, we get 

\[
W_{i}=\begin{pmatrix}1 & \bar{Z}_{i} & \lambda\bar{B}_{2,i}'(\hat{Q}_{1,i}+\lambda\tilde{D}_{1,i})^{-1}\tilde{D}_{1,i}\\
0 & 0 & 0\\
0 & 0 & (\hat{Q}_{1,i}+\lambda\tilde{D}_{1,i})^{-1}\hat{Q}_{1,i},
\end{pmatrix}
\]
where $\bar{Z}_{i}=\frac{1}{T_{i}}\sum_{t=1}^{T_{i}}Z_{i}$ and $\bar{B}_{2,i}=\frac{1}{T_{i}}\sum_{t=1}^{T_{i}}B_{2,it}$.
And so, taking the limit, we obtain $\lim_{\lambda\to0}W_{i}=\begin{pmatrix}1 & \bar{Z}_{i} & 0\\
0 & 0 & 0\\
0 & 0 & I
\end{pmatrix}$. On the other hand, if $C_{i}=0$ and so $Q_{i}$ is non-singular,
then it is easy to see that $\lim_{\lambda\to0}W_{i}$ is the identity.

Now, let us consider the limit of the individual ridge estimates.
Note we can write these estimates as $\hat{\beta}_{i}=\begin{pmatrix}\bar{Y}_{i}-(\bar{Z}_{i},\bar{B}_{2,i}')\hat{\beta}_{1,i}\\
\hat{\beta}_{1,i}
\end{pmatrix}$, where $\hat{\beta}_{1,i}$ are slope parameters given by 
\[
\hat{\beta}_{i,1}=(\tilde{Q}_{i}+\lambda D_{1,i})^{-1}\frac{1}{T_{i}}\sum_{t=1}^{T_{i}}(B_{1,it}-\bar{B}_{i})Y_{i}.
\]
If $Z_{it}$ is constant, that is $C_{i}=1$, then this becomes 
\begin{align*}
\hat{\beta}_{i,1} & =\begin{pmatrix}\frac{1}{\lambda d_{1,i}} & 0\\
0 & (\hat{Q}_{1,i}+\lambda\tilde{D}_{1,i})^{-1}
\end{pmatrix}\begin{pmatrix}0\\
\frac{1}{T_{i}}\sum_{t=1}^{T_{i}}(B_{2,it}-\bar{B}_{2,i})Y_{i}
\end{pmatrix}\\
 & =\begin{pmatrix}0\\
(\hat{Q}_{1,i}+\lambda\tilde{D}_{1,i})^{-1}\frac{1}{T_{i}}\sum_{t=1}^{T_{i}}(B_{2,it}-\bar{B}_{2,i})Y_{i}
\end{pmatrix}.
\end{align*}
Because $\hat{Q}_{1,i}$ is non-singular we have 
\begin{align*}
\lim_{\lambda\to0}(\hat{Q}_{1,i}+\lambda\tilde{D}_{1,i})^{-1}\frac{1}{T_{i}}\sum_{t=1}^{T_{i}}(B_{2,it}-\bar{B}_{2,i})Y_{i}=\hat{Q}_{1,i}^{-1}\frac{1}{T_{i}}\sum_{t=1}^{T_{i}}(B_{2,it}-\bar{B}_{2,i})Y_{i}=\tilde{\beta}_{i,3},
\end{align*}
and so, we see that $\lim_{\lambda\to0}\hat{\beta}_{i}=(\bar{Y}_{i}-\bar{B}_{2,i}'\tilde{\beta}_{i,3},0,\tilde{\beta}_{i,3}')'$.
However, if $C_{i}=0$ then, letting $\tilde{\beta}_{i,2}$ and $\tilde{\beta}_{i,3}$
be the coefficients from individual OLS regression, $\lim_{\lambda\to0}\hat{\beta}_{i}$
is equal to $(\bar{Y}_{i}-\bar{Z}_{i}\tilde{\beta}_{i,2}-\bar{B}_{2,i}'\tilde{\beta}_{i,3},\tilde{\beta}_{i,2},\tilde{\beta}_{i,3}')'$.

Let $a_{0,i}$ and $a_{1,i}$ be the first and second components of
$a_{i}$, and $a_{2,i}$ the vector of remaining components. Substituting
everything, we see 

\begin{align*}
\lim_{\lambda\to0}\hat{\theta}= & \frac{1}{n}\sum_{i=1}^{n}a_{i}'\begin{pmatrix}\frac{1}{n}\sum_{i=1}^{n}a_{0,i} & \frac{1}{n}\sum_{i=1}^{n}C_{i}a_{0,i}\bar{Z}_{i}+\frac{1}{n}\sum_{i=1}^{n}(1-C_{i})a_{1,i} & \frac{1}{n}\sum_{i=1}^{n}a_{2,i}'\\
0 & \hat{p} & 0\\
0 & 0 & I
\end{pmatrix}^{-1}\\
\times & \begin{pmatrix}x\\
\frac{1}{n}\sum_{i=1}^{n}(1-C_{i})\tilde{\beta}_{i,2}\\
\frac{1}{n}\sum_{i=1}^{n}\tilde{\beta}_{i,3}
\end{pmatrix}\\
= & x+\frac{1}{n}\sum_{i=1}^{n}C_{i}(a_{1,i}'-a_{0,i}\bar{Z}_{i})\frac{1}{n\hat{p}}\sum_{j=1}^{n}(1-C_{j})\tilde{\beta}_{j,2}.
\end{align*}

where $x$ is given below: 
\begin{align*}
x= & \frac{1}{n}\sum_{i=1}^{n}C_{i}\bigg(a_{0,i}(\bar{Y}_{i}-\bar{B}_{2,i}'\tilde{\beta}_{i,3})+a_{2,i}'\tilde{\beta}_{i,3}\bigg)\\
+ & \frac{1}{n}\sum_{i=1}^{n}(1-C_{i})\bigg(a_{0,i}(\bar{Y}_{i}-\bar{Z}_{i}\tilde{\beta}_{i,2}-\bar{B}_{2,i}'\tilde{\beta}_{i,3})+a_{1,i}\tilde{\beta}_{i,2}+a_{2,i}'\tilde{\beta}_{i,3}\bigg).
\end{align*}
Substituting for $x$ and simplifying using the variables defined
in the theorem, we can rewrite the limit of $\hat{\theta}$ succinctly
as $\lim_{\lambda\to0}\hat{\theta}=\frac{1}{n}\sum_{i=1}^{n}a_{i}'\tilde{\beta}_{i}^{*}.$
\end{proof}

\begin{proof}[Proof of Proposition 6]
First note that 
\begin{align*}
(\tilde{Q}_{i}+\lambda D_{1,i})^{-1}\tilde{Q}_{i} & =D_{1,i}^{-1/2}(\tilde{B}_{i}'\tilde{B}_{i}+\lambda I)^{-1}\tilde{B}_{i}'\tilde{B}_{i}D_{1,i}^{1/2}.
\end{align*}
By the properties of the Moore-Penrose pseudo-inverse $
\lim_{\lambda\to0}(\tilde{B}_{i}'\tilde{B}_{i}+\lambda I)^{-1}\tilde{B}_{i}'=\tilde{B}_{i}^{\dagger}
$. And so 
$
\lim_{\lambda\to0}(\tilde{Q}_{i}+\lambda D_{1,i})^{-1}\tilde{Q}_{i}=P_{i}
$. By definition of $W_{i}$ we then get 
\begin{align*}
\lim_{\lambda\to0}W_{i} & =\lim_{\lambda\to0}\begin{pmatrix}1 & \bar{B}_{i}'\big(I-(\tilde{Q}_{i}+\lambda D_{1,i})^{-1}\tilde{Q}_{i}\big)\\
0 & (\tilde{Q}_{i}+\lambda D_{1,i})^{-1}\tilde{Q}_{i}
\end{pmatrix}
  =\begin{pmatrix}1 & \bar{B}_{i}'(I-P_{i})\\
0 & P_{i}
\end{pmatrix}.
\end{align*}
In addition, using the definition of $\hat{\beta}_{i,2}$, we have
\begin{align*}
\lim_{\lambda\to0}\hat{\beta}_{i,2} & =\lim_{\lambda\to0}D_{1,i}^{-1/2}\big(\tilde{B}_{i}'\tilde{B}_{i}+\lambda I\big)^{-1}\tilde{B}_{i}'Y_{i}/T_{i}
  =D_{1,i}^{-1/2}\tilde{B}_{i}^{\dagger}Y_{i}/T_{i}
  =\tilde{\beta}_{i,2}^{\circ}.
\end{align*}
And so, since $\hat{\beta}_{i}=(\bar{Y}_{i}-\bar{B}_{i}'\hat{\beta}_{i,2},\hat{\beta}_{i,2}')'$,
we see that $
\lim_{\lambda\to0}\hat{\beta}_{i}=(\bar{Y}_{i}-\bar{B}_{i}'\tilde{\beta}_{i,2}^{\circ},\tilde{\beta}_{i,2}^{\circ}{}')'
$. 

Combining and using the definition of $A_{i}$, after some
work simplifying we obtain 
\begin{align*}
\lim_{\lambda\to0}\hat{\theta} & =\frac{1}{n}\sum_{i=1}^{n}a_{it}'\bigg(\frac{1}{n}\sum_{i=1}^{n}A_{i}\begin{pmatrix}1 & \bar{B}_{i}'(I-P_{i})\\
0 & P_{i}
\end{pmatrix}\bigg)^{-1}\frac{1}{n}\sum_{i=1}^{n}A_{i}\begin{pmatrix}\bar{Y}_{i}-\bar{B}_{i}'\tilde{\beta}_{i,2}^{\circ}\\
\tilde{\beta}_{i,2}^{\circ}
\end{pmatrix}\\
 & =\frac{1}{n}\sum_{i=1}^{n}a_{i}'\tilde{\beta}_{i}^{*}.
\end{align*}
\end{proof}

\theoremstyle{plain} \newtheorem*{L1}{Lemma 1 (General Asymptotics)} 
\begin{L1}
Suppose that Assumptions 1, 2, and 3.i-vii hold and $W_{i}Q_{i}^{\dagger}Q_{i}=W_{i}$.
Define $\kappa_{n}=E[\|W_{i}-I\|]$ and $\gamma_{n}=E[\|W_{i}(Q_{i}^{\dagger})^{1/2}\|^{2}]$, 
and suppose $\kappa_{n}=o(1)$. Then
\begin{align*}
\hat{\theta}-\theta_{0}= & O_{p}\bigg(n^{-1/2}+\kappa_{n}+\sqrt{\frac{\gamma_{n}}{nT}}+\sqrt{\frac{\gamma_{n}J\kappa_{n}^{2}}{nT}}+(1+\sqrt{\gamma_{n}})\ell\bigg).
\end{align*}
If in addition, $\kappa_{n},\ell,\ell\sqrt{\gamma_{n}}=o(\sqrt{\frac{1}{n}})$,
$\frac{\gamma_{n}J\kappa_{n}^{2}}{T}=o(1)$, and Assumptions 4.i and 4.ii hold 
for $\delta,v,q>0$ with $(v-2)(q-2)>4$ and:
\[
n^{(\frac{1}{v}+\frac{1}{q}-\frac{1}{2})}E[\|W_{i}(Q_{i}^{\dagger})^{1/2}\|^{q}]^{1/q}(\gamma_{n}/T)^{\delta}\to0,
\]
then $\hat{\theta}-\theta_{0}=O_{p}(n^{-1/2})$, and we have $\sqrt{n}\sigma_{n}^{-1}(\hat{\theta}-\theta_{0})\sim^{a}N(0,1)$,  
where the variance $\sigma_{n}^{2}$ is equal to $E[|\frac{1}{T_{i}}a'_{i}W_{i}Q_{i}^{\dagger}B_{i}'u_{i}|^{2}]+Var(a_{i}'\beta_{i})$.
\end{L1}

\begin{proof}
Recall that $h(x,X_i)=b(x)'\beta_i+r(x,X_i)$. Substituting this and using the definition of $a_i$, we obtain
\begin{align*}
 & E\big[\frac{1}{T}\sum_{t=1}^{T}\big(H_{it}^{+}h(X_{it}^{+},X_{i})-H_{it}^{-}h(X_{it}^{-},X_{i})\big)\big|X_{i}\big]\\
= & E[a_{i}|X_i]'\beta_{i}
+  E[\frac{1}{T}\sum_{t=1}^{T}H_{it}^{+}r(X_{it}^{+},X_{i})|X_{i}]-E[\frac{1}{T}\sum_{t=1}^{T}H_{it}^{-}r(X_{it}^{-},X_{i})|X_{i}].
\end{align*}
By Assumption 3, the terms on the RHS are bounded above in magnitude,
and so by iterated expectations, and Theorem 1, we have:
\begin{align*}
\theta_{0}= & E[a_{i}'\beta_{i}]
+  \frac{1}{T}\sum_{t=1}^{T}E[H_{t}^{+}(X_{i})r(X_{it}^{+},X_{i})]-\frac{1}{T}\sum_{t=1}^{T}E[H_{t}^{-}(X_{i})r(X_{it}^{-},X_{i})].
\end{align*}

It will be convenient to define the mean-zero random variable $\epsilon_{i}$
as follows
\[
\epsilon_{i}=a_{i}'\beta_{i}-E[a_{i}'\beta_{i}]+a_{i}'W_{i}Q_{i}^{\dagger}\frac{1}{T_{i}}B_{i}'u_{i}.
\]

Using the expression for $\theta_{0}$ above, we decompose the estimation
error $\hat{\theta}-\theta_{0}$ into a zero-mean part $\frac{1}{n}\sum_{i=1}^{n}\epsilon_{i}$,
and a number of remainder terms which are generally not mean-zero:
\begin{align*}
  \hat{\theta}-\theta_{0}-\frac{1}{n}\sum_{i=1}^{n}\epsilon_{i}
= & \frac{1}{n}\sum_{i=1}^{n}a_{i}'(I-W_{i})\big(\frac{1}{n}\sum_{i=1}^{n}A_{i}W_{i}\big)^{-1}\frac{1}{n}\sum_{i=1}^{n}A_{i}W_{i}\beta_{i}
+  \frac{1}{n}\sum_{i=1}^{n}a_{i}'(W_{i}-I)\beta_{i}\\
+ & \frac{1}{n}\sum_{i=1}^{n}a_{i}'(I-W_{i})\big(\frac{1}{n}\sum_{i=1}^{n}A_{i}W_{i}\big)^{-1}\frac{1}{n}\sum_{i=1}^{n}A_{i}W_{i}Q_{i}^{\dagger}\frac{1}{T_{i}}B_{i}'u_{i}\\
+ & \frac{1}{n}\sum_{i=1}^{n}a_{i}'\big(\frac{1}{n}\sum_{i=1}^{n}A_{i}W_{i}\big)^{-1}\frac{1}{n}\sum_{i=1}^{n}A_{i}W_{i}Q_{i}^{\dagger}\frac{1}{T_{i}}B_{i}'r_{i}\\
- & \frac{1}{T}\sum_{t=1}^{T}E\big[H_{t}^{+}(X_{i})r(X_{it}^{+},X_{i})\big]+\frac{1}{T}\sum_{t=1}^{T}E\big[H_{t}^{-}(X_{i})r(X_{it}^{-},X_{i})\big],
\end{align*}
where we have used that $a_{i}=A_{i}'v$ for some vector $v$ and
that $W_{i}Q_{i}^{\dagger}Q_{i}=W_{i}$. By the triangle inequality
and properties of the matrix norm and $\|a_{i}\|\leq c$ by Assumption
3.i. we get:
\begin{align*}
  \|\hat{\theta}-\theta_{0}-\frac{1}{n}\sum_{i=1}^{n}\epsilon_{i}\|
\leq & \|\frac{1}{n}\sum_{i=1}^{n}a_{i}'(I-W_{i})\|\|\big(\frac{1}{n}\sum_{i=1}^{n}A_{i}W_{i}\big)^{-1}\|\|\frac{1}{n}\sum_{i=1}^{n}A_{i}W_{i}\beta_{i}\|\\
+ & \|\frac{1}{n}\sum_{i=1}^{n}a_{i}'(W_{i}-I)\beta_{i}\|\\
+ & \|\frac{1}{n}\sum_{i=1}^{n}a_{i}'(I-W_{i})\|\|\big(\frac{1}{n}\sum_{i=1}^{n}A_{i}W_{i}\big)^{-1}\|\|\frac{1}{n}\sum_{i=1}^{n}A_{i}W_{i}Q_{i}^{\dagger}\frac{1}{T_{i}}B_{i}'u_{i}\|\\
+ & c\|\big(\frac{1}{n}\sum_{i=1}^{n}A_{i}W_{i}\big)^{-1}\|\|\frac{1}{n}\sum_{i=1}^{n}A_{i}W_{i}Q_{i}^{\dagger}\frac{1}{T_{i}}B_{i}'r_{i}\|\\
+ & \|\frac{1}{T}\sum_{t=1}^{T}E\big[H_{t}^{+}(X_{i})r(X_{it}^{+},X_{i})\big]\|+\|\frac{1}{T}\sum_{t=1}^{T}E\big[H_{t}^{-}(X_{i})r(X_{it}^{-},X_{i})\big]\|
\end{align*}

The RHS above contains a number of objects which we derive rates for
below.

\subsection*{Step 1: Derive Rates for the Remainder}

\subsubsection*{1. $\|\frac{1}{n}\sum_{i=1}^{n}a_{i}'(W_{i}-I)\|,\|\frac{1}{n}\sum_{i=1}^{n}A_{i}(W_{i}-I)\|=O_{p}(\kappa_{n})$ }

Under Assumption 3.i, $\|a_{i}\|,\|A_{i}\|\leq c$, and so by the triangle
inequality and definition of the matrix norm, 
\begin{align*}
\|\frac{1}{n}\sum_{i=1}^{n}a_{i}'(W_{i}-I)\| & \leq\frac{1}{n}\sum_{i=1}^{n}\|a_{i}'(W_{i}-I)\|\leq c\frac{1}{n}\sum_{i=1}^{n}\|W_{i}-I\|=O_{p}(\kappa_{n})\\
\|\frac{1}{n}\sum_{i=1}^{n}A(W_{i}-I)\| & \leq\frac{1}{n}\sum_{i=1}^{n}\|A_{i}(W_{i}-I)\|\leq c\frac{1}{n}\sum_{i=1}^{n}\|W_{i}-I\|=O_{p}(\kappa_{n}),
\end{align*}
where the final equalities both follow by Markov's inequality. By
supposition, $\kappa_n=o(1)$, and so we see that the terms on
the LHSs above are also $o_{p}(1)$.

\subsubsection*{2. $\|(\frac{1}{n}\sum_{i=1}^{n}A_{i}W_{i})^{-1}\|=O_{p}(1)$ }

By Assumption 3.ii, $\|(\frac{1}{n}\sum_{i=1}^{n}A_{i})^{-1}\|=O_{p}(1)$,
and we have already shown that $\|\frac{1}{n}\sum_{i=1}^{n}A(W_{i}-I)\|=O_{p}(\kappa_{n})=o_{p}(1)$,
so we have:
\begin{align*}
\|\big(\frac{1}{n}\sum_{i=1}^{n}A_{i}W_{i}\big)^{-1}-(\frac{1}{n}\sum_{i=1}^{n}A_{i})^{-1}\| & =O_{p}(\kappa_n) =o_{p}(1).
\end{align*}

From the above and the fact that $\|(\frac{1}{n}\sum_{i=1}^{n}A_{i})^{-1}\|=O_{p}(1)$,
we get from the triangle inequality that $\|\big(\frac{1}{n}\sum_{i=1}^{n}A_{i}W_{i}\big)^{-1}\|=O_{p}(1)$.

\subsubsection*{3. $\|\frac{1}{n}\sum_{i=1}^{n}A_{i}W_{i}\beta_{i}\|=O_{p}(1)$}

By the triangle inequality and properties of the matrix norm,  
\begin{align*}
\|\frac{1}{n}\sum_{i=1}^{n}A_{i}W_{i}\beta_{i}\| & \leq\frac{1}{n}\sum_{i=1}^{n}\|A_{i}\|\|W_{i}\|\|\beta_{i}\| \leq c^{2}\frac{1}{n}\sum_{i=1}^{n}\|W_{i}\|.
\end{align*}
The second inequality follows from $\|A_{i}\|,\|\beta_{i}\|\leq c$
by Assumption 3.i and 3.iii. Recall that $E[\|W_{i}-I\|]=o(1)$, and
so by the triangle inequality $E[\|W_{i}\|]=O(1)$. Therefore, by
Markov's inequality, $
\|\frac{1}{n}\sum_{i=1}^{n}A_{i}W_{i}\beta_{i}\|=O_{p}(1)$.

\subsubsection*{4.$|\frac{1}{n}\sum_{i=1}^{n}a_{i}'(W_{i}-I)\beta_{i}|=O_{p}(\kappa_{n})$}

Note that by the triangle inequality and
properties of the matrix norm, 
\begin{align*}
|\frac{1}{n}\sum_{i=1}^{n}a_{i}'(W_{i}-I)\beta_{i}|  \leq c^{2}\frac{1}{n}\sum_{i=1}^{n}\|W_{i}-I\| =O_{p}(\kappa_n),
\end{align*}
where the inequality follows from Assumptions 3.i and 3.iii,
 and the equality follows from 
Markov's inequality.

\subsubsection*{5. $\|\frac{1}{n}\sum_{i=1}^{n}A_{i}W_{i}Q_{i}^{\dagger}\frac{1}{T_{i}}B_{i}'u_{i}\|=O_{p}\bigg(\sqrt{\frac{J\gamma_{n}}{Tn}}\bigg)$}

Note that $E[u_{it}|X_{i}]=0$. Therefore,  using Assumption 2, $A_{i}W_{i}Q_{i}^{\dagger}B_{i}'u_{i}$
is mean zero. Moreover, using standard trace inequalities and properties
of the psuedo-inverse we get:
\begin{align*}
  E\big[\|\frac{1}{T_{i}}A_{i}W_{i}Q_{i}^{\dagger}B_{i}'u_{i}\|^{2}\big]
= & E\big[\frac{1}{T_{i}^{2}}tr(A_{i}W_{i}Q_{i}^{\dagger}B_{i}'E[u_{i}u_{i}'|X_{i}]B_{i}Q_{i}^{\dagger}W_{i}'A_{i}')\big]\\
\leq & cE[\frac{1}{T_{i}}tr(A_{i}W_{i}Q_{i}^{\dagger}W_{i}'A_{i}')]\\
\leq & JcE[\|A_{i}W_{i}Q_{i}^{\dagger}W_{i}'A_{i}'\|/T_{i}]\\
\leq & Jc^{3}E[\|W_{i}(Q_{i}^{\dagger})^{1/2}\|^{2}/T_{i}],
\end{align*}
where the first equality uses Assumption 2, and the second line uses Assumption 3.iv, which states that $\|E[u_{i}u_{i}'|X_{i}]\|\leq c$
almost surely. And so, by the law of large numbers:
\begin{align*}
\frac{1}{n}\sum_{i=1}^{n}A_{i}W_{i}Q_{i}^{\dagger}\frac{1}{T_{i}}B_{i}'u_{i} & =O_{p}\bigg(\sqrt{\frac{JE[\|W_{i}(Q_{i}^{\dagger})^{1/2}\|^{2}/T_{i}]}{n}}\bigg) =O_{p}\bigg(\sqrt{\frac{J\gamma_{n}}{Tn}}\bigg),
\end{align*}
where the final equality follows by Assumption 3.vii.

\subsubsection*{6. $\|\frac{1}{n}\sum_{i=1}^{n}A_{i}W_{i}Q_{i}^{\dagger}\frac{1}{T_{i}}B_{i}'r_{i}\|=O_{p}(\ell\sqrt{\gamma_{n}})$}

Again, using properties of
the matrix norm, the triangle inequality, and $\|A_{i}\|\leq c$,
\begin{align*}
  \|\frac{1}{n}\sum_{i=1}^{n}A_{i}W_{i}Q_{i}^{\dagger}\frac{1}{T_{i}}B_{i}'r_{i}\|
\leq  c\frac{1}{n}\sum_{i=1}^{n}\|W_{i}Q_{i}^{\dagger}\frac{1}{T_{i}}B_{i}'r_{i}\|.
\end{align*}
Using the properties of the matrix norm and pseudo-inverse, we get 
\begin{align*}
\|W_{i}Q_{i}^{\dagger}\frac{1}{T_{i}}B_{i}'r_{i}\|^{2}  =\|W_{i}Q_{i}^{\dagger}Q_{i}Q_{i}^{\dagger}\frac{1}{T_{i}}B_{i}'r_{i}\|^{2}& \leq\|W_{i}Q_{i}^{\dagger}Q_{i}^{1/2}\|^{2}\|Q_{i}^{1/2}Q_{i}^{\dagger}\frac{1}{T_{i}}B_{i}'r_{i}\|^{2}\\
 & =\|W_{i}(Q_{i}^{\dagger})^{1/2}\|^{2}\frac{1}{T_{i}}\sum_{t=1}^{T_{i}}|b(X_{it})'Q_{i}^{\dagger}\frac{1}{T_{i}}B_{i}'r_{i}|^{2}.
\end{align*}
By the properties of least squares projections we have
\begin{align*}
\frac{1}{T_{i}}\sum_{t=1}^{T_{i}}|b(X_{it})'Q_{i}^{\dagger}\frac{1}{T_{i}}B_{i}'r_{i}|^{2} & \leq\frac{1}{T_{i}}\sum_{t=1}^{T_{i}}r_{it}^{2} \leq\ell^{2},
\end{align*}
and so:
\begin{align*}
\frac{1}{n}\sum_{i=1}^{n}\|W_{i}Q_{i}^{\dagger}\frac{1}{T_{i}}B_{i}'r_{i}\| & \leq\ell\frac{1}{n}\sum_{i=1}^{n}\|W_{i}(Q_{i}^{\dagger})^{1/2}\|.
\end{align*}

By Markov's inequality $\frac{1}{n}\sum_{i=1}^{n}\|W_{i}(Q_{i}^{\dagger})^{1/2}\|=O(E[\|W_{i}(Q_{i}^{\dagger})^{1/2}\|])$
and by Jensen's ineqality $E[\|W_{i}(Q_{i}^{\dagger})^{1/2}\|]\leq\sqrt{\gamma_{n}}$.
And so, using $\|\big(\frac{1}{n}\sum_{i=1}^{n}A_{i}W_{i}\big)^{-1}\|=O_{p}(1)$
as established earlier, we see that 
$
 \|\frac{1}{n}\sum_{i=1}^{n}A_{i}W_{i}Q_{i}^{\dagger}\frac{1}{T_{i}}B_{i}'r_{i}\|
=  O_{p}(\ell\sqrt{\gamma_{n}})
$.

\subsubsection*{7. $\|\frac{1}{T}\sum_{t=1}^{T}E\big[H_{t}^{+}(X_{i})r(X_{it}^{+},X_{i})\big]\|,  \|\frac{1}{T}\sum_{t=1}^{T}E\big[H_{t}^{-}(X_{i})r(X_{it}^{-},X_{i})\big]\|=O(\ell)$}

Note
that $H_{t}(X_{i}^{+})$ is uniformly bounded by Assumption 3.v, and
$|r(X_{it}^{+},X_{i})|\leq\ell$ by Assumption 3.vi. Therefore, 
$\|\frac{1}{T}\sum_{t=1}^{T}E\big[H_{t}^{+}(X_{i})r(X_{it}^{+},X_{i})\big]\|=O(\ell)$.
Similar reasoning shows the same rate applies for $\frac{1}{T}\sum_{t=1}^{T}E\big[H_{t}^{-}(X_{i})r(X_{it}^{-},X_{i})\big]$.

\subsubsection*{Combining:}

Using all of the rates derived above, we get
\begin{align*}
 & \hat{\theta}-\theta_{0}-\frac{1}{n}\sum_{i=1}^{n}\epsilon_{i}
=  O_{p}\bigg(\kappa_{n}+\kappa_{n}\sqrt{\frac{J\gamma_{n}}{nT}}+\ell+\ell\sqrt{\gamma_{n}}\bigg).
\end{align*}

\subsection*{Step 2: Derive a Rate for $\frac{1}{n}\sum_{i=1}^{n}\epsilon_{i}$}

Recall $E[\epsilon_{i}]=0$. We will derive a rate for the variance
of $\epsilon_{i}$. First note that $u_{it}$ is a function of $X_{i}$
and $\eta_{it}$, the latter of which is independent of $a_{it}$ given
$X_{i}$ by Assumption 2, and $E[u_{i}|X_{i}]=0$, so we have:
\begin{align*}
E[\epsilon_{i}^{2}] & =E[(\frac{1}{T_{i}}a_{i}'W_{i}Q_{i}^{\dagger}B_{i}'u_{i}\big)^{2}]+Var(a_{i}'\beta_{i})
\end{align*}

Using properties of the matrix norm and that $\|a_{i}\|\leq c$ and
$\|E[u_{i}u_{i}'|X_{i}]\|\leq c$ by Assumptions 3.i and 3.iv, we have:
\begin{align}
  E[(\frac{1}{T_{i}}a_{i}'W_{i}Q_{i}^{\dagger}B_{i}'u_{i})^{2}]
= & E[\|E[u_{i}u_{i}'|X_{i}]^{1/2}\frac{1}{T_{i}}B_{i}Q_{i}^{\dagger}W_{i}'a_{i}\|^{2}]\nonumber \\
\leq & cE[\|\frac{1}{T_{i}}B_{i}Q_{i}^{\dagger}W_{i}'a_{i}\|^{2}]\nonumber \\
= & cE[\frac{1}{T_{i}}a_{i}'W_{i}Q_{i}^{\dagger}W_{i}'a_{i}]\nonumber \\
\leq & c^{3}E[\|W_{i}(Q_{i}^{\dagger})^{1/2}\|^{2}/T_{i}],\label{eq:ubound}
\end{align}
where the second line uses the law of iterated expectations. In addition,
note that by Assumptions 3.i and 3.iii, $Var(a_{i}'\beta_{i})\leq2c^{2}$, so 
$
E[\epsilon_{i}^{2}]\leq c^{3}E[\|W_{i}(Q_{i}^{\dagger})^{1/2}\|^{2}/T_{i}]+2c^{2}$. So by the law of large numbers:
\begin{align*}
\frac{1}{n}\sum_{i=1}^{n}\epsilon_{i} & =O_{p}\bigg(\sqrt{\frac{E[\|W_{i}(Q_{i}^{\dagger})^{1/2}\|^{2}/T_{i}]}{n}}+n^{-1/2}\bigg) =O_{p}\bigg(\sqrt{\frac{\gamma_{n}}{nT}}+n^{-1/2}\bigg),
\end{align*}
where the final equality follows by Assumption 3.vii.

\subsection*{Convergence Rate}

Putting everything together,  we see that
\begin{align*}
 & \hat{\theta}-\theta_{0}
=  O_{p}\bigg(n^{-1/2}+\kappa_{n}+(1+\sqrt{J}\kappa_{n})\sqrt{\frac{\gamma_{n}}{nT}}+(1+\sqrt{\gamma_{n}})\ell\bigg).
\end{align*}
So for root-$n$ convergence it suffices that $\kappa_{n},\ell,\ell\sqrt{\gamma_{n}}=o(\sqrt{\frac{1}{n}})$,
$\frac{\gamma_{n}J\kappa_{n}^{2}}{T},\sqrt{\frac{\gamma_{n}}{T}}=o(1)$.

\subsection*{Applying the Central Limit Theorem}

Finally, in order to obtain the final result in the lemma, we apply
the central limit theorem to $\frac{1}{\sqrt{n}}\sum_{i=1}^{n}\epsilon_{i}$
using a Lyapunov condition. Given individuals are drawn iid from the
population, the Lyapunov condition requires that for some $\delta>0$, 
$
n^{-\delta/2}E\big[\|\sigma_{n}^{-1}\epsilon_{i}\|^{2+\delta}\big]$ goes to zero with $n$, 
where $\sigma_{n}$ is the square root of the variance of $\epsilon_{i}$
and is given by:
\begin{align*}
\sigma_{n}^{2} & =E[|\frac{1}{T_{i}}a'_{i}W_{i}Q_{i}^{\dagger}B_{i}'u_{i}|^{2}]+Var(a_{i}'\beta_{i}).
\end{align*}

From Assumption 4.ii, we then see $\sigma_{n}^{-1}=O(1)$.
By Jensen's inequality, 
\begin{align*}
n^{-\delta/2}E\big[\|\sigma_{n}^{-1}\epsilon_{i}\|^{2+\delta}\big] & \leq2^{1+\delta}n^{-\delta/2}E\big[|\sigma_{n}^{-1}(a_{i}'\beta_{i}-E[a_{i}'\beta_{i}])|^{2+\delta}\big]\\
 & +2^{1+\delta}n^{-\delta/2}E\big[|\sigma_{n}^{-1}\frac{1}{T_{i}}a'_{i}W_{i}Q_{i}^{\dagger}B_{i}'u_{i}|^{2+\delta}\big].
\end{align*}
And so, it suffices to show that the following two conditions hold:
\begin{align}
n^{-\delta/2}E\big[|a_{i}'\beta_{i}-E[a_{i}'\beta_{i}]|^{2+\delta}\big] & \to0\label{eq:conv1}\\
n^{-\delta/2}E\big[|\frac{1}{T_{i}}a'_{i}W_{i}Q_{i}^{\dagger}B_{i}'u_{i}|^{2+\delta}\big] & \to0\label{eq:conv2}
\end{align}

The first condition follows trivially from Assumptions 3.i and 3.iii.
The second condition requires more work. By H\"older's inequality we
have that for any $0<\delta\leq\alpha$:
\begin{align*}
  E\big[|\frac{1}{T_{i}}a'_{i}W_{i}Q_{i}^{\dagger}B_{i}'u_{i}|^{2+\delta}\big]
\leq & E\big[|\frac{1}{T_{i}}a'_{i}W_{i}Q_{i}^{\dagger}B_{i}'u_{i}|^{2+\alpha}\big]^{\delta/\alpha} E\big[|\frac{1}{T_{i}}a'_{i}W_{i}Q_{i}^{\dagger}B_{i}'u_{i}|^{2}\big]^{1-\delta/\alpha}\\
\leq & E\big[|\frac{1}{T_{i}}a'_{i}W_{i}Q_{i}^{\dagger}B_{i}'u_{i}|^{2+\alpha}\big]^{\delta/\alpha} c^{2(1-\delta/\alpha)}E[\|W_{i}(Q_{i}^{\dagger})^{1/2}\|^{2}/T_{i}]^{1-\delta/\alpha},
\end{align*}
where we have used our earlier result that $E\big[|\frac{1}{T_{i}}a'_{i}W_{i}Q_{i}^{\dagger}B_{i}'u_{i}|^{2}\big]$ is bounded above by $c^{2}E[\|W_{i}(Q_{i}^{\dagger})^{1/2}\|^{2}/T_{i}]$,
which is $O(E[\|W_{i}(Q_{i}^{\dagger})^{1/2}\|^{2}]/T)$ by Assumption 3.vii. Thus for (\ref{eq:conv2}), it is sufficient that:
\[
n^{-1/2}E\big[|\frac{1}{T_{i}}a'_{i}W_{i}Q_{i}^{\dagger}B_{i}'u_{i}|^{2+\alpha}\big]^{1/\alpha}E[\|W_{i}(Q_{i}^{\dagger})^{1/2}\|^{2}/T]^{\frac{1}{\delta}-\frac{1}{\alpha}}\to0
\]
Note that $0<\delta\leq\alpha\iff\frac{1}{\delta}-\frac{1}{\alpha}\geq0$,
and so the above holds for some $0<\delta\leq\alpha$ if for some
$\delta,\alpha>0$:
\[
n^{-1/2}E\big[|\frac{1}{T_{i}}a'_{i}W_{i}Q_{i}^{\dagger}B_{i}'u_{i}|^{2+\alpha}\big]^{1/\alpha}E[\|W_{i}(Q_{i}^{\dagger})^{1/2}\|^{2}/T]^{\delta}\to0
\]
Applying Cauchy-Schwartz we get:
\begin{align*}
  |\frac{1}{T_{i}}a'_{i}W_{i}Q_{i}^{\dagger}B_{i}'u_{i}|^{2}
\leq & \|a_{i}'W_{i}(Q_{i}^{\dagger})^{1/2}\|^{2}\frac{1}{T_{i}}\|u_{i}\|^{2}
\leq  c^{2}\|W_{i}(Q_{i}^{\dagger})^{1/2}\|^{2}\frac{1}{T_{i}}\|u_{i}\|^{2},
\end{align*}
where the final inequality uses Assumption 3.i. And so, again applying H\"older's
inequality, for any $p>1$,
\begin{align*}
  E\big[|\frac{1}{T_{i}}a'_{i}W_{i}Q_{i}^{\dagger}B_{i}'u_{i}|^{2+\alpha}\big]
\leq & c^{2+\alpha}E\bigg[\|W_{i}(Q_{i}^{\dagger})^{1/2}\|^{2+\alpha}(\frac{1}{T_{i}}\|u_{i}\|^{2})^{\frac{2+\alpha}{2}}\bigg]\\
\leq & c^{2+\alpha}E\bigg[\|W_{i}(Q_{i}^{\dagger})^{1/2}\|^{(2+\alpha)p}\bigg]^{\frac{1}{p}}E\bigg[(\frac{1}{T_{i}}\|u_{i}\|^{2})^{(\frac{2+\alpha}{2})\frac{p}{p-1}}\bigg]^{\frac{p-1}{p}}.
\end{align*}
Reparameterizing by $q=(2+\alpha)p$ and $v=(2+\alpha)\frac{p}{p-1}$,
in which case $\alpha=\frac{vq-2q-2v}{q+v}$, we get:
\[
E\big[|\frac{1}{T_{i}}a'_{i}W_{i}Q_{i}^{\dagger}B_{i}'u_{i}|^{2+\alpha}\big]^{\frac{1}{\alpha}}\leq\bigg(cE\bigg[\|W_{i}(Q_{i}^{\dagger})^{1/2}\|^{q}\bigg]^{\frac{1}{q}}E\bigg[(\frac{1}{T_{i}}\|u_{i}\|^{2})^{\frac{v}{2}}\bigg]^{\frac{1}{v}}\bigg)^{\frac{vq}{vq-2q-2v}}.
\]
By Jensen's inequality and using $r>2$ we have:
\begin{align*}
E[(\frac{1}{T_{i}}\|u_{i}\|^{2})^{\frac{r}{2}}] & =E[(\frac{1}{T_{i}}\sum_{t=1}^{T_{i}}u_{it}^{2})^{\frac{r}{2}}]\leq E[\frac{1}{T_{i}}\sum_{t=1}^{T_{i}}u_{it}^{r}].
\end{align*}
By Assumption 4.i the quantity on the RHS is bounded above by $c$,
and so:
\begin{align*}
  n^{-1/2}E\big[|\frac{1}{T_{i}}a'_{i}W_{i}Q_{i}^{\dagger}B_{i}'u_{i}|^{2+\alpha}\big]^{\frac{1}{\alpha}}
\leq & n^{-1/2}\bigg[cE[\|W_{i}(Q_{i}^{\dagger})^{1/2}\|^{q}]^{1/q}\bigg]^{\frac{vq}{vq-2q-2v}}
\end{align*}

So for (\ref{eq:conv2}) it suffices that:
\[
n^{(\frac{1}{v}+\frac{1}{q}-\frac{1}{2})}E[\|W_{i}(Q_{i}^{\dagger})^{1/2}\|^{q}]^{1/q}E[\|W_{i}(Q_{i}^{\dagger})^{1/2}\|^{2}/T]^{\delta(1-\frac{2}{v}-\frac{2}{q})}\to0
\]

Now, recall that $q=(2+\alpha)p$ and $v=(2+\alpha)\frac{p}{p-1}$,
where $\alpha>0$ and $p>1$. We will show that for a given $q$ and
$v$, such an $\alpha$ and $p$ exist if $q,v>0$ and $(q-2)(v-2)>4$.
Fix $q,v>0$ and consider some $\alpha>0$. From the expression for
$q$, we have $p=q/(2+\alpha)$. Substituting out $p$ from the expression
for $v$ and solving for $\alpha$, we get $\alpha=\frac{qv-2q-2v}{q+v}$,
and so, given $q,v>0$ , $\alpha>0$ if and only if $qv-2q-2v>0$, or
equivalently, $(q-2)(v-2)>4$. Moreover, plugging our expression for
$\alpha$ back into the expression for $p$, we get $p=(q+v)/v$,
which is strictly greater than $1$ because $q$ and $v$ are both
strictly positive.

Note also that because $qv-2q-2v>0$ and $v,q>0$, we have that $1-\frac{2}{v}-\frac{2}{q}>0$.
And since the convergence to zero needs only hold for some fixed $\delta>0$, we
can reparameterize again and we see that it suffices that for some
$\delta$: 
\[
n^{(\frac{1}{v}+\frac{1}{q}-\frac{1}{2})}E[\|W_{i}(Q_{i}^{\dagger})^{1/2}\|^{q}]^{1/q}E[\|W_{i}(Q_{i}^{\dagger})^{1/2}\|^{2}/T]^{\delta}\to0,
\]
which holds by supposition.

Having confirmed the Lyapunov condition, we can apply the central limit
theorem to $\frac{1}{\sqrt{n}}\sum_{i=1}^{n}\epsilon_{i}$. We just
need that the squares of the remainder terms go to zero strictly faster
than the variance of $\frac{1}{n}\sum_{i=1}^{n}\epsilon_{i}$, i.e.,
each remainder term must be $o\big(\sqrt{\frac{1}{n}}\big)$. Recall
the remainders have the rate $
\kappa_{n}+\sqrt{\frac{J\gamma_{n}\kappa_{n}^{2}}{nT}}+\ell+\ell\sqrt{\gamma_{n}}
$. 
Hence it suffices that $\kappa_{n},\ell,\ell\sqrt{\gamma_{n}}=o(\sqrt{\frac{1}{n}})$,
$\frac{\gamma_{n}J\kappa_{n}^{2}}{T}=o(1)$.
\end{proof}

\begin{proof}[Proof of Theorem 2]
For our choice of $W_{i}$ we have:
\begin{align*}
\|W_{i}-I\|  =\lambda\|(Q_{i}+\lambda D_{i})^{-1}D_{i}\|
 & \leq\lambda\|(Q_{i}+\lambda D_{i})^{-1}\|\|D_{i}\| \leq c\lambda\|(Q_{i}+\lambda D_{i})^{-1}\|,
\end{align*}
and so $E[\|W_{i}-I\|]=O(\lambda\delta)$. Moreover, we have:
\begin{align*}
\|W_{i}(Q_{i}^{\dagger})^{1/2}\|^{2} & =\|(Q_{i}+\lambda D_{i})^{-1}Q_{i}(Q_{i}+\lambda D_{i})^{-1}\| \leq\|(Q_{i}+\lambda D_{i})^{-1}Q_{i}\|\|(Q_{i}+\lambda D_{i})^{-1}\|
\end{align*}

Now, with some work one can show that:
\begin{align*}
(Q_{i}+\lambda D_{i})^{-1}Q_{i} & =\begin{pmatrix}1 & -\bar{B}_{i}'\\
0 & I
\end{pmatrix}\begin{pmatrix}1 & 0\\
0 & (\tilde{Q}_{i}+\lambda D_{1,i})^{-1}\tilde{Q}_{i}
\end{pmatrix}\begin{pmatrix}1 & \bar{B}_{i}'\\
0 & I
\end{pmatrix},
\end{align*}
and so by properties of the matrix norm:
\begin{align*}
\|(Q_{i}+\lambda D_{i})^{-1}Q_{i}\| & \leq(1+\|\bar{B}_{i}\|)^{2}\big(1+\|(\tilde{Q}_{i}+\lambda D_{1,i})^{-1}\tilde{Q}_{i}\|\big)
\end{align*}

Now, using that $D_{1,i}$ is symmetric and strictly positive definite
by Assumption 3.viii,
\begin{align*}
\|(\tilde{Q}_{i}+\lambda D_{1,i})^{-1}\tilde{Q}_{i}\| & =\|D_{1,i}^{-1/2}\big(D_{1,i}^{-1/2}\tilde{Q}_{i}D_{1,i}^{-1/2}+\lambda I\big)^{-1}D_{1,i}^{-1/2}\tilde{Q}_{i}D_{1,i}^{-1/2}D_{1,i}^{1/2}\|\\
 & \leq\|D_{1,i}^{-1/2}\|\|\big(D_{1,i}^{-1/2}\tilde{Q}_{i}D_{1,i}^{-1/2}+\lambda I\big)^{-1}D_{1,i}^{-1/2}\tilde{Q}_{i}D_{1,i}^{-1/2}\|\|D_{1,i}^{1/2}\|\\
 & \leq\|D_{1,i}^{-1/2}\|\|D_{1,i}^{1/2}\|.
\end{align*}
Where the final line uses that $\|(A+\lambda I)^{-1}A\|\leq1$ for
any positive definite matrix $A$, and $\lambda>0$. By Assumptions
3.i and 3.viii, $\|\bar{B}_{i}\|$ $\|D_{1,i}^{-1/2}\|$, $\|D_{1,i}^{1/2}\|$
are all uniformly bounded, and so, for some constant $C$, $
\|W_{i}(Q_{i}^{\dagger})^{1/2}\|^{2}  \leq C\|(Q_{i}+\lambda D_{i})^{-1}\|$. 
Therefore, $E[\|W_{i}-I\|]=O(\lambda\delta)$, $E[\|W_{i}(Q_{i}^{\dagger})^{1/2}\|^{2}]=O(\delta)$,
and $E[\|W_{i}(Q_{i}^{\dagger})^{1/2}\|^{q}]^{1/q}=O(E[\|(Q_{i}+\lambda D_{i})^{-1}\|^{q/2}]^{1/q})$.
Substituting into Lemma 1 then gives the result.
\end{proof}

\theoremstyle{plain} \newtheorem*{L2}{Lemma 2} 
\begin{L2}
Suppose $\lambda>0$ and with probability $1$, the eigenvalues of
$D_{1,i}$ are all bounded above by $c$ and below by $1/c$, and
$\|\bar{B}_{i}\|\leq c$. Then for any $\alpha>0$:
\[
E[\|(Q_{i}+\lambda D_{i})^{-1}\|^{\alpha}]^{1/\alpha}=O(1+E[(\mu_{\min}(\tilde{Q}_{i})+\lambda)^{-\alpha}]^{1/\alpha}\big).
\]

Thus if $E[\mu_{\min}(\tilde{Q}_{i})^{-\alpha}]^{1/\alpha}<c$ then
$E[\|(Q_{i}+\lambda D_{i})^{-1}\|^{\alpha}]^{1/\alpha}=O(1)$.
\end{L2}
\begin{proof}

We can decompose $(Q_{i}+\lambda D_{i})^{-1}$ into the product of
block matrices as follows:
\begin{align*}
  (Q_{i}+\lambda D_{i})^{-1}
= & \begin{pmatrix}1 & -\bar{B}_{i}'\\
0 & I
\end{pmatrix}\begin{pmatrix}1 & 0\\
0 & (\tilde{Q}_{i}+\lambda D_{1,i})^{-1}
\end{pmatrix}\begin{pmatrix}1 & 0\\
-\bar{B}_{i} & I
\end{pmatrix}.
\end{align*}

By the properties of the Euclidean matrix norm, we then have:
\begin{align*}
\|(Q_{i}+\lambda D_{i})^{-1}\| & \leq(1+\|\bar{B}_{i}\|)^{2}(1+\|(\tilde{Q}_{i}+\lambda D_{1,i})^{-1}\|).
\end{align*}

Note that:
\begin{align*}
\|(\tilde{Q}_{i}+\lambda D_{1,i})^{-1}\| & =\|D_{1,i}^{-1/2}\big(D_{1,i}^{-1/2}\tilde{Q}_{i}D_{1,i}^{-1/2}+\lambda I\big)^{-1}D_{1,i}^{-1/2}\|\\
 & \leq\|\big(D_{1,i}^{-1/2}\tilde{Q}_{i}D_{1,i}^{-1/2}+\lambda I\big)^{-1}\|\|D_{1,i}^{-1/2}\|^{2}\\
 & \leq\frac{1}{\mu_{\min}\big(D_{1,i}^{-1/2}\tilde{Q}_{i}D_{1,i}^{-1/2}\big)+\lambda}\times\frac{1}{\mu_{\min}(D_{1,i}^{1/2})^{2}}\\
 & \leq\frac{1}{\mu_{\min}(D_{1,i}^{-1/2})^{2}\mu_{\min}(\tilde{Q}_{i})+\lambda}\times\frac{1}{\mu_{\min}(D_{1,i}^{1/2})^{2}}\\
 & =\frac{\mu_{\max}(D_{1,i})/\mu_{\min}(D_{1,i})}{\mu_{\min}(\tilde{Q}_{i})+\lambda\mu_{\max}(D_{1,i})}
\end{align*}
Where we have used that $D_{1,i}$ is symmetric and so $\mu_{\min}(D_{1,i}^{-1/2})^{2}=1/\mu_{\max}(D_{1,i})$
and $\mu_{\min}(D_{1,i}^{1/2})^{2}=\mu_{\min}(D_{1,i})$. Combining, we get:
\begin{align*}
\|(Q_{i}+\lambda D_{i})^{-1}\| & \leq(1+\|\bar{B}_{i}\|)^{2}\bigg(1+\frac{\mu_{\max}(D_{1,i})/\mu_{\min}(D_{1,i})}{\mu_{\min}(\tilde{Q}_{i})+\lambda\mu_{\max}(D_{1,i})}\bigg)\\
 & \leq(1+c)^{2}\bigg(1+\frac{c^{2}}{\mu_{\min}(\tilde{Q}_{i})+\lambda/c}\bigg)\\
 & \leq(1+c)^{2}\bigg(1+\frac{c^{3}}{\mu_{\min}(\tilde{Q}_{i})+\lambda}\bigg)
\end{align*}

Where the final line uses $\|\bar{B}_{i}\|\leq c$, $\mu_{\max}(D_{1,i})\leq c$
and $\mu_{\min}(D_{1,i})\geq1/c$ by supposition and we can take $c\geq1$
without loss of generality. So for any $0<\alpha$:
\begin{align*}
E[\|(Q_{i}+\lambda D_{i})^{-1}\|^{\alpha}]^{1/\alpha} & \leq(1+c)^{2}
  +(1+c)^{2}E\bigg[\bigg(\frac{c^{3}}{\mu_{\min}(\tilde{Q}_{i})+\lambda}\bigg)^{\alpha}\bigg]^{1/\alpha},
\end{align*}
and hence:
\[
E[\|(Q_{i}+\lambda D_{i})^{-1}\|^{\alpha}]^{1/\alpha}=O(1+E[(\mu_{\min}(\tilde{Q}_{i})+\lambda)^{-\alpha}]^{1/\alpha}\big)
\]

For the final step simply note that $(\mu_{\min}(\tilde{Q}_{i})+\lambda)^{-\alpha}\leq\mu_{\min}(\tilde{Q}_{i})^{-\alpha}$.
\end{proof}

\begin{proof}[Proof of Corollary 1]
By Lemma 2, $E[\|(Q_{i}+\lambda D_{i})^{-1}\|]=O(1)$. Applying Theorem
2 with $\delta=O(1)$ then gives the result.
\end{proof}

\theoremstyle{plain} \newtheorem*{L3}{Lemma 3} 
\begin{L3}
Suppose $\lambda>0$ and with probability $1$, the eigenvalues of
$D_{1,i}$ are all bounded above by $c$ and below by $1/c$. Suppose
for some $1\leq\alpha$, $\|\bar{B}_{i}\|\leq c$ and $E[\mu_{i}^{-\alpha}]^{1/\alpha}=O(1)$
for some individual-specific random scalar $\mu_{i}$, then for any
fixed $\varepsilon>0$:
\[
E[\|(Q_{i}+\lambda D_{i})^{-1}\|^{\alpha}]^{1/\alpha}\leq O\bigg(1+\lambda^{-1}P[\mu_{\min}(\tilde{Q}_{i})\leq(1-\varepsilon)\mu_{i}]^{1/\alpha}\bigg).
\]
\end{L3}
\begin{proof}
From Lemma 2:
\[
E[\|(Q_{i}+\lambda D_{i})^{-1}\|^{\alpha}]^{1/\alpha}\leq O(1+E[(\mu_{\min}(\tilde{Q}_{i})+\lambda)^{-\alpha}]^{1/\alpha}\big).
\]

Let $\varepsilon\in[0,1]$ and define the binary random variable $\epsilon_{i}$
by $
\epsilon_{i}=1\{\mu_{\min}(\tilde{Q}_{i})\geq(1-\varepsilon)\mu_{i}\}$. 
Using this random variable we get:
\begin{align*}
\frac{1}{\mu_{\min}(\tilde{Q}_{i})+\lambda} & \leq\frac{1}{(1-\varepsilon)\mu_{i}}+(1-\epsilon_{i})\frac{1}{\lambda}.
\end{align*}
Using the triangle inequality and the fact that $\epsilon_{i}$ is
binary we have:
\begin{align*}
E[(\mu_{\min}(\tilde{Q}_{i})+\lambda)^{-\alpha}]^{1/\alpha} & \leq(1-\varepsilon)^{-1}E[\mu_{i}^{-\alpha}]^{1/\alpha}+\frac{E[1-\epsilon_{i}]^{1/\alpha}}{\lambda}.
\end{align*}
By definition, we have $
E[1-\epsilon_{i}]=P[\mu_{\min}(\tilde{Q}_{i})\leq(1-\varepsilon)\mu_{i}]$. 
Substituting gives the result.
\end{proof}

\theoremstyle{plain} \newtheorem*{L4}{Lemma 4} 
\begin{L4}
Suppose $b(X_{it})=(1,X_{it})'$ where $X_{it}$ is binary. Suppose
$P(X_{it}=1|\pi_{i})=\pi_{i}$ and the entries of the sequence $\{X_{it}\}_{t=1}^{T_i}$
are jointly independent conditional on $\pi_{i}$ and $T_i\geq T$ for all $i$. Let $\pi_{i}$
admit a probability density $f_{\pi}$ so that $f_{\pi}(x)\leq C(1-x)^{\omega}x{}^{\omega}$
for some $C$. If $\omega>0$ then $E[\pi_{i}^{-1}(1-\pi_{i})^{-1}]<\infty$
and
\[
E[\mu_{\min}(\tilde{Q}_{i})\leq(1-\varepsilon)\pi_{i}(1-\pi_{i})]=O(T^{-(1+\omega)}).
\]

If in addition, $\omega>q/2-1$ for some $q>2$, then $E[\pi_{i}^{-q/2}(1-\pi_{i})^{-q/2}]<\infty$.
\end{L4}
\begin{proof}
Let $\bar{X}_{i}=\frac{1}{T_{i}}\sum_{t=1}^{T_{i}}X_{it}$. In this
case $\mu_{\min}(\tilde{Q}_{i})=\bar{X}_{i}(1-\bar{X}_{i})$ so we
have:
\begin{align*}
 & P[\mu_{\min}(\tilde{Q}_{i})\leq(1-\varepsilon)\pi_{i}(1-\pi_{i})|\pi_{i}]\\
= & P[\bar{X}_{i}(1-\bar{X}_{i})\leq(1-\varepsilon)\pi_{i}(1-\pi_{i})|\pi_{i}]\\
\leq & P[\bar{X}_{i}\leq\sqrt{1-\varepsilon}\pi_{i}|\pi_{i}]+P[(1-\bar{X}_{i})\leq\sqrt{1-\varepsilon}(1-\pi_{i})|\pi_{i}].
\end{align*}

Let $\tilde{\varepsilon}=1-\sqrt{1-\varepsilon}$. By the multiplicative
Chernoff bound, for $0\leq\tilde{\varepsilon}\leq1$:
\[
P[\bar{X}_{i}\leq(1-\tilde{\varepsilon})\pi_{i}|\pi_{i}]\leq \exp(-\tilde{\varepsilon}^{2}\pi_{i}T_{i}/2).
\]

By supposition, the pdf of $\pi_{i}$ is bounded above by $C(1-\pi)^{\omega}\pi{}^{\omega}$
with $\omega>0$. It is easy to see that $E[\pi_{i}^{-1}(1-\pi_{i})^{-1}]<\infty$
and if $\omega>q/2-1$, then $E[\pi_{i}^{-q/2}(1-\pi_{i})^{-q/2}]<\infty$.
Moreover, from the above, and $T_i\geq T$, we get:
\begin{align*}
E\big[P[\bar{X}_{i}\leq\sqrt{1-\varepsilon}\pi_{i}|\pi_{i}]\big] & \leq C\int_{0}^{1}y^{\omega}(1-y)^{\omega}exp(-\tilde{\varepsilon}^{2}yT/2)dy\\
 & \leq C\int_{0}^{1}y^{\omega}exp(-\tilde{\varepsilon}^{2}yT/2)dy\\
 & =CT^{-(1+\omega)}(\tilde{\varepsilon}^{2}/2)^{-(1+\omega)}\int_{0}^{\tilde{\varepsilon}^{2}T/2}u^{\omega}exp(-u)du\\
 & \leq CT^{-(1+\omega)}(\tilde{\varepsilon}^{2}/2)^{-(1+\omega)}\int_{0}^{\infty}u^{\omega}exp(-u)du.
\end{align*}

The integral $\int_{0}^{\infty}u^{\omega}exp(-u)du$ is finite for
$\omega>-1$ (it is the gamma function evaluated at $\omega+1$),
and so we see $
E\big[P[\bar{X}_{i}\leq\sqrt{1-\varepsilon}\pi_{i}|\pi_{i}]\big]=O(T^{-(1+\omega)})
$.

We can apply the same reasoning for $E\big[P[(1-\bar{X}_{i})\leq\sqrt{1-\varepsilon}(1-\pi_{i})|\pi_{i}]\big]$.
Combining and using iterated expectations, we get that:
\begin{align*}
P\big(\mu_{\min}(\tilde{Q}_{i})\leq(1-\varepsilon)\mu_{i}\big) & =E\big[P[\mu_{\min}(\tilde{Q}_{i})\leq(1-\varepsilon)\mu_{i}|\pi_{i}]\big] =O(T^{-(1+\omega)}).
\end{align*}
\end{proof}

\begin{proof}[Proof of Corollary 2 (Binary Regressor)]

From Lemma 4 we see $E[\pi_{i}^{-1}(1-\pi_{i})^{-1}]\leq C$. Applying
Lemma 3 with $\alpha=1$, we then get:
\[
E[\|(Q_{i}+\lambda D_{i})^{-1}\|]\leq O\bigg(1+\lambda^{-1}P[\mu_{\min}(\tilde{Q}_{i})\leq(1-\varepsilon)\pi_{i}(1-\pi_{i})]\bigg).
\]

Using Lemma 4, we then have $
E[\|(Q_{i}+\lambda D_{i})^{-1}\|]= O\big(1+\lambda^{-1}T^{-(1+\omega)}\big)
$. Therefore, we get that if $\lambda=o(1)$ and $T\to\infty$, then
$\lambda\delta=o(1)$. $J$ is fixed and clearly $r_{it}=0$ almost
surely because the model is exhaustive (so $\ell=0$). And so, combining
with Theorem 2 gives the first result (where we have simplified the
rate using the fact that $\lambda\kappa_{n}\sqrt{\frac{\kappa_{n}}{nT}}$
is dominated). By Lemma 4, if $\omega>q/2$, we have $E[\pi_{i}^{-q/2}(1-\pi_{i})^{-q/2}]\leq C$.
Then applying Lemma 3 with $\alpha=q/2$, we get:
\[
E[\|(Q_{i}+\lambda D_{i})^{-1}\|^{q/2}]^{\frac{1}{q}}\leq O\bigg(1+\big(\lambda^{-1}T^{-(1+\omega)}\big)^{q/2}\bigg).
\]

The above is $O(1)$ by supposition, and thus the condition in Assumption
4.iii becomes $
n^{(\frac{1}{v}+\frac{1}{q}-\frac{1}{2})}(1/T)^{\delta}=o(1)
$. But this holds trivially. In addition, we assume $\frac{T^{-(1+v)}}{\lambda}=O(1)$
and thus $\delta=O(1)$ and  $T^{-(1+v)},\lambda=o(\sqrt{1/n})$,
so applying Theorem 2 we are done.

\end{proof}

%% file: panel_cites.bib
@misc{moon2026,
      title={Panel Data Estimation of Individual Demand in Markets with Many Consumers}, 
      author={Sarah Moon and Whitney K. Newey},
      year={2026},
      eprint={2606.11047},
      archivePrefix={arXiv},
      primaryClass={econ.EM},
      url={https://arxiv.org/abs/2606.11047}, 
}

@article{Jin2026,
    author = {Jin, Sainan and Lu, Xun and Su, Liangjun},
    title = {Two-Way Mean Group Estimators for Heterogeneous Panel Models with Fixed T},
    journal = {The Econometrics Journal},
    pages = {utag010},
    year = {2026},
    month = {04},
    issn = {1368-4221},
    doi = {10.1093/ectj/utag010},
    url = {https://doi.org/10.1093/ectj/utag010},
    eprint = {https://academic.oup.com/ectj/advance-article-pdf/doi/10.1093/ectj/utag010/67725565/utag010.pdf},
}

@misc{newey2025,
      title={Identification of Treatment Effects under Limited Exogenous Variation}, 
      author={Whitney K. Newey and Sami Stouli},
      year={2025},
      eprint={1811.09837},
      archivePrefix={arXiv},
      primaryClass={econ.EM},
      url={https://arxiv.org/abs/1811.09837}, 
}

@book{Jackson1911,
author = {Jackson, Dunham},
language = {ger},
location = {Göttingen},
publisher = {Dieterich},
title = {Über die Genauigkeit der Annäherung stetiger Funktionen durch ganze rationale Funktionen gegebenen Grades und trigonometrische Summen gegebener Ordnung},
url = {http://eudml.org/doc/204248},
year = {1911},
}

@article{Fernandez2021,
    author = {Fernández-Val, Iván and Freeman, Hugo and Weidner, Martin},
    title = {Low-rank approximations of nonseparable panel models},
    journal = {The Econometrics Journal},
    volume = {24},
    number = {2},
    pages = {C40-C77},
    year = {2021},
    month = {05},
    issn = {1368-4221},
    doi = {10.1093/ectj/utab007},
    url = {https://doi.org/10.1093/ectj/utab007},
    eprint = {https://academic.oup.com/ectj/article-pdf/24/2/C40/46560665/utab007.pdf},
}

@article{Bonhomme2022,
author = {Bonhomme, Stéphane and Lamadon, Thibaut and Manresa, Elena},
title = {Discretizing Unobserved Heterogeneity},
journal = {Econometrica},
volume = {90},
number = {2},
pages = {625-643},
doi = {https://doi.org/10.3982/ECTA15238},
url = {https://onlinelibrary.wiley.com/doi/abs/10.3982/ECTA15238},
eprint = {https://onlinelibrary.wiley.com/doi/pdf/10.3982/ECTA15238},
year = {2022}
}

@article{Bonhomme2015,
 ISSN = {00129682, 14680262},
 URL = {http://www.jstor.org/stable/43616962},
 author = {Stéphane Bonhomme and Elena Manresa},
 journal = {Econometrica},
 number = {3},
 pages = {1147--1184},
 publisher = {[Wiley, The Econometric Society]},
 title = {GROUPED PATTERNS OF HETEROGENEITY IN PANEL DATA},
 urldate = {2026-06-21},
 volume = {83},
 year = {2015}
}

@misc{blomquist2024,
      title={Panel Estimation of Taxable Income Elasticities with Heterogeneity and Endogenous Budget Sets}, 
      author={Soren Blomquist and Anil Kumar and Whitney K. Newey},
      year={2024},
      eprint={2501.00633},
      archivePrefix={arXiv},
      primaryClass={econ.EM},
      url={https://arxiv.org/abs/2501.00633}, 
}

@Article{Manski1987,
  author  = {Charles F. Manski},
  journal = {Econometrica},
  title   = {Semiparametric Analysis of Random Effects Linear Models from Binary Panel Data},
  year    = {1987},
  issn    = {0012-9682},
  pages   = {357-362},
  volume  = {55},
  doi     = {10.2307/1913240},
}

@article{Pesaran,
author = {M. Hashem Pesaran and Yongcheol Shin and Ron P. Smith},
title = {Pooled Mean Group Estimation of Dynamic Heterogeneous Panels},
journal = {Journal of the American Statistical Association},
volume = {94},
number = {446},
pages = {621--634},
year = {1999},
publisher = {ASA Website},
doi = {10.1080/01621459.1999.10474156},


URL = { 
    
    
        https://www.tandfonline.com/doi/abs/10.1080/01621459.1999.10474156
    

},
eprint = { 
    
    
        https://www.tandfonline.com/doi/pdf/10.1080/01621459.1999.10474156
    

}

}

@misc{kwon2025,
      title={Estimating Treatment Effects Under Bounded Heterogeneity}, 
      author={Soonwoo Kwon and Liyang Sun},
      year={2025},
      eprint={2510.05454},
      archivePrefix={arXiv},
      primaryClass={econ.EM},
      url={https://arxiv.org/abs/2510.05454}, 
}

@misc{fernandezval2025,
      title={Dynamic Heterogeneous Distribution Regression Panel Models, with an Application to Labor Income Processes}, 
      author={Ivan Fernandez-Val and Wayne Yuan Gao and Yuan Liao and Francis Vella},
      year={2025},
      eprint={2202.04154},
      archivePrefix={arXiv},
      primaryClass={econ.EM},
      url={https://arxiv.org/abs/2202.04154}, 
}

@Article{Birge2001,
 ISSN = {07492170},
 URL = {http://www.jstor.org/stable/4356108},
 author = {Lucien Birgé},
 journal = {Lecture Notes-Monograph Series},
 pages = {113--133},
 publisher = {Institute of Mathematical Statistics},
 title = {An Alternative Point of View on Lepski's Method},
 urldate = {2025-09-24},
 volume = {36},
 year = {2001}
}

@Article{Chamberlain1982,
  author  = {Gary Chamberlain},
  journal = {Journal of Econometrics},
  title   = {Multivariate regression models for panel data},
  year    = {1982},
  issn    = {0304-4076},
  pages   = {5-46},
  volume  = {18},
  doi     = {10.1016/0304-4076(82)90094-x},
}

@Article{Chamberlain1992,
  author  = {Gary Chamberlain},
  journal = {Econometrica},
  title   = {Efficiency Bounds for Semiparametric Regression},
  year    = {1992},
  issn    = {0012-9682},
  pages   = {567-596},
  volume  = {60},
  doi     = {10.2307/2951584},
}

@Article{Chernozhukov2013,
  author   = {Chernozhukov, Victor and Fern{\'a}ndez-Val, Iv{\'a}n and Hahn, Jinyong and Newey, Whitney},
  journal  = {Econometrica},
  title    = {Average and quantile effects in nonseparable panel models},
  year     = {2013},
  issn     = {0012-9682},
  number   = {2},
  pages    = {535--580},
  volume   = {81},
  doi      = {10.3982/ECTA8405},
  language = {English},
  zbl      = {1274.62580},
  zbmath   = {6222247},
}

@Article{Pesaran1995,
  author   = {Pesaran, M. Hashem and Smith, Ron},
  journal  = {Journal of Econometrics},
  title    = {Estimating long-run relationships from dynamic heterogeneous panels},
  year     = {1995},
  issn     = {0304-4076},
  number   = {1},
  pages    = {79--113},
  volume   = {68},
  doi      = {10.1016/0304-4076(94)01644-F},
  language = {English},
  zbl      = {0832.62104},
  zbmath   = {803337},
}

@Article{Graham2012,
  author   = {Graham, Bryan S. and Powell, James L.},
  journal  = {Econometrica},
  title    = {Identification and estimation of average partial effects in ``irregular'' correlated random coefficient panel data models},
  year     = {2012},
  issn     = {0012-9682},
  number   = {5},
  pages    = {2105--2152},
  volume   = {80},
  doi      = {10.3982/ECTA8220},
  language = {English},
  zbl      = {1274.62368},
  zbmath   = {6224050},
}

@Article{Wooldridge2005,
  author   = {Wooldridge, Jeffrey M.},
  journal  = {Econometric Theory},
  title    = {Instrumental variables estimation with panel data},
  year     = {2005},
  issn     = {0266-4666},
  number   = {4},
  pages    = {865--869},
  volume   = {21},
  doi      = {10.1017/S0266466605050437},
  language = {English},
  zbl      = {1083.62118},
  zbmath   = {5014418},
}

@Article{Arellano,
  author  = {Manuel Arellano and Stéphane Bonhomme},
  journal = {Review of Economic Studies},
  title   = {Identifying distributional characteristics in random coefficients panel data models},
  year    = {2012},
  pages   = {987–1020},
  volume  = {79},
  doi     = {10.1920/wp.cem.2009.2209},
}

@Article{Honore1992,
  author  = {Bo E. Honore},
  journal = {Econometrica},
  title   = {Trimmed Lad and Least Squares Estimation of Truncated and Censored Regression Models with Fixed Effects},
  year    = {1992},
  issn    = {0012-9682},
  pages   = {533-565},
  volume  = {60},
  doi     = {10.2307/2951583},
}

@Article{Abrevaya2000,
  author  = {Jason Abrevaya},
  journal = {Journal of Econometrics},
  title   = {Rank estimation of a generalized fixed-effects regression model},
  year    = {2000},
  issn    = {0304-4076},
  pages   = {1-23},
  volume  = {95},
  doi     = {10.1016/s0304-4076(99)00027-5},
}

@Article{Hoderlein2012,
  author  = {Stefan Hoderlein and Halbert White},
  journal = {Journal of Econometrics},
  title   = {Nonparametric identification in nonseparable panel data models with generalized fixed effects},
  year    = {2012},
  issn    = {0304-4076},
  pages   = {300-314},
  volume  = {168},
  doi     = {10.1016/j.jeconom.2012.01.033},
}

@Article{Shi,
  author  = {Xiaoxia Shi and Matthew Shum and Wei Song},
  journal = {Econometrica},
  title   = {Estimating Semi-Parametric Panel Multinomial Choice Models Using Cyclic Monotonicity},
  year    = {2018},
  issn    = {0012-9682},
  pages   = {737-761},
  volume  = {86},
  doi     = {10.3982/ecta14115},
}

@Article{Pakes,
  author  = {Ariel Pakes and Jack Porter},
  journal = {Quantitative Economics},
  title   = {Moment inequalities for multinomial choice with fixed effects},
  year    = {2024},
  issn    = {1759-7323},
  pages   = {1-25},
  volume  = {15},
  doi     = {10.3982/qe1776},
}

@Article{Altonji2005,
  author  = {Joseph G. Altonji and Rosa L. Matzkin},
  journal = {Econometrica},
  title   = {Cross Section and Panel Data Estimators for Nonseparable Models with Endogenous Regressors},
  year    = {2005},
  issn    = {0012-9682},
  pages   = {1053-1102},
  volume  = {73},
  doi     = {10.1111/j.1468-0262.2005.00609.x},
}

@Article{Semenova,
  author  = {Vira Semenova and Matt Goldman and Victor Chernozhukov and Matt Taddy},
  journal = {Quantitative Economics},
  title   = {Inference on heterogeneous treatment effects in high‐dimensional dynamic panels under weak dependence},
  year    = {2023},
  issn    = {1759-7323},
  pages   = {471-510},
  volume  = {14},
  doi     = {10.3982/qe1670},
}

@Article{Imbens,
  author  = {Guido Imbens and Whitney Newey},
  journal = {Econometrica},
  title   = {Identification and Estimation of Triangular Simultaneous Equations Models Without Additivity},
  year    = {2009},
  pages   = {1481–1512},
  volume  = {77},
  doi     = {10.3386/t0285},
}

@Article{Torgovitsky,
  author  = {Alexander Torgovitsky},
  journal = {Econometrica},
  title   = {Nonparametric Inference on State Dependence in Unemployment},
  year    = {2019},
  issn    = {1556-5068},
  pages   = {1475–1505},
  volume  = {87},
  doi     = {10.2139/ssrn.3332088},
}

@Article{Hausman,
  author  = {Jerry A. Hausman and Whitney K. Newey},
  journal = {Econometrica},
  title   = {Individual Heterogeneity and Average Welfare},
  year    = {2016},
  issn    = {0012-9682},
  pages   = {1225-1248},
  volume  = {84},
  doi     = {10.3982/ecta11899},
}

@Book{Deaton1980,
  author    = {Angus Deaton and John Muellbauer},
  publisher = {Cambridge:Cambridge University Press},
  title     = {Economics and Consumer Behavior},
  year      = {1980},
  doi       = {10.1017/cbo9780511805653},
}

@Article{Chaudhuri2006,
  author  = {Shubham Chaudhuri and Pinelopi K. Goldberg and Panle Jia},
  journal = {American Economic Review},
  title   = {Estimating the Effects of Global Patent Protection in Pharmaceuticals: A Case Study of Quinolones in India},
  year    = {2006},
  issn    = {0002-8282},
  pages   = {1477-1514},
  volume  = {96},
  doi     = {10.1257/aer.96.5.1477},
}

@article{hsiao2021,
author = {Hsiao, Allan},
title = {Coordination and Commitment in International Climate Action: Evidence From Palm Oil},
journal = {Econometrica},
volume = {94},
number = {1},
pages = {1-33},
doi = {https://doi.org/10.3982/ECTA20608},
url = {https://onlinelibrary.wiley.com/doi/abs/10.3982/ECTA20608},
eprint = {https://onlinelibrary.wiley.com/doi/pdf/10.3982/ECTA20608},
year = {2026}
}

@Article{McFadden2005,
  author  = {Daniel L. McFadden},
  journal = {Economic Theory},
  title   = {Revealed stochastic preference: a synthesis},
  year    = {2005},
  issn    = {0938-2259},
  pages   = {245-264},
  volume  = {26},
  doi     = {10.1007/s00199-004-0495-3},
}

@Article{Kitamura,
  author  = {Yuichi Kitamura and Jörg Stoye},
  journal = {Econometrica},
  title   = {Nonparametric Analysis of Random Utility Models},
  year    = {2018},
  issn    = {0012-9682},
  pages   = {1883-1909},
  volume  = {86},
  doi     = {10.3982/ecta14478},
}

@Article{Lewbel2001,
  author  = {Arthur Lewbel},
  journal = {American Economic Review},
  title   = {Demand Systems With and Without Errors},
  year    = {2001},
  issn    = {0002-8282},
  pages   = {611-618},
  volume  = {91},
  doi     = {10.1257/aer.91.3.611},
}

@Article{Blomquista,
  author  = {Soren Blomquist and Anil Kumar and Che-Yuan Liang and Whitney K. Newey},
  journal = {CEMMAP working paper 21/14},
  title   = {Individual Heterogeneity, Nonlinear Budget Sets, and Taxable Income},
  year    = {2014},
  issn    = {1556-5068},
  doi     = {10.2139/ssrn.2603189},
}

@Article{Blundell2014,
  author  = {Richard Blundell and Dennis Kristensen and Rosa Matzkin},
  journal = {Journal of Econometrics},
  title   = {Bounding quantile demand functions using revealed preference inequalities},
  year    = {2014},
  issn    = {0304-4076},
  pages   = {112-127},
  volume  = {179},
  doi     = {10.1016/j.jeconom.2014.01.005},
}

@Article{Hoderlein2014,
  author  = {Stefan Hoderlein and Jörg Stoye},
  journal = {Review of Economics and Statistics},
  title   = {Revealed Preferences in a Heterogeneous Population},
  year    = {2014},
  issn    = {0034-6535},
  pages   = {197-213},
  volume  = {96},
  doi     = {10.1162/rest_a_00397},
}

@Article{Bhattacharya,
  author  = {Debopam Bhattacharya},
  journal = {Econometrica},
  title   = {Nonparametric Welfare Analysis for Discrete Choice},
  year    = {2015},
  issn    = {0012-9682},
  pages   = {617-649},
  volume  = {83},
  doi     = {10.3982/ecta12574},
}

@Article{Dette2016,
  author  = {Holger Dette and Stefan Hoderlein and Natalie Neumeyer},
  journal = {Journal of Econometrics},
  title   = {Testing multivariate economic restrictions using quantiles: The example of Slutsky negative semidefiniteness},
  year    = {2016},
  issn    = {0304-4076},
  pages   = {129-144},
  volume  = {191},
  doi     = {10.1016/j.jeconom.2015.07.004},
}

@Article{Berry1994,
  author  = {Steven T. Berry},
  journal = {RAND Journal of Economics},
  title   = {Estimating Discrete-Choice Models of Product Differentiation},
  year    = {1994},
  issn    = {0741-6261},
  pages   = {242},
  volume  = {25},
  doi     = {10.2307/2555829},
}

@Article{Berry1995,
  author  = {Steven Berry and James Levinsohn and Ariel Pakes},
  journal = {Econometrica},
  title   = {Automobile Prices in Market Equilibrium},
  year    = {1995},
  issn    = {0012-9682},
  pages   = {841},
  volume  = {63},
  doi     = {10.2307/2171802},
}

@InCollection{Chamberlain1984,
  author    = {Gary Chamberlain},
  title   = {Chapter 22 Panel data},
  editor    = {Zvi Griliches and Michael D. Intriligator},
  pages     = {1247-1318},
  publisher = {North-Holland},
  booktitle     = {Handbook of Econometrics, Volume 2},
  year      = {1984},
  doi       = {10.1016/s1573-4412(84)02014-6},
  issn      = {1573-4412},
}

@InCollection{Hausmana,
  author    = {Jerry Hausman},
  title   = {Valuation of New Goods under Perfect and Imperfect Competition},
  editor    = {Timothy F. Bresnahan and Robert J. Gordon},
  pages     = {209-237},
  publisher = {University of Chicago Press},
  booktitle     = {The Economics of New Goods},
  year      = {1997},
  doi       = {10.3386/w4970},
}

@Misc{crawford2019,
  author  = {Ian Crawford},
  title   = {Nonparametric Analysis of Labour Supply Using Random Fields},
  year    = {2019},
  journal = {Economics Papers No 2019-W06, Economics Group, Nuffield College, University of Oxford},
  url     = {https://EconPapers.repec.org/RePEc:nuf:econwp:1906},
}

@Article{Hendel2006,
  author  = {Igal Hendel and Aviv Nevo},
  journal = {Econometrica},
  title   = {Measuring the Implications of Sales and Consumer Inventory Behavior},
  year    = {2006},
  issn    = {0012-9682},
  pages   = {1637-1673},
  volume  = {74},
  doi     = {10.1111/j.1468-0262.2006.00721.x},
}

@Article{Hausman1981,
  author  = {Hausman, Jerry A.},
  journal = {American Economic Review},
  title   = {Exact consumer's surplus and deadweight loss},
  year    = {1981},
  pages   = {662-676},
  volume  = {4},
  ppn_gvk = {484620916},
}

@Article{Hoderlein2012a,
  author  = {Stefan Hoderlein and Arthur Lewbel},
  journal = {Econometric Theory},
  title   = {REGRESSOR DIMENSION REDUCTION WITH ECONOMIC CONSTRAINTS: THE EXAMPLE OF DEMAND SYSTEMS WITH MANY GOODS},
  year    = {2012},
  issn    = {0266-4666},
  pages   = {1087-1120},
  volume  = {28},
  doi     = {10.1017/s0266466612000412},
}

@Article{Blundell2000,
  author  = {Richard Blundell and Jean-Marc Robin},
  journal = {Econometrica},
  title   = {Latent Separability: Grouping Goods without Weak Separability},
  year    = {2000},
  issn    = {0012-9682},
  pages   = {53-84},
  volume  = {68},
  doi     = {10.1111/1468-0262.00093},
}

@Article{Gorman1959,
  author  = {W. M. Gorman},
  journal = {Econometrica},
  title   = {Separable Utility and Aggregation},
  year    = {1959},
  issn    = {0012-9682},
  pages   = {469},
  volume  = {27},
  doi     = {10.2307/1909472},
}

@InCollection{Gorman1981,
  author    = {W. M. Gorman},
  title  = {Some Engel curves},
  editor    = {Angus Deaton},
  pages     = {7-30},
  publisher = {Cambridge:Cambridge University Press},
  booktitle     = {Essays in the Theory and Measurementof Consumer Behaviour in Honor of Sir Richard Stone},
  year      = {1981},
  doi       = {10.1017/cbo9780511984082.003},
}

@Article{Burda2012,
  author  = {Burda, Martin and Harding, Matthew and Hausman, Jerry A.},
  journal = {Journal of Econometrics},
  title   = {A poisson mixture model of discrete choice},
  year    = {2012},
  pages   = {184-203},
  volume  = {166},
  ppn_gvk = {688520294},
}

@Article{Burda2008,
  author  = {Martin Burda and Matthew Harding and Jerry Hausman},
  journal = {Journal of Econometrics},
  title   = {A Bayesian mixed logit–probit model for multinomial choice},
  year    = {2008},
  issn    = {0304-4076},
  pages   = {232-246},
  volume  = {147},
  doi     = {10.1016/j.jeconom.2008.09.029},
}

@Article{Diewert1976,
  author  = {W. E. Diewert},
  journal = {Journal of Econometrics},
  title   = {Exact and superlative index numbers},
  year    = {1976},
  issn    = {0304-4076},
  pages   = {115-145},
  volume  = {4},
  doi     = {10.1016/0304-4076(76)90009-9},
}

@Article{Harding2017,
  author  = {Matthew Harding and Michael Lovenheim},
  journal = {Journal of Health Economics},
  title   = {The effect of prices on nutrition: Comparing the impact of product- and nutrient-specific taxes},
  year    = {2017},
  issn    = {0167-6296},
  pages   = {53-71},
  volume  = {53},
  doi     = {10.1016/j.jhealeco.2017.02.003},
}

@Article{Allcott2019,
  author  = {Hunt Allcott and Benjamin B. Lockwood and Dmitry Taubinsky},
  journal = {Journal of Economic Perspectives},
  title   = {Should We Tax Sugar-Sweetened Beverages? An Overview of Theory and Evidence},
  year    = {2019},
  issn    = {0895-3309},
  pages   = {202-227},
  volume  = {33},
  doi     = {10.1257/jep.33.3.202},
}

@Article{Dubois2020,
  author  = {Pierre Dubois and Rachel Griffith and Martin O’Connell},
  journal = {American Economic Review},
  title   = {How Well Targeted Are Soda Taxes?},
  year    = {2020},
  issn    = {0002-8282},
  pages   = {3661-3704},
  volume  = {110},
  doi     = {10.1257/aer.20171898},
}

@InCollection{Richter1990,
  author    = {Daniel McFadden and Marcel K. Richter},
  title  = {Stochastic rationality and revealed stochastic preference},
  pages     = {161-186},
  publisher = {Westview Press, Boulder, CO},
  booktitle     = {Preferences, Uncertainty, and Optimality, Essays in Honor of Leo Hurwicz},
  year      = {1990},
}

@Article{Deaton1980a,
  author  = {Deaton, Angus and Muellbauer, John},
  journal = {American Economic Review},
  title   = {An almost ideal demand system},
  year    = {1980},
  pages   = {312-326},
  volume  = {3},
  ppn_gvk = {388232404},
}

@Article{Andreyeva2010,
  author  = {Tatiana Andreyeva and Michael W. Long and Kelly D. Brownell},
  journal = {American Journal of Public Health},
  title   = {The Impact of Food Prices on Consumption: A Systematic Review of Research on the Price Elasticity of Demand for Food},
  year    = {2010},
  issn    = {0090-0036},
  pages   = {216-222},
  volume  = {100},
  doi     = {10.2105/ajph.2008.151415},
}

@Book{DeVore1993,
  author    = {Ronald A. DeVore and George G. Lorentz},
  publisher = {Springer Berlin, Heidelberg},
  title     = {Constructive Approximation},
  year      = {1993},
  doi       = {10.1007/978-3-662-02888-9},
  issn      = {0072-7830},
}
